\documentclass[journal=jacsat,manuscript=article]{achemso}

\usepackage[version=3]{mhchem}
\usepackage{chemformula}
\usepackage[flushleft]{threeparttable}
\usepackage{lineno}
\usepackage{multirow}
\usepackage{array}
\usepackage{dcolumn}
\usepackage{adjustbox}
\usepackage{multirow}
\usepackage{caption}
\usepackage{chemfig}
\usepackage{amsmath}  
\usepackage{helvet}         
\usepackage{sansmath}        

\sansmath                     

\newcommand{\fade}[1]{{\color{gray} #1}}

\author{S. B. White}
\email{sbw33@cam.ac.uk}
\affiliation{\footnotesize Cavendish Laboratory, University of Cambridge, JJ Thomson Avenue, Cambridge, CB3 0HE, UK.}
\author{P. B. Rimmer}
\email{pbr27@cam.ac.uk}
\affiliation{\footnotesize Cavendish Laboratory, University of Cambridge, JJ Thomson Avenue, Cambridge, CB3 0HE, UK.}
\author{Z. Liu}
\affiliation{\footnotesize Department of Earth Sciences, University of Cambridge, Downing St., Cambridge, CB2 3EQ, UK.}

\title[An \textsf{achemso} demo]
  {Shedding a Light on the Kinetics of the Carboxysulfitic Scenario}

\abbreviations{NMR, UV}

\begin{document}

\begin{abstract}

One way in which we can attempt to relate chemical pathways to geochemical environments is by studying the kinetics of a given sequence of reactions and identifying the conditions under which this chemistry is the most productive. Many prebiotic reactions rely on a source of fixed carbon, therefore chemical pathways that suggest prebiotically plausible ways of fixing carbon are of significant interest. One such pathway is the carboxysulfitic reaction network which uses solvated electrons, produced as a result of electron photodetachment from sulfite, to reduce carbon. In this work we explore carboxysulfitic chemistry at three different pH values: 6, 9, and 12. We utilise a new light source, that matches the broadband spectral shape of the young Sun, to irradiate a mixture of bicarbonate and sulfite. We determine the rate equation for the production of formate from these compounds and find the order to be 0.71 $\pm$ 0.12 with respect to bicarbonate and -0.60 $\pm$ 0.10 with respect to sulfite. Following this, we determine rate constants for the production of formate considering two different mechanisms. We find this chemistry to be feasible at all three of the pH values tested, with the magnitude of the rate constants being highly dependent on the assumed mechanism. We suggest that these results may have implications for Mars Sample Return owing to Jezero Crater having had lakes similar to those in which we propose carboxysulfitic chemistry to have been the most productive. Due to Mars' relatively unaltered surface, we propose that Mars Sample Return missions could look for preserved tracers of this chemistry, shedding light on Mars' past conditions and its potential for having hosted life.\\

    
\end{abstract}

\section{Introduction}

The question of how life began has traditionally been considered a matter within the realm of chemistry \cite{Lazcano-1996, Orgel-1998, Cleaves-2013}, however, there is increasing evidence that a more interdisciplinary approach is required \cite{Cockell-2002, Kiang-2018, Preiner-2020, Muller-2022}. Despite this evidence, many laboratory experiments aimed at synthesising life's building blocks have not yet been attempted under prebiotically plausible conditions. This makes it challenging to draw conclusions regarding the feasibility of these mechanisms. Being able to place these complex reactions within the narrative of origin of life research requires knowledge of the environments in which this chemistry is most efficient \cite{Powner-2011}, as well as how it behaves under less optimal conditions. By utilising geochemical conditions to inform laboratory experiments and vice versa, we can attempt to identify promising areas of the parameter space and better focus our research efforts \cite{Sasselov-2020}.

In order to better determine the interplay between prebiotic chemistry and the geochemical environment, we must take a closer look at the physiochemical properties of plausible environments. Included in these environments are pumice rafts \cite{Brasier-2011, Brasier-2013}, subaerial geysers \cite{Westall-2018}, impact craters \cite{Osinski-2020}, volcanic pools \cite{Damer-2020}, alkaline lakes \cite{Toner-2019, Toner-2020}, mudflats \cite{Clark-2020}, deep seated fault zones \cite{Schreiber-2012} and hydrothermal vents \cite{Martin-2008}, including shallow and surface hydrothermal vents \cite{Rimmer-2019, Barge-2022, Rimmer-2024}. One parameter that can be used to compare these environments is pH. The pH of any given system has a role in driving key reactions by facilitating redox processes and determining the speciation of chemical components \cite{Krumbein-1952, Mariani-2018, Barge-2019, Hudson-2020}. Thus, by studying the pH of a system, we can make deductions regarding the chemical landscape of a given environment. Table \ref{tbl:Environmental-pHs} shows the pH range across different environments.

\begin{table}[H]
  \centering
  \begin{threeparttable}
    \caption{\textbf{Environments Proposed as Potential Sites for the Origin of Life and their pH Ranges.}}
    \label{tbl:Environmental-pHs}
    \renewcommand{\arraystretch}{1.5}
    \begin{tabular}{ccc}
      \hline
      \textbf{environment} & \textbf{pH range} & \textbf{reference} \\
      \hline
      Alkaline Lakes & 6.5 -- 9.0 & Toner et al. \cite{Toner-2020} \\
      Marine Hydrothermal Vents (Black Smokers)$^\dagger$ & 3.8 -- 6.3  & Deamer et al. \cite{Deamer-2019} \\
      Marine Hydrothermal Vents (White Smokers)$^\dagger$ & 9.0 -- 9.8  & Deamer et al. \cite{Deamer-2019} \\
      Glaciovolcanic (Surface) Hydrothermal Vents$^\dagger$ & 2.0 -- 7.5  & Cousins et al. \cite{Cousins-2013} \\
      Archean Sea water  & 6.5 -- 7.0 & Halevy \& Bachan \cite{Halevy-2017} \\
      Mudflats$^\dagger$ & 1.4 -- 5.8 & Johnson et al. \cite{Johnson-2020} \\
      \hline
    \end{tabular}
    \begin{tablenotes}
      \item[$\dagger$] Based on a series of measurements from collected samples.
    \end{tablenotes}
  \end{threeparttable}
\end{table}

\begin{figure}[H]
    \centering
    \includegraphics[width=1.0\textwidth]{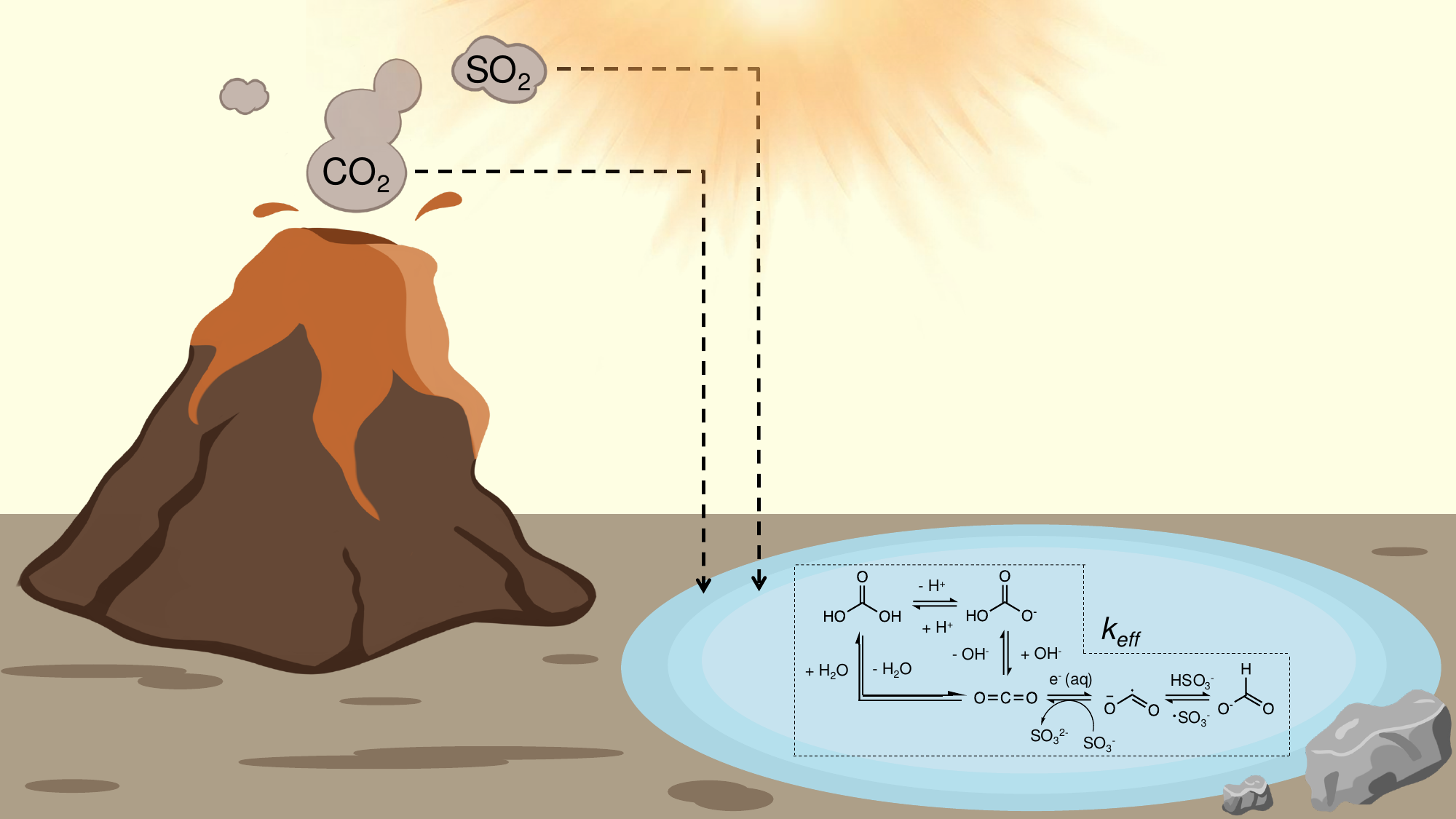}
    \caption{\textbf{The Carboxysulfitic Scenario.} A conceptual diagram linking the carboxysulfitic reaction network and a geochemical environment. Volcanic outgassing supplies \ch{SO2} and \ch{CO2} to the atmosphere which can then dissolve in mildly alkaline lakes.  Dissolution leads to the formation of bicarbonate (\ch{HCO3-}) and sulfite (\ch{SO3^2-}). Photochemical reactions involving UV photons and \ch{SO3^2-} can lead to the production of solvated electrons. These solvated electrons can then be used to reduce \ch{CO2}. The effective rate constant for this multi-step reaction is denoted as $k_{eff}$.}
    \label{fig:CBS}
\end{figure}

One universal feature of all these environments is their access to a source of energy \cite{Deamer-2010}. There are multiple sources of energy thought to have been relevant to prebiotic chemistry on the early Earth \cite{Templeton-2020, Lin-2005, Pastorek-2020, Adam-2018, Segura-2005, Ferus-2015}. In particular, ultraviolet (UV) radiation can drive aqueous photochemistry and has been found to be prebiotically productive in both theory \cite{Pascal-2013} and experiment \cite{Green-2021}, making scenarios that utilise UV radiation very promising \cite{Cnossen-2007}. One application of UV radiation is the generation of highly reactive solvated electrons \cite{Armas-2021, Todd-2022, Todd-2024}. The generation of solvated electrons via UV radiation has featured in several prebiotic scenarios \cite{Sagan-1957, Ponnamperuma-1963, Ritson-2012, Xu-2018}. The plausibility of these scenarios has been investigated, both in terms of the flux of ultraviolet light \cite{Ranjan-2017, Rimmer-2018, Ranjan-2022} and the availability of anions in solution whose electrons can be photodetached \cite{Ranjan-2018,Ranjan-2023}. Sulfite is a notably plausible anion owing to UV photodetachment of its electron being efficient in prebiotic environments \cite{Xu-2018,Rimmer-2018}. In particular, sulfite has been shown to serve as a source of solvated electrons for the reduction of carbon dioxide (\ch{CO2}) in what has been called the carboxysulfitic scenario \cite{Liu-2021}. A network of the reactions involved in this scenario has been mapped highlighting a critical pathway involving glycolate production \cite{Liu-2021}. This pathway proceeds through carboxysulfitic chemistry, yielding essential compounds crucial for central carbon metabolism \cite{Liu-2021}. Carbonate-rich alkaline lakes exposed to UV radiation and fed by redox-neutral volcanic gases are likely to contain carbonate and sulfite making these environments ideal for the carboxysulfitic scenario. A schematic of the mechanism for the reduction of \ch{CO2}, and the environment in which we propose this chemistry to occur, can be seen in Figure \ref{fig:CBS}.

The exploration of prebiotically plausible \ch{CO2} reduction reactions, and how they adapt in response to the acquisition or synthesis of simple organic molecules, has emerged as a key area of research \cite{Nakashima-2018}, with the reduction of \ch{CO2} being of considerable interest to the wider chemistry community \cite{He-2010, Appel-2013, Kornienko-2015, Wang-2021}. In order to relate these chemical pathways discovered in the lab to prebiotically relevant environments, we need to know the rates at which these reactions occur. This approach, which relies on a better understanding of chemical kinetics, has proven invaluable, as evidenced by several studies \cite{Miyakawa-2002, Rimmer-2018, Todd-2019, Rimmer-2021, Todd-2022, Todd-2024}. The plausibility of a prebiotic scenario within a specific geochemical environment will depend on whether the approximate timescale required for a reaction to occur in that environment exceeds the timescales for other reactions or processes responsible for removing a reactant. Ultimately, balancing these reactions will require navigating a complex network of timescales which can only be done by measuring the rate constants associated with each reaction involved in a chemical pathway. Therefore, by determining rate constants we should be able to predict the outcome of our proposed chemistry within a given environmental context.

In this work, we measure the rate constant for the production of formate from the UV photodetachment of electrons from sulfite, followed by their subsequent reaction with carbon species. This represents the first step of the carboxysulfitic scenario \cite{Liu-2021}. In order to investigate this photochemical reaction, we use a broadband light source designed to mimic the spectral properties of the young Sun's light as it would have reached the surface of either the Archean Earth or early Mars. This light source is used to irradiate a mixture of sulfite and bicarbonate. We determine the order of this reaction with respect to each reactant and measure the rate constant for the production of formate at three different pH values: 6, 9, and 12, based on two different mechanisms. We discuss our results in the context of analogous environments, such as Jezero Crater on Mars, which could potentially host preserved tracers of this chemistry. We also propose avenues for future research in anticipation of Mars Sample Return missions.

\section{Methods}
\label{sec:methods}

\subsection{Experimental Setup for Sample Irradiation}

\begin{figure}[H]
    \centering
    \includegraphics[width=1.0\textwidth]{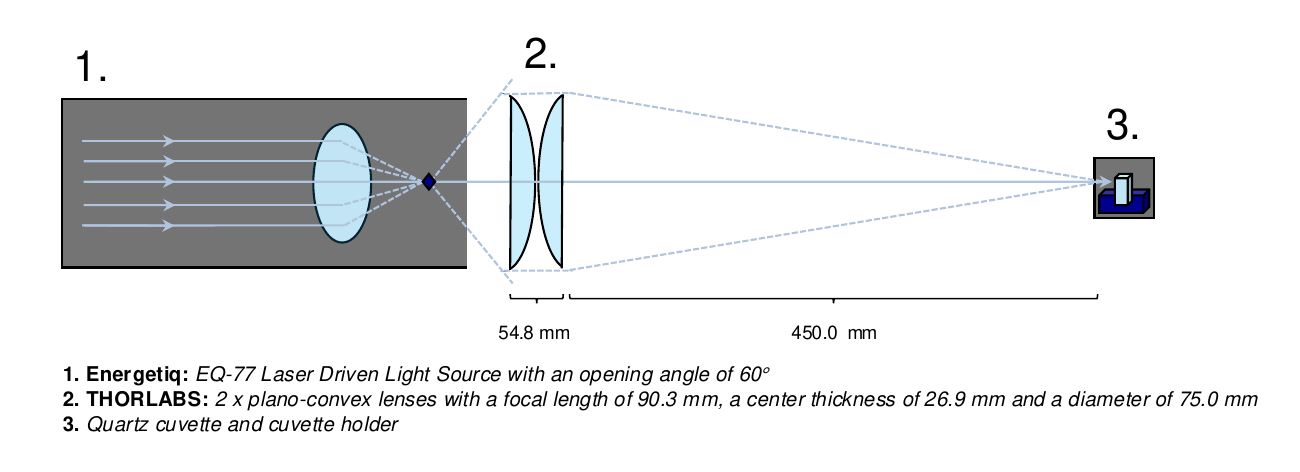}
    \caption{\textbf{The Broadband Laser Driven Light Source Setup.} The light source (EQ-77 LDLS) has a wavelength range from 170 -- 2100 nm with the source windows blocking UV light below 200 nm. The lenses used to focus the light on the sample, which was held in a 2 mL quartz cuvette, are shown.}
    \label{fig:Lights}
\end{figure}

In all experiments a broadband Laser Driven Light Source (LDLS), the Energetiq EQ-77, was used to irradiate the reaction mixture. This source has a wavelength range from 170 -- 2100 nm, with windows designed to transmit light between 200 -- 800 nm, blocking much of the infrared (IR) light above 800 nm and virtually all of the UV light below 200 nm. A schematic for the setup of this light source can be seen in Figure \ref{fig:Lights}.

\begin{figure}[H]
    \centering
    \includegraphics[width=1.0\textwidth]{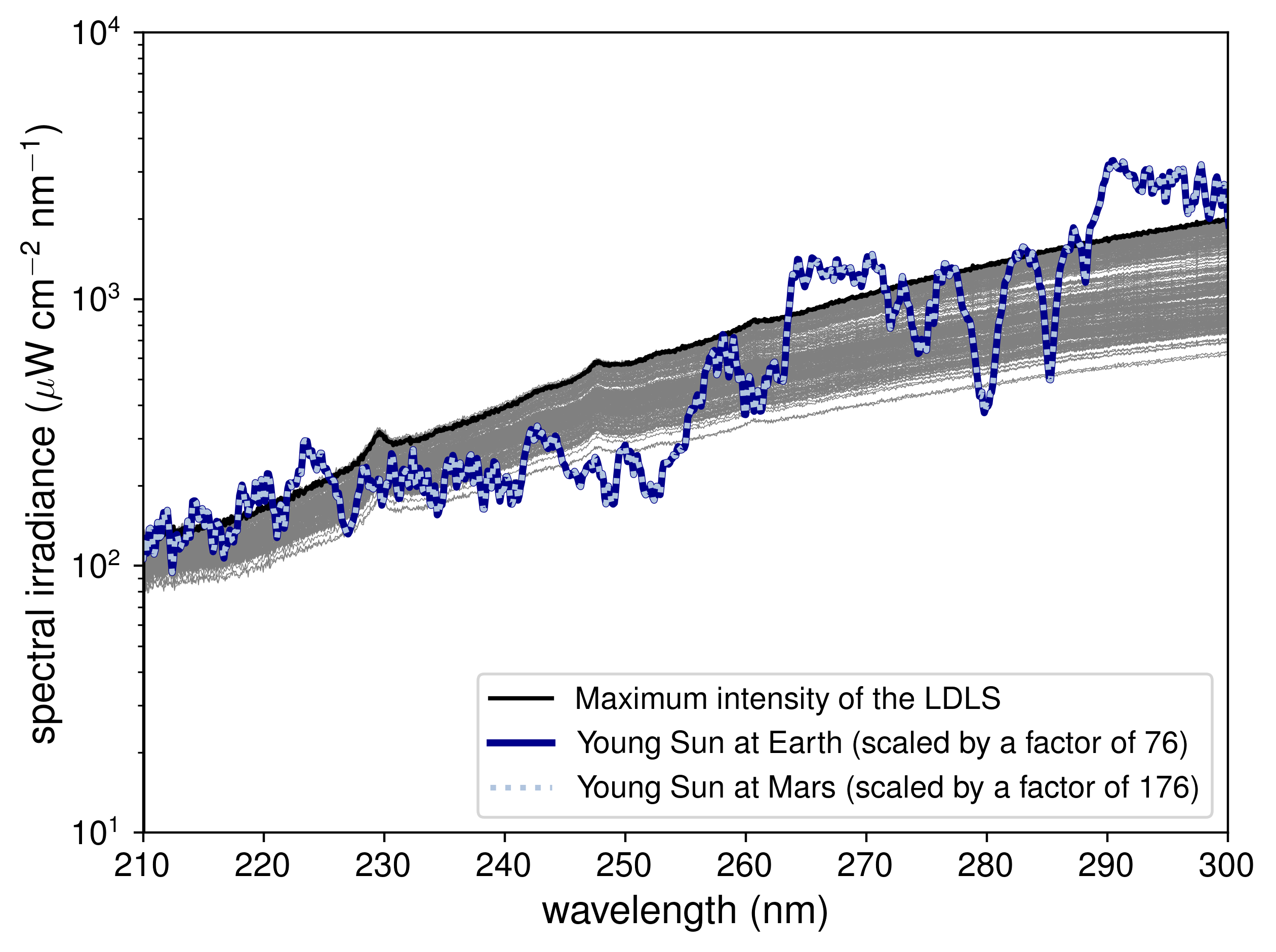}
    \caption{\textbf{Comparative Spectral Irradiance of the LDLS versus the Young Sun}. The measured spectral irradiance (${\rm \mu W \, cm^{-2} \, nm^{-1}}$) as a function of wavelength (nm) of the LDLS EQ-77 at the center of the focus to the edge of the cuvette (see Figure \ref{fig:Lights}) is shown in grey, with the maximum measured irradiance shown in black. We determine scaling factors to place our light source in the context of the young Sun on early Mars and on the Archean Earth. The predicted spectral irradiance for the young Sun at a distance of $1.52 \, {\rm AU}$, scaled by a factor of 176, is shown in light blue. The predicted spectral irradiance for the young Sun at a distance of $1.00 \, {\rm AU}$, scaled by a factor of 76, is shown in dark blue. Both lines match the spectral irradiance of the LDLS as a function of wavelength.}
    \label{fig:LDLS_Spectra}
\end{figure}

The spectral irradiance of the light source was measured using a calibrated OceanInsight Flare-S Spectrometer. The spectrometer was calibrated at a wavelength range from 210 -- 800 nm using a DH-3P-CAL Radiometrically Calibrated Light Source. The calibration protocol does not extend below 210 nm. The spectral irradiance (${\rm \mu W \, cm^{-2} \, nm^{-1}}$) can be compared to the spectral irradiance from the young Sun at $1.52 \, {\rm AU}$, the distance from the Sun to Mars, and at $1.00 \, {\rm AU}$, the distance from the Sun to Earth. To make these comparisons scaling factors were approximated using a predicted solar spectrum \cite{Ribas-2010, Claire-2012}. Scaling factors were determined by matching the spectral irradiance of the LDLS to the solar irradiance from the predicted solar spectrum. Scaling to the Sun on the Archean Earth was done by multiplying the LDLS flux by a factor of 76. Scaling to early Mars required adjusting the solar spectrum to account for the distance from the Sun to Mars and then multiplying the LDLS flux by a factor of 176. The full spectra for the light source and the young Sun can be seen in Figure \ref{fig:LDLS_Spectra}.

\subsection{General Experimental Procedure}
\label{sec:general-procedure}

A solution of \ch{D2O}/\ch{H2O} (1:4) was degassed by bubbling through argon for 30 minutes. A given mass of \ch{Na2SO3} and either $^{13}$C-labelled or non-labelled \ch{NaHCO3} was then dissolved into the degassed solution to create a 2.5 mL solution of known concentrations of \ch{Na2SO3} and \ch{NaHCO3}. The pH of the resultant mixture was tested using an Orion VestaStar Pro Advanced Electrochemistry Meter from Thermo Scientific. For experiments conducted at pH 6 and pH 12, a 150 mM phosphate buffer was used. A phosphate buffer stock solution (150 mM phosphate) was prepared by dissolving \ch{NaH2PO4} (1039.8 mg) and \ch{Na2HPO4} (485.1 mg) in a degassed \ch{D2O}/\ch{H2O} (1:4, 80 mL) solution. To prepare the stock solution for experiments conducted at pH 6, the pH was adjusted to 2.28 by adding degassed \ch{HCl} (6 M) solution. Similarly, to prepare the stock solution for experiments conducted at pH 12, the pH was adjusted to 12.17 by adding a degassed \ch{NaOH} (3 M) solution. After checking the pH of the resultant mixture, an aliquot of 2.0 mL was transferred to a quartz cuvette and irradiated for a given amount of time using the LDLS setup (see Figure \ref{fig:Lights}).

\subsection{Determination of a Rate Equation}

The rate equation for the production of formate from the initial reactants, \ch{Na2SO3} and \ch{NaHCO3}, can be described as follows:
\begin{equation}
\dfrac{d[\ce{HCO2-}]}{dt} = k_{eff}[\ch{Na2SO3}]^n[\ch{NaHCO3}]^m
\label{eq:rate_eq}
\end{equation}
where $d[\ch{HCO2-}]/dt \; {\rm (mM \, s^{-1})}$ refers to the rate of production of formate, $k_{eff} \, {\rm(mM^{1-m-n} s^{-1})}$ is the effective rate constant for this reaction, [$\ch{Na2SO3}$] and [$\ch{NaHCO3}$] are the concentrations of sodium sulfite and sodium bicarbonate respectively (mM), and $m$ and $n$ refer to the order with respect to each reactant.

To calculate rate constants, the order of reaction with respect to each reactant was determined. This was done by varying the concentrations of \ch{Na2SO3} and \ch{NaHCO3}, and measuring $d[\ch{HCO2-}]/dt$. Initial concentrations of 100 mM and 10 mM were chosen for \ch{Na2SO3} and \ch{NaHCO3} respectively in line with the works of Liu et al \cite{Liu-2021}. These concentrations were then changed by up to a factor of 4. The total irradiation time was also varied to ensure the reaction network did not progress past the production of formate (see Table \ref{tbl:orders}).

\begin{table}[H]
  \centering
  \caption{\textbf{Experimental Parameters for Determining Reaction Order.}}
  \label{tbl:orders}
  \renewcommand{\arraystretch}{1.25} 
  \begin{tabular}{ccc}
    \hline
    \multirow{1}{*}{\textbf{[\ch{Na2SO3}]}} & \multirow{1}{*}{\textbf{[\ch{NaHCO3}]}} & \multirow{1}{*}{\textbf{{irradiation time}}} \\
    \textbf{(mM)} & \textbf{(mM)} & \textbf{(minutes)} \\
    \hline
    \phantom{0}25.0 & 10.0 & 120 \\
    \phantom{0}50.0 & 10.0 & \phantom{0}90 \\
    100.0 & 10.0 & 105 \\
    200.0 & 10.0 & \phantom{0}90 \\
    400.0 & 10.0 & \phantom{0}45 \\
    \hline
    100.0 & \phantom{0}2.5 & 120 \\
    100.0 & \phantom{0}5.0 & 120 \\
    100.0 & 20.0 & \phantom{0}75 \\
    100.0 & 40.0 & \phantom{0}60 \\
    \hline
  \end{tabular}
\end{table}

By keeping the concentration of \ch{Na2SO3} constant and varying the concentration of \ch{NaHCO3}, Equation \ref{eq:rate_eq} can be rewritten as follows:
\begin{equation}
\ln\Bigg(\dfrac{d[\ch{HCO2-}]}{dt}\Bigg) = \ln(k_{eff}) + m \ln([\text{NaHCO}_3])
\end{equation}
This equation can then be used to determine the order of reaction with respect to \ch{NaHCO3}. This process was repeated, holding the concentration of \ch{NaHCO3} constant and varying the concentration of \ch{Na2SO3}, to determine the order of reaction with respect to \ch{Na2SO3}. 

\subsection{NMR Spectroscopy}

For each of these experiments, rates of formate production could be determined by extracting concentrations of formate from analysis of $^{13}$C and $^{1}$H Nuclear Magnetic Resonance (NMR) spectra. Concentrations of formate were determined by taking 0.8 mL of the irradiated sample and spiking it with 15 $\mu$L of a 50 mM pentaerythritol standard. 

NMR spectroscopy measurements were conducted using a high-field NMR spectrometer. For most of the experiments the AVIII 500 MHz Spectrometer was used, equipped with a $^{13}$C/$^1$H ``Dual'' helium-cooled cryoprobe, and operated under the Microsoft Windows operating system running Topspin 3.7. The temperature of the probe was consistently set to 298 K for each measurement. Relative integration shows the integral of the standard peak to be directly proportional to the concentration of the standard in solution. This enabled comparison of the intensity of the standard's signal with its known concentration, and the establishment of a proportionality constant denoted as $\kappa \, {\rm (mM)}$. Once determined, this constant was used to calculate the concentration of any products indicated by the $^{1}$H NMR spectrum. The relationship between the integration and concentration, C (mM), can be defined as follows:
\begin{equation}
    C = \kappa \frac{I}{\eta}
\end{equation}
where $I$ is the integral of the peak of interest and $\eta$ is the number of protons in the species of interest, which in this case is the pentaerythritol standard.
Both $^1$H and $^{13}$C NMR spectroscopy were performed. Further details of these techniques and subsequent analysis of NMR spectra can be found in the Supporting Information.

\subsection{Determination of Rate Constants}

After obtaining the order of reaction with respect to each reactant, the rate constant for the production of formate, $k_{eff}$, could be calculated. A stock solution of \ch{NaHCO3} (100 mM) was prepared by dissolving \ch{NaHCO3} (235.2 mg) in a degassed solution of \ch{D2O}/\ch{H2O} (1:4, 28 mL). The reactant solution (2.5 mL) was then prepared by diluting the \ch{NaHCO3} stock solution (10 mM) and adding \ch{Na2SO3} (31.5 mg, 100 mM). The pH of this solution was naturally buffered at a pH of $\sim 9$. The general procedure, as outlined previously, was then followed, irradiating for up to 135 minutes starting from 30 minutes and gathering data points every 30 minutes. These measurements were then used to determine the rate of formate production. To confirm formate as the sole product of this reaction $^{13}$C NMR spectra were analysed (see Figures S22, S31 and S40 in the Supporting Information). These rates were then applied to the previously determined rate equation, along with the measured orders of reaction, to determine the value of $k_{eff}$.  This procedure was carried out for pH values of 6, 9, and 12. When discussing rate constants units are converted from mM$^{1-m-n}$ s$^{-1}$ to M$^{1-m-n}$ yr$^{-1}$ to better contextualise these numbers within this work.

\subsection{Error Estimation}

Most experiments were carried out in triplicate to minimise random error. In all figures the average of these runs was plotted. Error bars represent the standard deviation of each data set. When performing linear regression analysis, all of the acquired data was included with a fixed intercept of zero. To determine the reaction order, the slope of a plot of the natural logarithm of the rate against the natural logarithm of the reactant concentration was used. The standard error of the slope provided the error associated with the reaction order in each case. For the calculation of the rate of formate production, the slope determined from a plot of formate concentration versus time was used. The error in the rate was taken as the standard error of the slope. This method for estimating the standard error of the slope relies on several assumptions: a linear relationship between variables, independence of observations, constant variance of residuals across all values of $x$, and normally distributed residuals. While these assumptions are typically met in such analyses, any deviations could influence the estimated errors.

The rate constants calculated in this work were reported along with a range, representing the absolute minimum and maximum value for each rate constant. This range was determined by assuming that each variable in the rate law could have been erroneous in a manner that maximised the deviation from the true value. The errors associated with the concentration of each reactant in the rate law were not propagated as their impact was assumed to be minimal compared to other sources of error. This method for estimating error allows for the impact of skew and non-Guassian errors in linear space to be accounted for, however, it significantly overestimates the error, as the probability of each variable varying in a way that causes the rate constant to lie at either end of the range is low. This method also ignores any expected codependencies between variables, such as the order with respect to each reactant. This approach highlights the need for research into more accurate methods for uncertainty estimation in this context.

The errors associated with NMR spectroscopy are not quantified, however, three water suppression techniques are employed that are suggested to be sufficient in minimising errors (see the Supporting Information). It is noted that there may be other errors associated with NMR spectroscopy that are not accounted for, however, the accuracy of the quantitative NMR spectroscopy procedure is high as long as the concentrations of the species of interest are sufficiently high ($\geq$ 0.1 mM). It is noted that for experiments producing low quantities of detectable products, there will be an increase in the error due to the detection limits associated with NMR spectroscopy.

\section{Results}
\label{sec:results}

\subsection{The Rate Equation}

\begin{figure}[H]
    \centering
    \includegraphics[width=1.0\textwidth]{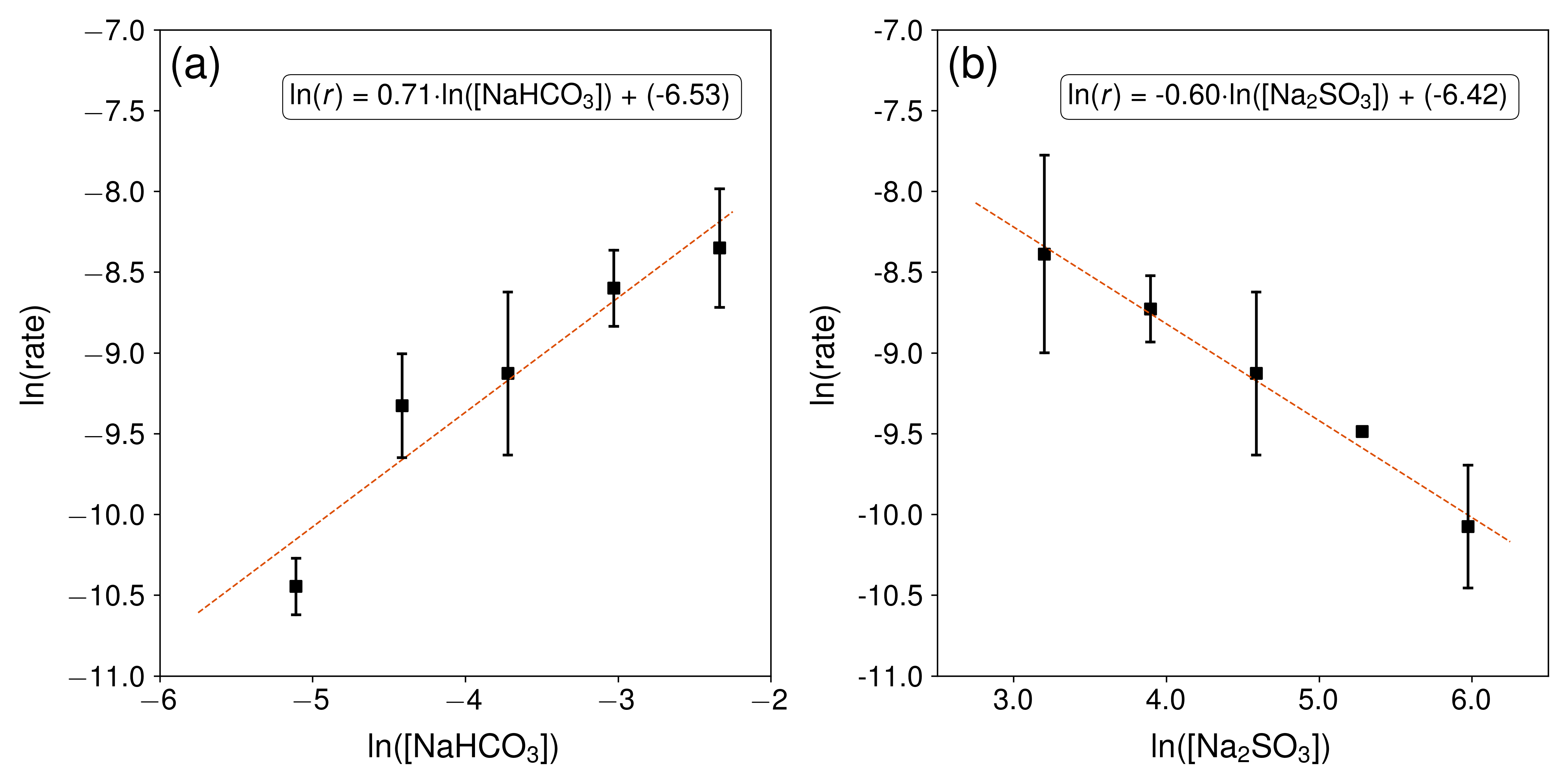}
    \caption{\textbf{The Order of Reaction with Respect to Sodium Bicarbonate and Sodium Sulfite.} (a) the order of reaction with respect to \ch{NaHCO3}, determined from the slope of the line of best fit to the data. (b) the order of reaction with respect to \ch{Na2SO3}, determined from the slope of the line of best fit to the data. Axes are shown as the natural logarithm of the rate of production of formate (mM s$^{-1}$) against the natural logarithm of the concentration of the reactant (mM). All data points are plotted as an average of the runs with error bars representing the standard deviation of the data set. The best fit is shown by the orange dashed line with a corresponding equation in the top right corner of each plot.}
    \label{fig:orders}
\end{figure}

First, we determine the rate equation by elucidating the order of reaction with respect to both \ch{Na2SO3} and \ch{NaHCO3}. The results of these experiments can be seen in Figure \ref{fig:orders}.

A linear relationship, consistent with the data shown in Figure \ref{fig:orders}, is adopted here. By performing linear regression analysis, we determine the order of reaction with respect to \ch{NaHCO3} to be 0.71 $\pm$ 0.12. Looking at the dependence of the rate on \ch{Na2SO3} concentration we see a negative correlation, with increasing concentrations of \ch{Na2SO3} leading to decreased rates of reaction. Using the concentrations of \ch{Na2SO3} shown in Figure \ref{fig:orders}, we determine the order with respect to \ch{Na2SO3} to be -0.60 $\pm$ 0.10.

Based on these results we define our rate equation as follows:
\begin{equation}
\dfrac{d[\ce{HCO2-}]}{dt} = k_{eff}[\ch{Na2SO3}]^{-0.60}[\ch{NaHCO3}]^{0.71}
\label{eq:rate_eq_full}
\end{equation}
where $k_{eff}$ will have units of ${\rm mM^{0.89} \, s^{-1}}$.
We use variations of this rate equation to determine the rate constants for formate production at three pH values: 6, 9, and 12.

\subsection{Rate Constants as a Function of pH}

\begin{figure}[H]
    \centering
    \includegraphics[width=1.0\textwidth]{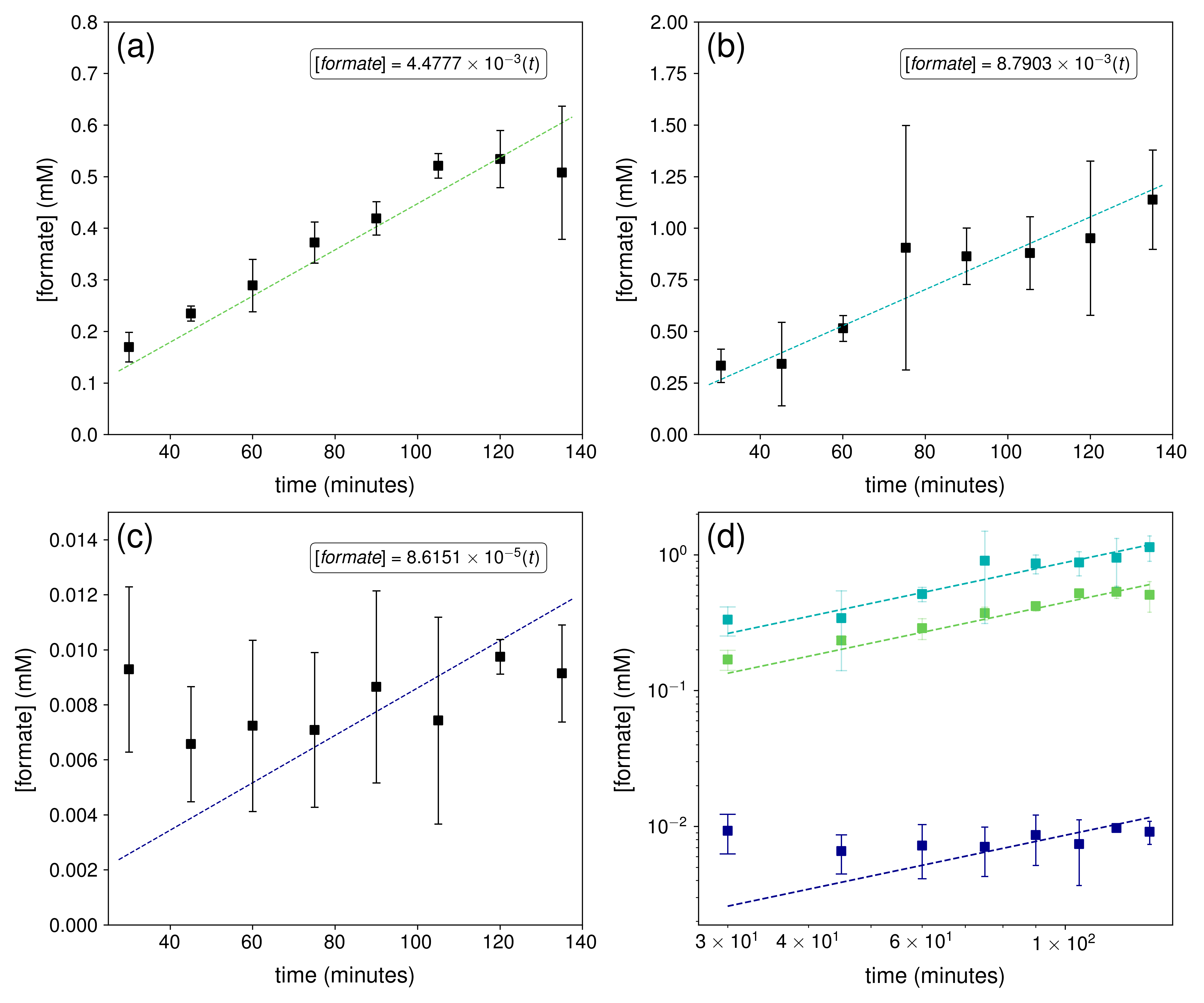}
    \caption{\textbf{The Rate of Formate Production as a Function of pH.} (a) experiments carried out at pH 6. (b) experiments carried out at pH 9. (c) experiments carried out at pH 12. (d) all 3 sets of experiments for comparison with green denoting pH 6, teal denoting pH 9, and blue denoting pH 12. All data points are plotted as an average of the triplicate runs with error bars representing the standard deviation of the data set. Fits for this data are shown as dashed lines with corresponding equations where $[formate]$ is the concentration of formate (mM), $t$ is time (minutes), and the slope is the rate of formate production (mM min$^{-1}$). We calculate the standard error of the slope in each case. The rate at pH 6 is $4.4777 \times 10^{-3} \pm 1.5217 \times 10^{-4}$ mM min$^{-1}$, the rate at pH 9 is $8.7903 \times 10^{-3} \pm 5.9528 \times 10^{-4}$ mM min$^{-1}$, and the rate at pH 12 is $8.6151 \times 10^{-5} \pm 8.5511 \times 10^{-6}$ mM min$^{-1}$.}
    \label{fig:rates}
\end{figure}

We measure the rate of production of formate using constant concentrations of \ch{Na2SO3} (100 mM) and \ch{NaHCO3} (10 mM). We measure rates in intervals of 15 minutes, starting from 30 minutes and irradiating up to 135 minutes. We do not irradiate past this time point so as to avoid further reaction of formate which would propagate the chemical network and thus alter the previously derived rate equation. The results of these experiments can be seen in Figure \ref{fig:rates}.

We perform linear regression analysis at each pH and determine that the fastest rate of production of formate is at pH 9. We utilise the results shown here to determine rate constants. We calculate rate constants using three versions of the rate equation: one based on an empirical rate constant, and two based on potential mechanisms for the production of formate. The rate laws can be seen in Table \ref{tbl:mechanisms}. Further description of the reactions involved in each mechanism, and the derivation of each rate equation, can be found in the Supporting Information.

\begin{table}[htbp]
  \centering
  \scriptsize
  \caption{\textbf{Rate Laws Derived Based on Different Mechanisms for the Production of Formate.}}
  \label{tbl:mechanisms}
  \renewcommand{\arraystretch}{1.75}
  \begin{tabular}{>{\centering\arraybackslash}m{1.7cm}|>{\centering\arraybackslash}m{7cm}|>{\centering\arraybackslash}m{6cm}}
    \hline
    \textbf{mechanism} & \textbf{reactions} & \textbf{rate law} \\
    \hline
    Empirical & N/A & $\frac{d[\ce{HCO2-}]}{dt} = k_{\text{eff}}[\ce{Na2SO3}]^{-0.60}[\ce{NaHCO3}]^{0.71}$ \\
    \hline
    Mechanism 1 &
      $\begin{array}{rcl}
        \ch{SO3^{2-}} + h\nu & \ch{->} & \ch{.SO3- + e-} \\
        \ch{CO2 + e-} & \ch{->} & \ch{.CO2-} \\
        \ch{.CO2- + HSO3-} & \ch{->} & \ch{HCO2- + .SO3^{-} \phantom{+ X}}
      \end{array}$
      &
      $\frac{d[\ce{HCO2-}]}{dt} = k_{\text{eff}}[\ce{SO3^{2-}}]^{-0.60}[\ce{CO2}]^{0.71}\phantom{NaH}$ \\
    \hline
    Mechanism 2 &
      $\begin{array}{rcl}
        \ch{SO3^{2-}} + h\nu & \ch{->} & \ch{.SO3- + e-} \\
        \ch{HCO3- + e-} & \ch{->} & \ch{HCO3^{2-}} \\
        \ch{HCO3^{2-} + HSO3-} & \ch{->} & \ch{HCO2- + OH- + .SO3-}
      \end{array}$
      &
      $\frac{d[\ce{HCO2-}]}{dt} = k_{\text{eff}}[\ce{SO3^{2-}}]^{-0.60}[\ce{HCO3-}]^{0.71}\phantom{Na}$ \\
    \hline
  \end{tabular}
\end{table}

To calculate rate constants we apply the rate laws from Table \ref{tbl:mechanisms}. For the empirical rate law we use concentrations of 100 mM and 10 mM for \ch{Na2SO3} and \ch{NaHCO3}, respectively. For Mechanism 1 and 2 we use equilibrium concentrations of \ch{CO2}, \ch{HCO3^-}, and \ch{SO3^{2-}}. These are calculated based on the initial concentrations of \ch{NaHCO3} and \ch{Na2SO3} and the pH of the solution. This was done using a simple model that accounts for the dissociation constants of carbonic and sulfurous acids. Table \ref{tbl:R_Ks} shows the pH range across which rate constants were measured using the average starting and ending pH of triplicate runs for pH 6, 9, and 12. Rate constants are shown scaled for the young Sun on the Archean Earth and early Mars. For the Archean Earth, the rate constants are divided by 76, and for early Mars, they are divided by 176. This scaling is based on the scale factors determined from Figure \ref{fig:LDLS_Spectra}. Rate constants are shown at each pH, with the range representing the absolute minimum and maximum values calculated by assuming that each variable in the rate law could deviate in a way that maximises the potential error.

\begin{table}[H]
  \centering
  \scriptsize
  \begin{threeparttable}
    \caption{\textbf{Rate Constants for the Production of Formate Scaled for the Young Sun on both the Archean Earth and Early Mars}}
    \label{tbl:R_Ks}
    \renewcommand{\arraystretch}{1.5}
    \begin{tabular}{c|c|c|c|c|c|c}
      \hline
      & \multicolumn{6}{c}{\textbf{rate constant (M$^{0.89}$ yr$^{-1}$)}}\\
      \hline
      & \multicolumn{2}{c}{} & \multicolumn{4}{|c}{\textbf{scaled for the young Sun}}\\
      \hline
      & \multicolumn{2}{c|}{\textbf{LDLS}} & \multicolumn{2}{c|}{\textbf{on the Archean Earth}} & \multicolumn{2}{c}{\textbf{on early Mars}} \\
      \hline
      \textbf{pH range$^\dagger$} & \textbf{$k_{eff}$} & \textbf{range} & \textbf{$k_{eff}$} & \textbf{range} & \textbf{$k_{eff}$} & \textbf{range} \\
      \hline
      & \multicolumn{6}{c}{\textbf{empirical rate constants}} \\
      \hline
      6.02 -- 6.06 & $7.27(0)\phantom{-}$ & \fade{$[3.37(0), 1.57(1)]\phantom{--}$} & $9.57(-2)$ & \fade{$[4.43(-2), 2.06(-1)]$} & $4.13(-2)$ & \fade{$[1.91(-2), 8.91(-2)]$} \\
      8.92 -- 9.24 & $1.43(1)\phantom{-}$ & \fade{$[6.38(0), 3.18(1)]\phantom{--}$} & $1.88(-1)$ & \fade{$[8.40(-2), 4.19(-1)]$} & $1.84(-3)$ & \fade{$[7.95(-4), 4.22(-3)]$} \\
      11.92 -- 11.93 & $1.40(-1)$ & \fade{$[6.04(-2), 3.21(-1)]$} & $8.11(-2)$ & \fade{$[3.63(-2), 1.81(-1)]$} & $7.95(-4)$ & \fade{$[3.43(-4), 1.82(-3)]$} \\
      \hline
      & \multicolumn{6}{c}{\textbf{mechanism-based rate constants (assuming the reduction of \ch{CO2})}} \\
      \hline
      6.02 -- 6.06 & $1.80(0)\phantom{-}$ & \fade{$[1.16(0), 2.80(0)]\phantom{--}$} & $2.37(-2)$ & \fade{$[1.52(-2), 3.69(-2)]$} & $1.02(-2)$ & \fade{$[6.57(-3), 1.59(-2)]$} \\
      8.92 -- 9.24 & $1.02(3)\phantom{-}$ & \fade{$[3.87(2), 2.67(3)]\phantom{--}$} & $1.34(1)\phantom{-}$ & \fade{$[5.09(0), 3.52(1)]\phantom{--}$} & $5.79(0)\phantom{-}$ & \fade{$[2.20(0), 1.52(1)]\phantom{--}$} \\
      11.92 -- 11.93 & $1.31(4)\phantom{-}$ & \fade{$[1.48(3), 1.15(5)]\phantom{--}$} & $1.72(2)\phantom{-}$ & \fade{$[1.94(1), 1.51(3)]\phantom{--}$} & $7.42(1)\phantom{-}$ & \fade{$[8.38(0), 6.51(2)]\phantom{--}$} \\
      \hline
      & \multicolumn{6}{c}{\textbf{mechanism-based rate constants (assuming the reduction of \ch{HCO3-})}} \\
      \hline
      6.02 -- 6.06 & $3.09(0)\phantom{-}$ & \fade{$[2.17(0), 3.36(0)]\phantom{--}$} & $4.07(-2)$ & \fade{$[2.86(-2), 4.43(-2)]$} & $1.76(-2)$ & \fade{$[1.23(-2), 1.91(-2)]$} \\
      8.92 -- 9.24 &  $1.46(1)\phantom{-}$ & \fade{$[6.56(0), 3.22(1)]\phantom{--}$} & $1.92(-1)$ & \fade{$[8.63(-2), 4.24(-1)]$} & $8.28(-2)$ & \fade{$[3.73(-2), 1.83(-1)]$} \\
      11.92 -- 11.93 & $2.11(0)\phantom{-}$ & \fade{$[1.00(0), 4.42(0)]\phantom{--}$} & $2.78(-2)$ & \fade{$[1.32(-2), 5.82(-2)]$} & $1.20(-2)$ & \fade{$[5.77(-3), 2.51(-2)]$} \\
      \hline
    \end{tabular}
    \begin{tablenotes}
        \item[$\dagger$] Average starting and ending pH of triplicate runs
        \item[$\ddagger$] The format $w.xy(z)$ is used to denote $w.xy \times 10^{z}$
        \item[*] The range denotes the absolute minimum and maximum value of the rate constant when maximising the error
    \end{tablenotes}
  \end{threeparttable}
\end{table}

\section{Discussion}
\label{sec:discussion}

\subsection{Sulfite Self-Shielding and the Production of Formate}

The orders of reaction were -0.60 $\pm$ 0.10 and 0.71 $\pm$ 0.12 with respect to \ch{Na2SO3} and \ch{NaHCO3}, respectively. The negative order with respect to \ch{Na2SO3} implies that sulfite, at the concentrations used in this work, may inhibit the reaction rather than promote it. We suggest this may be due to self-shielding effects supported by the fact that, at the concentrations used here, the optical depth of sulfite at $\leq 270 \, {\rm nm}$ is $\gg 1$. We cannot explain the order of the reaction completely, however, investigation of the mechanism based on the prior work of Deister \& Warneck \cite{Deister-1990} allows us to predict what this order should be if it were due to self-shielding. We predict the order to be $-0.50$, which is within error of -0.60 $\pm$ 0.10. For the derivation of this order from the reaction mechanism see the Supporting Information.

Based on these findings, we postulate that at lower concentrations the sulfite will cease to effectively self-shield leading to a transition threshold, where at some concentration of sulfite, the trend will switch from negative to positive. Therefore, we suggest that exploring carboxysulfitic chemistry at lower sulfite concentrations should be a productive direction for future work. We also highlight the importance of validating our rate law by examining the order with respect to each reactant over a wider range of concentrations. This work could demonstrate the accuracy and reliability of our calculated rate constants across a broad range of conditions.

After determining the order of reaction, we calculate the rate of production of formate as a function of pH. Figure \ref{fig:rates} shows that the highest rate is at pH 9, with the rate at pH 6 being  $\sim$ 50 \% lower and the rate at pH 12 being an order of magnitude lower. The error associated with each rate is determined to be within 10 \% of the true value in each case. We note that the largest error is for the rate determined at pH 12. We attribute this to the low concentrations of formate present in our solutions at this pH. This is likely a result of the low concentrations of both \ch{CO2} and \ch{HCO3-} as would be expected from the speciation of bicarbonate in a high pH system \cite{Andersen-2002}. At such low concentrations of formate, the accuracy of quantitative  $^1$H NMR spectroscopy is greatly reduced leading to larger uncertainties and a contribution from systematic error. While the scatter around our fits at pH 6, 9, and 12 is generally consistent with random error, there may be a contribution from systematic error in each case, as is likely for the results at pH 12. Further investigation into potential systematic errors in these experiments would be important for improving the accuracy of the rate measurements, however, it is unlikely to affect our overall results or conclusions.

After establishing both the order of reaction with respect to each reactant and the rate of production of formate, we calculate rate constants as can be seen in Table \ref{tbl:R_Ks}. In our results we include three rate constants, reflecting the complexity in fully describing the mechanism for formate production. Two of these constants are based on distinct mechanisms for formate production, while the third is an empirical rate constant. Prior work by Liu et al \cite{Liu-2021} described the production of formate from \ch{CO2} by the direct reduction of \ch{CO2} to a radical species following electron photodetachment from \ch{SO3^{2-}}. This radical species then combines with a proton forming formate. Although we propose this to be the most likely mechanism, we also suggest that formate could be formed directly through the reduction for \ch{HCO3^-} \cite{Sreekanth-2014, Bonet-2021, Nguyen-2023}. Based on this, we posit that the production of formate might have a contribution from both of these reaction pathways. Differentiating between these two mechanisms presents a particular challenge that is beyond the scope of this paper, however, we suggest that future studies might be able to distinguish between these reaction schemes using refined experimental analysis techniques. Thus, we provide individual rate constants under the assumption that formate production arises from the reduction of either \ch{CO2} or \ch{HCO3^-}.

\subsection{The Impact of pH and Mechanism on the Rate Constant for the Production of Formate}

Table \ref{tbl:R_Ks} presents rate constants as a function of pH and the proposed mechanism of formate production, along with a range of absolute values for each rate constant. Based on the empirical rate law, the highest rate constant is found at pH 9. The rate constant at pH 6 is lower, however it overlaps with the lower end of the range determined at pH 9. In contrast, the rate constant at pH 12 is significantly reduced. Regardless of the mechanism used, we observe an increase in the rate constant going from pH 6 to pH 9. For Mechanism 1 we see a general trend of increasing rate constant with increasing pH, whereas for Mechanism 2 we observe an increase in the rate constant going from pH 6 to pH 9 and a subsequent decrease in the rate constant at pH 12. Overall, these observations show the rate constant to increase from pH 6 to pH 9, with the mechanism influencing whether the peak is at pH 9 or pH 12. This highlights the significant impact the mechanism has on the productivity of the reaction. Based on these observations we suggest that future experiments could better constrain a general trend by measuring the rate constant at more pH values, and emphasise the importance of future work to determine the exact mechanism by which formate is produced.

We note that in this work we have not determined the order of reaction as a function of pH. While the rate laws derived for Mechanism 1 and Mechanism 2 do not explicitly depend on pH, we observe that the rate constants calculated using equilibrium concentrations of reactants do vary with pH suggesting an indirect impact of pH on the reaction. Thus, our rate constants appear to be pH dependent. As in Zhang \& Millero \cite{Zhang-1991}, we cannot say whether this inconsistency in rate constants across different pH levels is due to unaccounted for pH dependent changes in the mechanism or the influence of a chemical species whose concentration varies with pH.

\subsection{Using Kinetics to Inform the Geochemical Environment}

We suggest these findings to have implications for bridging the gap between chemical processes and the geochemical environments in which they occur. In order for formate to be produced via carboxysulfitic chemistry we can begin to establish a set of criteria to define the environments that could be the most favourable for the progression of this reaction network. 

One prerequisite for this chemistry is an aqueous environment containing dissolved \ch{CO2} and \ch{SO2}. In much of our work, we use concentrations of 100 mM and 10 mM for \ch{SO3^{2-}} and \ch{CO2_{(aq)}}, respectively. While it is challenging to find reliable constraints for environmental conditions on the early Earth or pre-Noachian Mars, the concentrations we use in this study are significantly higher than the expected global average abundances of these substances in natural waters \cite{Ranjan-2018}. However, while the global average provides some context, it is not generally an accurate proxy for local environments, where variations are important \cite{Warr-2023}. Notably, several local environments on early Mars are expected to reach or exceed the concentrations used in our experiments; see our discussion below. \ch{CO2} and \ch{SO2} are typically sourced through volcanic outgassing, where significant quantities of \ch{CO2} and \ch{SO2} can condense out of the atmosphere and dissolve in aqueous solutions. Alternatively, high concentrations of dissolved \ch{SO2} can be found in hydrothermal waters, where \ch{SO2} is directly pumped from below. Alongside the composition of these aqueous solutions, our findings suggest that the solution should be relatively alkaline. We base this suggestion on pH 9 providing the highest rate of production of formate at $8.7903 \times 10^{-3} \pm 5.9528 \times 10^{-4}$ mM min$^{-1}$ as well as the highest rate constant when considering an empirical rate law.

In accordance with this criteria, we suggest that the previously proposed high carbonate alkalinity lakes may act as an ideal setting \cite{Toner-2019}. These lakes not only provide the relatively alkaline pH for carboxysulfitic chemistry but have also been proposed to provide many other favourable conditions making them prime candidates for prebiotic chemistry \cite{Toner-2019, Toner-2020}. 

The final requirement for the progression of this reaction network is UV radiation. As shown in our methods, the LDLS we use in this study has a spectral irradiance 76 times higher than the young Sun on the Archean Earth. Taking this into account we can predict that, at a spectral irradiance equivalent to the predicted young Sun \cite{Ribas-2010, Claire-2012}, and at a pH of $\sim$ 9, assuming Mechanism 1, we would expect to make $\sim$ $1.34 \times 10^{1}$ M$^{0.89}$ yr$^{-1}$ of formate. Similarly, scaling for the young Sun on an early Mars would result in $\sim$ $5.79 \times 10^{0}$ M$^{0.89}$ yr$^{-1}$ of formate being produced. This method of scaling assumes that all the UV radiation from the Sun would penetrate the water and be able to interact with sulfite to generate highly reactive solvated electrons. This approach also assumes a linear dependence on photon flux. We suggest this assumption to be reasonable based on other sulfite based UV chemistry having been assumed, or measured, to be linear \cite{Deister-1990, Rimmer-2018}. It is important to note that should the scaling not be linear it will be crucial to identify the functional dependence of the rate constant on the flux in order to correctly relate these results to alternative environments. Thus, we stress the importance of this type of work in the future. 

It was previously proposed that prebiotic waters would be largely transparent to UV radiation, however, more recent works have suggested that aqueous-phase UV absorbers may have existed in high concentrations and can block key wavelengths \cite{Ranjan-2022, Ranjan-2016}. For example, prebiotic freshwater environments have been proposed to be largely transparent to UV radiation up to tens of meters, however, high salinity waters such as carbonate lakes may be deficient in shortwave UV flux (220 -- 260 nm) \cite{Ranjan-2022}. In addition, prebiotic waters that contain high levels of ferrous iron (\ch{Fe^{2+}}) can be strongly UV shielded due to the speciation of iron as ferrocyanide. Although we only account for the potential shielding effect of sulfite in this work, we emphasise the importance of considering other such aqueous-phase UV absorbers and suggest that future work be done to determine the impact such species may have on rate constants associated with UV-driven reactions. Further to this, if our proposed mechanism dominates and the absorption cross-sections of the other species are known, our rate constants can be scaled accordingly, and can therefore be incorporated into kinetics models for natural waters of arbitrary composition.

In addition, other properties of these lakes can severely impact the chemistry occurring. For example, in all of our experiments we use a phosphate buffer to control the pH of our solution. We chose phosphate based on it's prebiotic plausibility. Recent work by Toner \& Catling \cite{Toner-2020} and Haas et al \cite{Haas-2024} has demonstrated that alkaline lakes might have provided conditions relatively rich in phosphate, which we and many others \cite{Powner-2011, Toner-2020} show is effective in buffering the pH, underscoring the capability of such water bodies to aid in \ch{CO2} reduction reactions. Despite these findings, we recommend investigating the impact of buffers by examining how various buffer solutions affect the rate constants determined at different pH levels. As well as the addition of the phosphate buffer, we pH control our solutions using both \ch{HCl} and \ch{NaOH}. The addition of these species will have changed the ionic strength of our solutions. This change could potentially impact the chemistry, as our original assumption was that the activity coefficients of each species would be unity. This assumption neglects the potential decrease in reactivity that may accompany high ionic strengths \cite{Sauer-2004}. Therefore, we suggest that future work should be done to determine the impact of ionic strength on this scenario. Alongside the impact of buffers and non-zero ionic strength, we suggest other factors such as temperature and oxygen levels should be explored. These future studies, combined with ours and others \cite{Ritson-2012,  Xu-2018, Liu-2021}, will further narrow down the parameter space and help to place better constraints on relevant geochemical environments. 

We emphasise that while our results do not definitively suggest that this chemistry is the most productive at pH 9, they demonstrate that the rate of production of formate is the highest at pH 9 and that the rate constants determined at pH 9 are higher than those measured at pH 6 and 12 when considering an empirical rate law. Further to this, our results indicate that any environment with a pH between 6 and 12, such as those shown in Table \ref{tbl:Environmental-pHs}, should be able to host this chemistry provided the aforementioned criteria are met. Therefore, we suggest that searching for tracers of this chemistry in these environments could be a productive avenue for future work.

\subsection{The Search for Analogous Environments and Implications for Mars Sample Return Missions}

Following these conclusions, we look to find analogous environments that may be conducive to the formation of formate via the carboxysulfitic scenario. We suggest the optimal environment in which we would expect this chemistry to occur to be an aqueous system, open to the atmosphere, and in the photic zone, with a relatively alkaline pH and access to a source of electrons from dissolved \ch{SO2}. We suggest that, based on this criteria, we may be able to introduce one further constraint relating to the alkalinity of the system. In an open system, under high partial pressures of \ch{CO2} it has been shown that increasing the pH above 7 is difficult due to the speciation of dissolved \ch{CO2} as carbonic acid \cite{Kissick-2021, Hurowitz-2023}. One way to overcome this problem is to have waters with high alkalinity which can act as a reservoir of buffering capacity. To get such high alkalinities requires extreme evaporation of solutions that were already relatively alkaline \cite{Tosca-2023}. As water evaporates, the system becomes highly saline, leading to the formation of high concentrations of carbonate minerals \cite{Tosca-2023}. Therefore, we suggest that the environment of interest should also have high alkalinity.

Finding environments on Earth that might preserve signatures of this prebiotic chemistry has previously proved challenging due to the alteration of Earth's rock record \cite{Bosak-2021}. One potential environment that might avoid this issue is that of Mars. Jezero Crater is of particular interest owing to it previously having hosted paleolakes where carbonates were deposited, thus this provides an environment analogous to one where we might expect the carboxysulfitic scenario to have taken place \cite{Horgan-2020}. This environment exhibits various geological features associated with ancient river delta deposits, which suggest the presence of a sustained water system in the past \cite{Mangold-2020}. Unlike on Earth, the sedimentary rocks on Mars present a distinct chance to investigate relatively unaltered deposits that might have retained organic compounds and potential signatures of prebiotic chemical processes. This is due to the cold temperatures, the loss of surface water, and the absence of tectonics \cite{Bosak-2021}. Due to this, we suggest that Mars Sample Return should focus on looking for traces of carboxysulfitic chemistry in the rock record. More specifically, we suggest that Mars Sample Return missions should focus on environments that have evidence of carbonate formation, as this would be indicative of relatively alkaline pH which we suggest is the pH at which the rate of production of formate is highest. These environments should also be host to sulfite or sulfate minerals which would imply high concentrations of sulfite in the past. We note that NASA's Perseverance Rover has, in fact, already discovered a blend of crystalline and amorphous hydrated Mg-sulfate minerals alongside anhydrous Ca-sulfate minerals \cite{Siljestrom-2024}. In addition, there has been the discovery of both Mg- and Fe-rich carbonates in the Nili Fossae region at Jezero Crater \cite{Tarnas-2021, Tice-2022}. Together, these facts imply that Jezero Crater could provide optimal conditions for facilitating carboxysulfitic chemistry. Should the Mars Sample Return mission discover lake sediments, deposited within the photic zone, with carbonate and sulfur-rich minerals; we suggest that these minerals should be investigated further to search for inclusions of formate alongside other stable intermediates, such as acetate and oxalate \cite{Liu-2021}, that are formed as a result of carboxysulfitic chemistry \cite{Schreiber-2017}. 

Despite Mars offering a higher preservation potential than Earth, it is important to note that compounds on the Martian surface will have been altered by atmospheric chemistry and radiative processes \cite{Bosak-2021}. This problem is compounded by the fact that the environments proposed for this chemistry would have required wet conditions, and hydrological activity poses further challenges for preservation. Moreover, even if these signals had been preserved, there is the possibility of false positives \cite{McMahon-2022}. Despite this, we suggest that the work presented here adds important paleoenvironmental constraints that must be fulfilled for this chemistry to operate. 

Before Mars Sample Return, the scientific community should prioritise experiments that focus on making predictions about what might be expected in terms of ``pre-biosignatures'' \cite{Rimmer-2021-life}. This work should involve wet lab experiments alongside modelling studies. Rate constants obtained from experiments, which account for varied geochemical conditions, should be utilised to model the steady state concentrations of prebiotically relevant compounds. Using these results we can aim to come up with a list of compounds we might predict to be preserved in the Martian rock record as well as their ratios relative to one another.

\section{Conclusions}
\label{sec:conclusion}

Having explored the production of formate from carboxysulfitic chemistry, we suggest the optimal pH for the production of formate to be at pH 9 owing to the high rate of reaction and the comparatively high rate constant when assuming an empirical rate law. These findings imply that, should this chemistry have been important in the origin of life, it may have been the most productive in relatively alkaline environments such as high carbonate alkalinity lakes which have previously been suggested to be important for the formation of other prebiotically relevant molecules. However, we show that rate constants at pH 6, 9, and 12 indicate carboxysulfitic chemistry to be feasible across a wide range of environments.

Should further studies prove carboxysulfitic chemistry not to be directly relevant to prebiotic chemical synthesis, it could still serve to trace environmental conditions and verify the occurrence of broader chemical processes, such as UV-induced electron photodetachment and the reduction of carbon. These chemical tracers may be accessible by future Mars Sample Return missions, however, more work is required to bring this chemistry further into the geological context of Jezero Crater, and to carry this chemistry from its ancient origins, through geological timescales, to the present day. In these future endeavors, kinetics will continue to lead the way.

\begin{acknowledgement}

S. B. W. and P. B. R. both acknowledge financial support from the IPLU grant on “Quantifying Aliveness” (Cambridge Planetary Science and Life in the Universe Research Grants Scheme;  2021-2022). S. B. W. acknowledges support from the Leverhulme Centre for Life in the Universe (University of Cambridge), the Department of Physics (University of Cambridge) and Newnham College (University of Cambridge). P.B.R. thanks the Cavendish Laboratory (University of Cambridge) for internal funding supporting this work. The authors appreciate the valuable input and assistance from Duncan Howe and all of the team at the Yusuf Hamied Department of Chemistry NMR Facility (University of Cambridge). The authors also thank Nicholas Tosca for providing feedback on the discussion regarding predictions for Mars Sample Return. Finally, the authors thank the reviewers for their insightful comments that have contributed to the improvement of this manuscript. A special thanks also goes to Scarlett T. White for her artistic contribution to Figure 1. 

\end{acknowledgement}

\begin{suppinfo}

Supporting Information (Shedding a Light on the Kinetics of the Carboxysulfitic Scenario): Additional details of the techniques used to obtain NMR spectra including the characterisation of compounds detected in $^{13}$C and $^1$H NMR spectra as well as a description of the reaction mechanisms and rate equations used in this study (pdf).

\end{suppinfo}

\bibliography{main-formate.bib}

\end{document}


\tableofcontents
\setcounter{figure}{0}
\renewcommand{\thefigure}{S\arabic{figure}} 

\newpage

\section{NMR Spectroscopy}
\addcontentsline{toc}{section}{Experimental Methods}

Here we include details of our NMR Spectroscopy techniques as well as all of our water suppression techniques. We also include information on how we assign peaks during the analysis of all of our NMR Spectra. We note that no unexpected or unusually high safety hazards were encountered whilst performing this work.

\subsection{NMR Spectroscopy Techniques}
\addcontentsline{toc}{subsection}{NMR Spectroscopy Techniques}

Both $^1$H and $^{13}$C NMR spectroscopy were performed, the spectra for which are shown in this section. For $^1$H NMR spectroscopy, a T1 relaxation analysis was performed on one of the initial samples. This analysis was conducted to determine what interpulse delay would be appropriate to ensure that the relative integral sizes in the results would be quantitative. The longest observed T1 value suggested that an interpulse delay exceeding 60 seconds would be necessary. The number of scans was set to 32, and a spectral width of either 20 or 30 ppm was used. Water suppression techniques were employed to reduce the size of the solvent peak; the exact techniques are covered in more detail below. Spectral processing and analysis were conducted using TopSpin. $^{13}$C NMR spectra were acquired from the same NMR tube containing the sample solution. For the $^{13}$C NMR, an operating frequency of $\sim$ 125 MHz was used. The Bruker standard library pulse program zgpg30 was used, which employs a 1D sequence, incorporating power-gated decoupling with a 30-degree flip angle. 1024 transients were recorded with an interpulse delay of $\sim$ 5 seconds. A spectral width of $\sim$ 280 ppm was used. To determine the chemical shifts reference compounds were used as internal standards. The tetramethylsilane (TMS) signal was set as the reference at 0 ppm for both $^1$H and $^{13}$C spectra. The chemical shifts of all signals were recorded in parts per million (ppm) relative to TMS.

\subsubsection{Water Suppression Techniques}
\addcontentsline{toc}{subsection}{Water Suppression Techniques}

Duirng analysis, three distinct water suppression techniques were used. These methods were employed to enhance the suppression of the water signal, ensuring that no signals were overlooked and to observe any potential effects of the pulse program on the relative integral ratios. The first technique used was the noesypr1d with presaturation during relaxation delay and mixing time from the Bruker standard library. Another version of this presaturation technique was employed in the form of the zgcppr pulse program. This program executes much faster than the other techniques, potentially reducing relaxation losses. The zgesgppe-cnst12 pulse program was also used, this uses excitation sculpting with gradients using perfect echo (please refer to Adams et al for more information$^1$). Minor modifications were implemented to enable the adjustment of the suppression pulse length to a predetermined value. Additionally, enhancements were made to facilitate the utilisation of the Bruker Topspin ``wavemaker'' software, which calculates an appropriate power level for this pulse. This was done to ensure that any relaxation losses were constant throughout the analysis. The data were accumulated under full automation with the IconNMR software within Topspin using a customised program called ``findwater'' which finds the water peak, calibrates the $^1$H pulse length, and executes the ``wavemaker'' Topspin module to ensure effective suppression of the water peak.

\subsection{Characterisation by NMR Spectroscopy}
\addcontentsline{toc}{subsection}{Characterisation by NMR Spectroscopy}

\vspace{10pt}

\begin{figure}[H]
    \centering
    \includegraphics[width=1.0\textwidth]{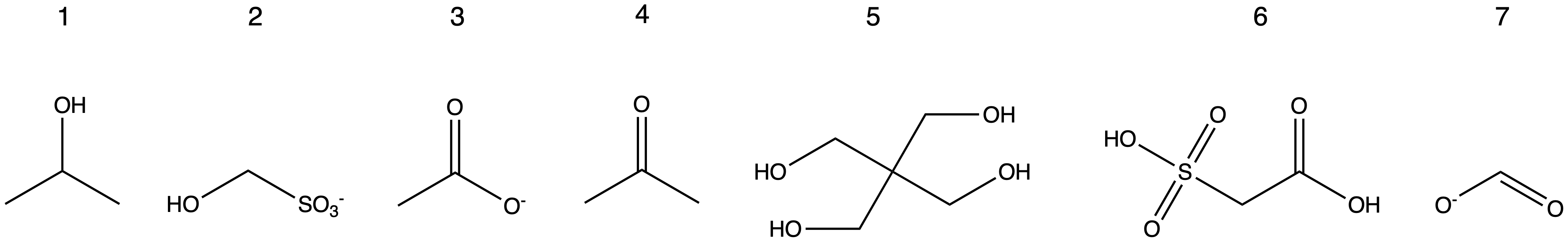}
    \caption{\textbf{Chemical Structures of Spectral Compounds.} Compounds showing up in $^1$H NMR spectra as reactants, products, and contaminants.}
    \label{fig:spectral-strcutures}
\end{figure}

\vspace{10pt}

\begin{table}[H]
  \centering
  \caption{\textbf{Compounds Identified in $^1$H NMR Spectra.}}
  \renewcommand{\arraystretch}{1.25} 
  \begin{threeparttable}
    \begin{tabular}{ccc}
      \hline
      \multirow{1}{*}{\textbf{approximate $^1$H}} & \multirow{1}{*}{\textbf{compound}} & \multirow{1}{*}{\textbf{compound}} \\
      \textbf{NMR shift (ppm)} & \textbf{name} & \textbf{structure$\dagger$} \\
      \hline
      1.10 & isopropanol & 1 \\
      1.43 & unknown & N/A \\
      1.79 & acetate & 3 \\
      2.11 & acetone & 4 \\
      2.60 & unknown & N/A \\
      3.24 & unknown & N/A \\
      3.48 & pentaerythritol & 5 \\
      3.58 & unknown & N/A \\
      3.69 & 2-sulfoacetic acid & 6 \\
      3.90 & isopropanol & 1 \\
      8.33 & formate & 7 \\
      \hline
    \end{tabular}
    \begin{tablenotes}
      \item[$\dagger$] Compound structures can be found by looking for the relevant number in Figure \ref{fig:spectral-strcutures}.
    \end{tablenotes}
  \end{threeparttable}
  \label{tab:supp_table1-proton}
\end{table}

\begin{table}[H]
  \centering
  \caption{\textbf{Compounds Identified in $^{13}$C NMR Spectra.}}
  \renewcommand{\arraystretch}{1.25} 
  \begin{threeparttable}
    \begin{tabular}{cc}
      \hline
      \multirow{1}{*}{\textbf{approximate $^{13}$C}} & \multirow{1}{*}{\textbf{compound}} \\
      \textbf{NMR shift (ppm)} & \textbf{name} \\
      \hline
      \phantom{0}24 & isopropanol \\
      \phantom{0}46 & pentaerythritol \\
      \phantom{0}61 & pentaerythritol / isopropanol \\
      124 & carbon dioxide \\
      162 & bicarbonate \\
      171 & formate \\
      \hline
    \end{tabular}
  \end{threeparttable}
  \label{tab:supp_table1-carbon}
\end{table}

\section{NMR Spectra}
\addcontentsline{toc}{section}{NMR Spectra}

We do not include all NMR spectra obtained in this study. We have conducted multiple runs of the same experiment, with each run employing three different water suppression techniques. We have chosen to include only one run utilising a specific water suppression technique here. The details of the run and the water suppression technique used are included in the figure captions. Note that across these runs we found no significant differences among the NMR spectra obtained.

\subsection{Control Experiments}
\addcontentsline{toc}{subsection}{Control Experiments}

Control experiments were performed to ensure that the formate being produced was due to carboxysulfitic chemistry. These controls also allowed for the accounting of any contaminant peaks, which were shown to be caused by acetone in the samples resulting from the washing of glassware.  Control experiments were run at pH 6, pH 9, and pH 12. These pH values were achieved by using the same buffers as outlined in the experimental section of the paper. To prepare the solution, 2250 $\mu$L of buffer solution or water, depending on the desired pH, was combined with 250 $\mu$L of a 10 mM acetone solution and 31.5 mg of \ch{Na2SO3} to obtain a 100 mM \ch{Na2SO3} solution. The solution was mixed, the pH was measured, and the mixture was transferred to a quartz cuvette. This solution was then irradiated for 30 minutes, and the results were analysed using NMR spectroscopy, as outlined in the experimental section of the paper. The results of these experiments can be seen below.

\begin{figure}[H]
    \centering
    \includegraphics[width=1.0\textwidth]{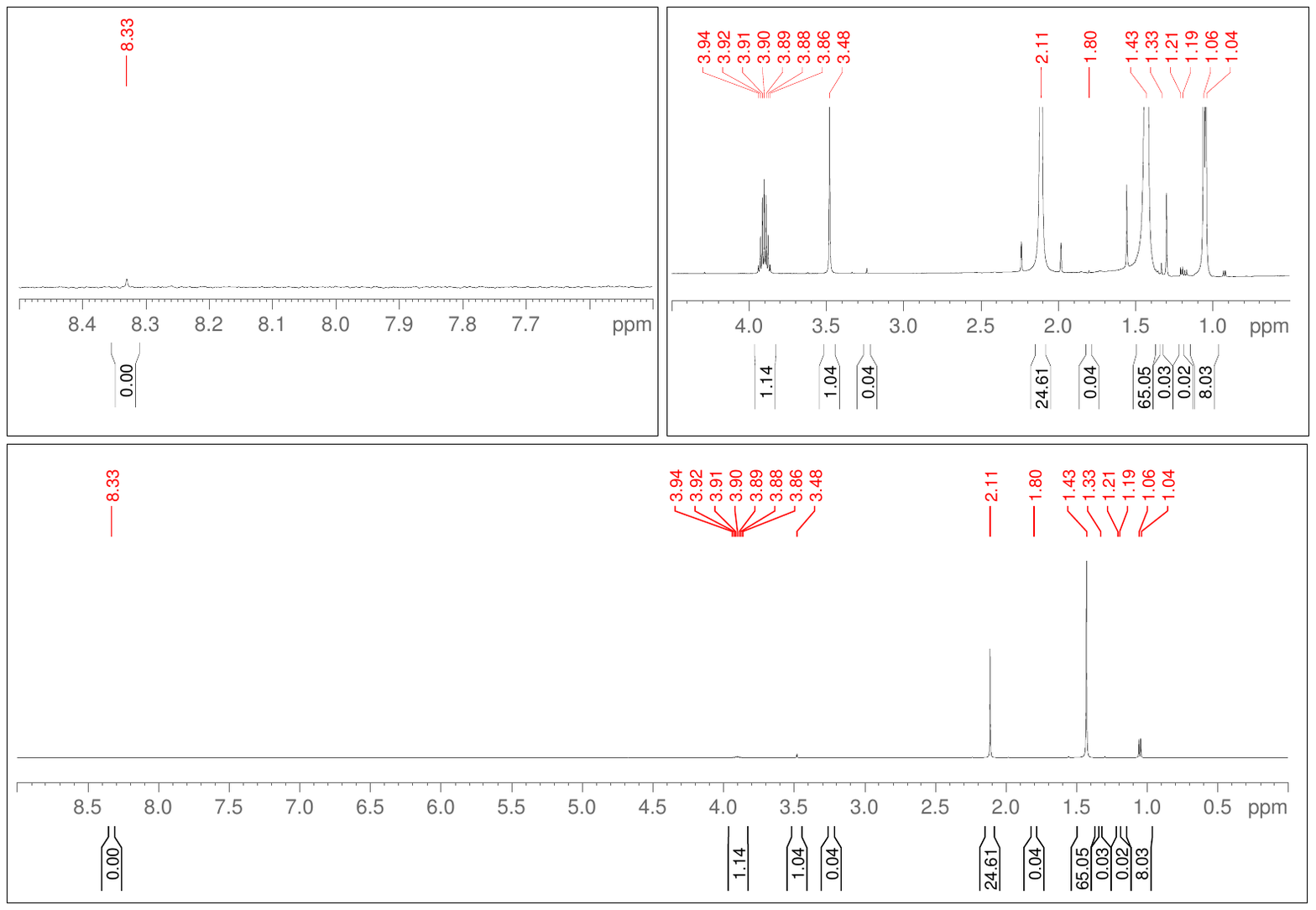}
    \caption{\textbf{pH 6 Control.} The experiment is carried out using 100 mM sodium sulfite and 10 mM acetone. The reaction mixture is irradiated for 30 minutes. A single NMR spectrum is shown here using the zgesgppe-cnst12 suppression sequence.}
    \label{fig:supp_figure_control1}
\end{figure}

\begin{figure}[H]
    \centering
    \includegraphics[width=1.0\textwidth]{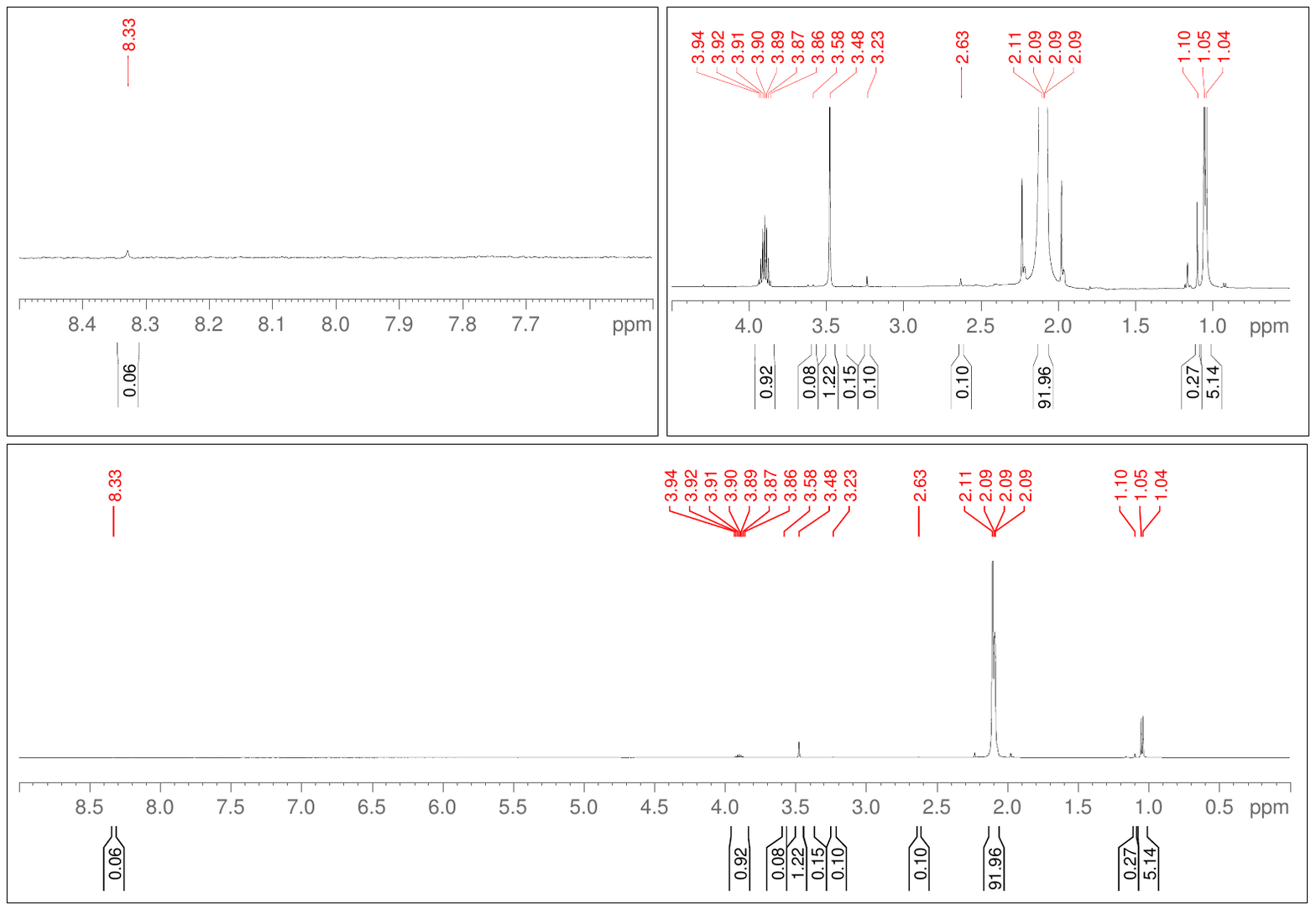}
    \caption{\textbf{pH 9 Control.} The experiment is carried out using 100 mM sodium sulfite and 10 mM acetone. The reaction mixture is irradiated for 30 minutes. A single NMR spectrum is shown here using the zgesgppe-cnst12 suppression sequence.}
    \label{fig:supp_figure_control2}
\end{figure}

\begin{figure}[H]
    \centering
    \includegraphics[width=1.0\textwidth]{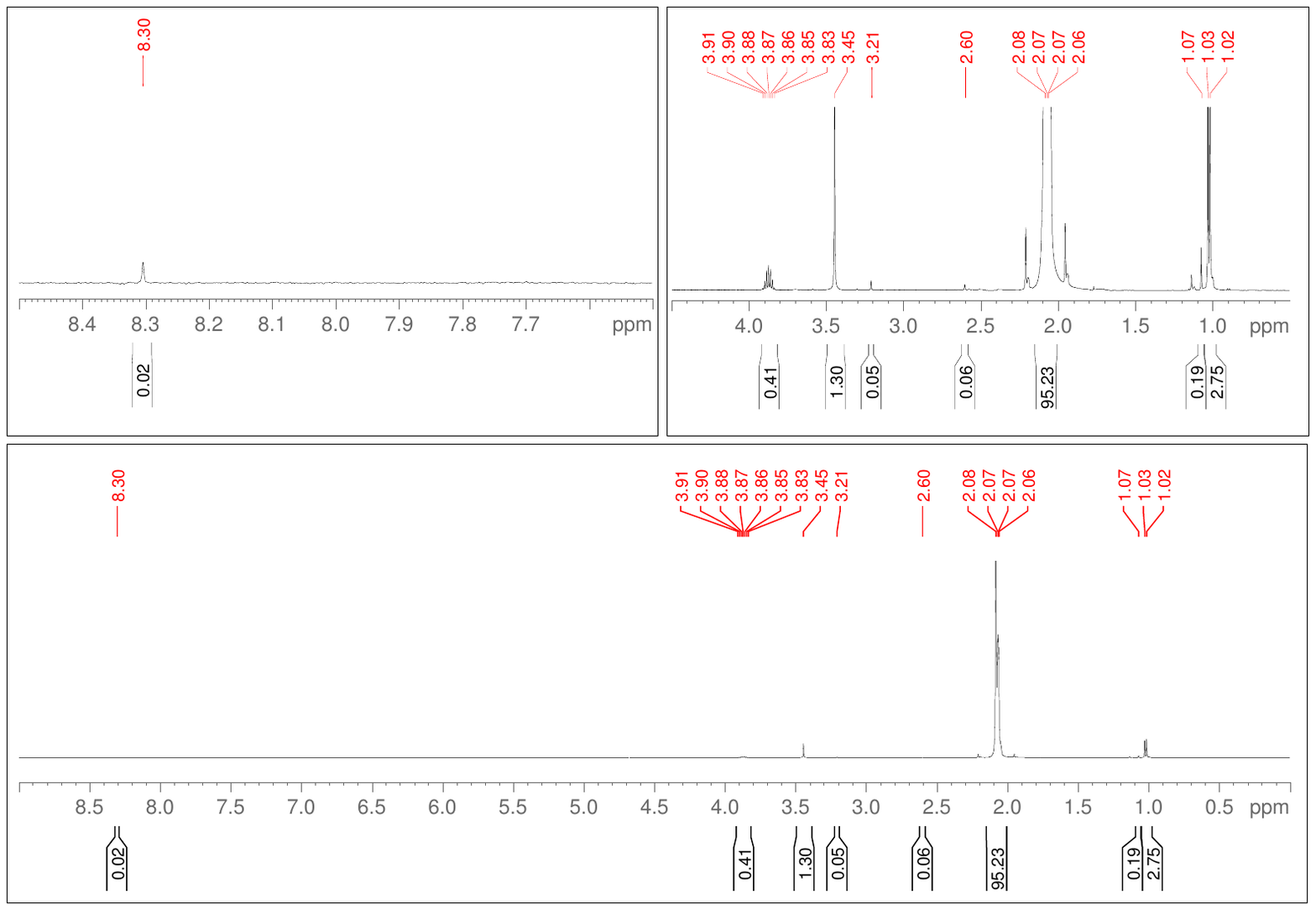}
    \caption{\textbf{pH 12 Control.} The experiment is carried out using 100 mM sodium sulfite and 10 mM acetone. The reaction mixture is irradiated for 30 minutes. A single NMR spectrum is shown here using the zgesgppe-cnst12 suppression sequence.}
    \label{fig:supp_control3}
\end{figure}

The peaks seen in these spectra account for isopropanol, acetate, acetone, and the four unknown peaks observed in our experiments. Although we do observe small amounts of formate in these experiments, as evidenced by the tiny peaks in our NMR spectra, we postulate that this peak is caused by dissolved \ch{CO2} from the air rather than from the acetone.

\subsection{Determining the Order of Reaction}
\begin{figure}[H]
    \centering
    \includegraphics[width=1.0\textwidth]{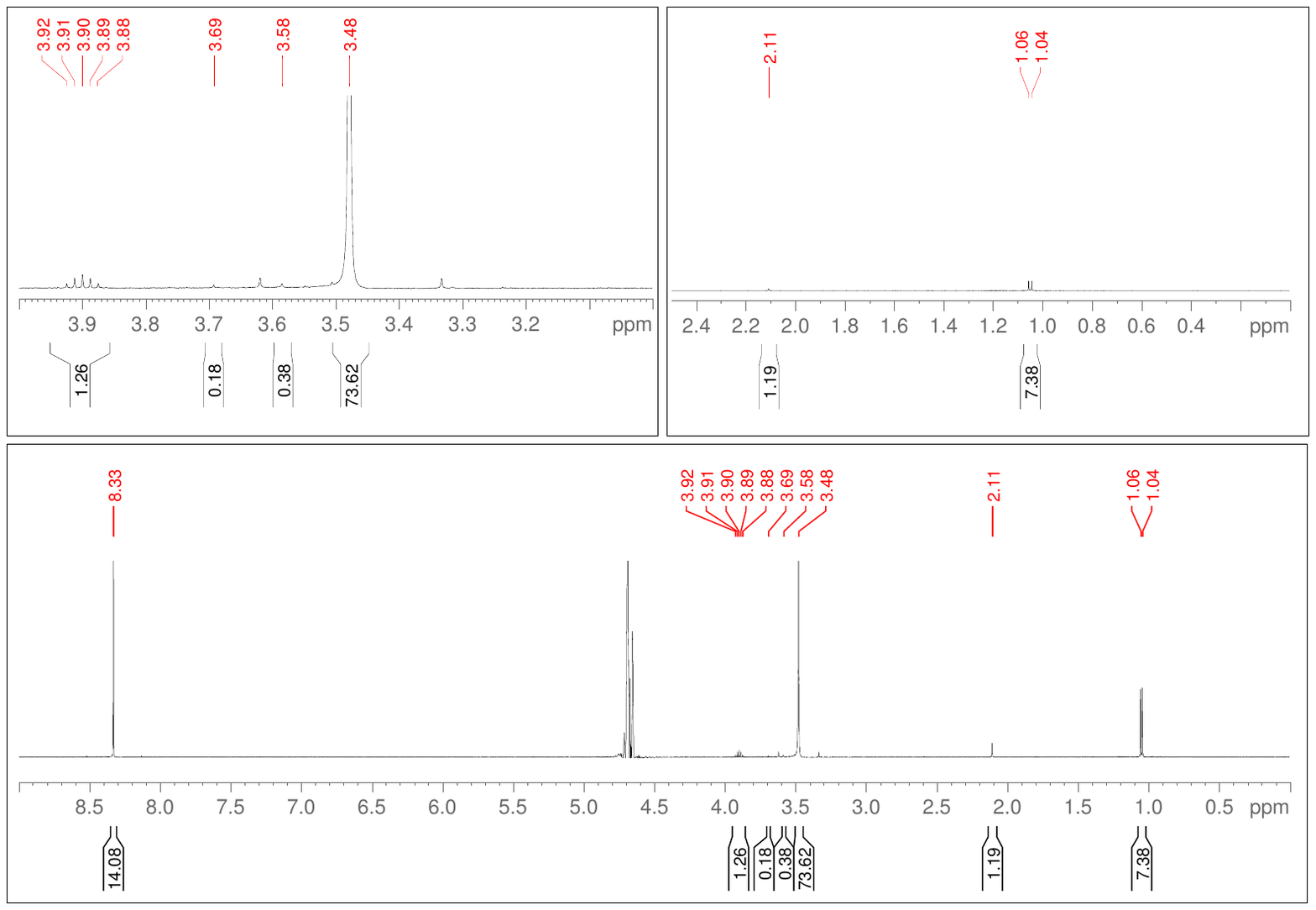}
    \caption{\textbf{Spectra 1.} Determining the Order of Reaction with Respect to Sulfite. The experiment is carried out as in the methods section using 25 mM sodium sulfite and 10 mM sodium bicarbonate. The reaction mixture is irradiated for 120 minutes. This experiment was repeated three times with three different water suppression techniques used for each run. A single NMR spectrum from the third run is shown here using the noesypr1d water suppression sequence.}
    \label{fig:supp_figure2}
\end{figure}

\begin{figure}[H]
    \centering
    \includegraphics[width=1.0\textwidth]{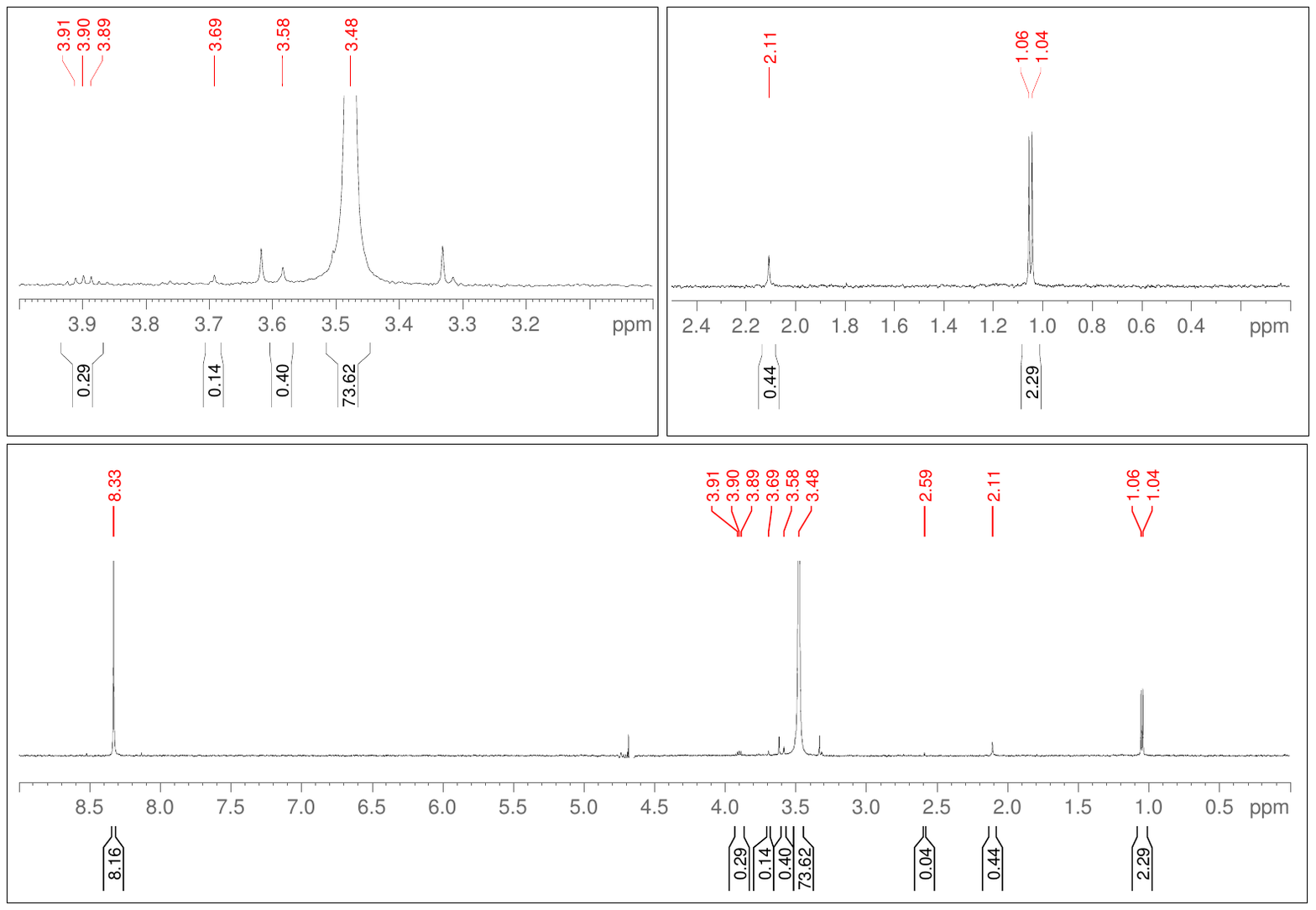}
    \caption{\textbf{Spectra 2.} Determining the Order of Reaction with Respect to Sulfite. The experiment is carried out as in the methods section using 50 mM sodium sulfite and 10 mM sodium bicarbonate. The reaction mixture is irradiated for 90 minutes. This experiment was repeated twice with three different water suppression techniques used for each run. A single NMR spectrum from the second run is shown here using the zgesgppe-cnst12 water suppression sequence.}
    \label{fig:supp_figure3}
\end{figure}

\begin{figure}[H]
    \centering
    \includegraphics[width=1.0\textwidth]{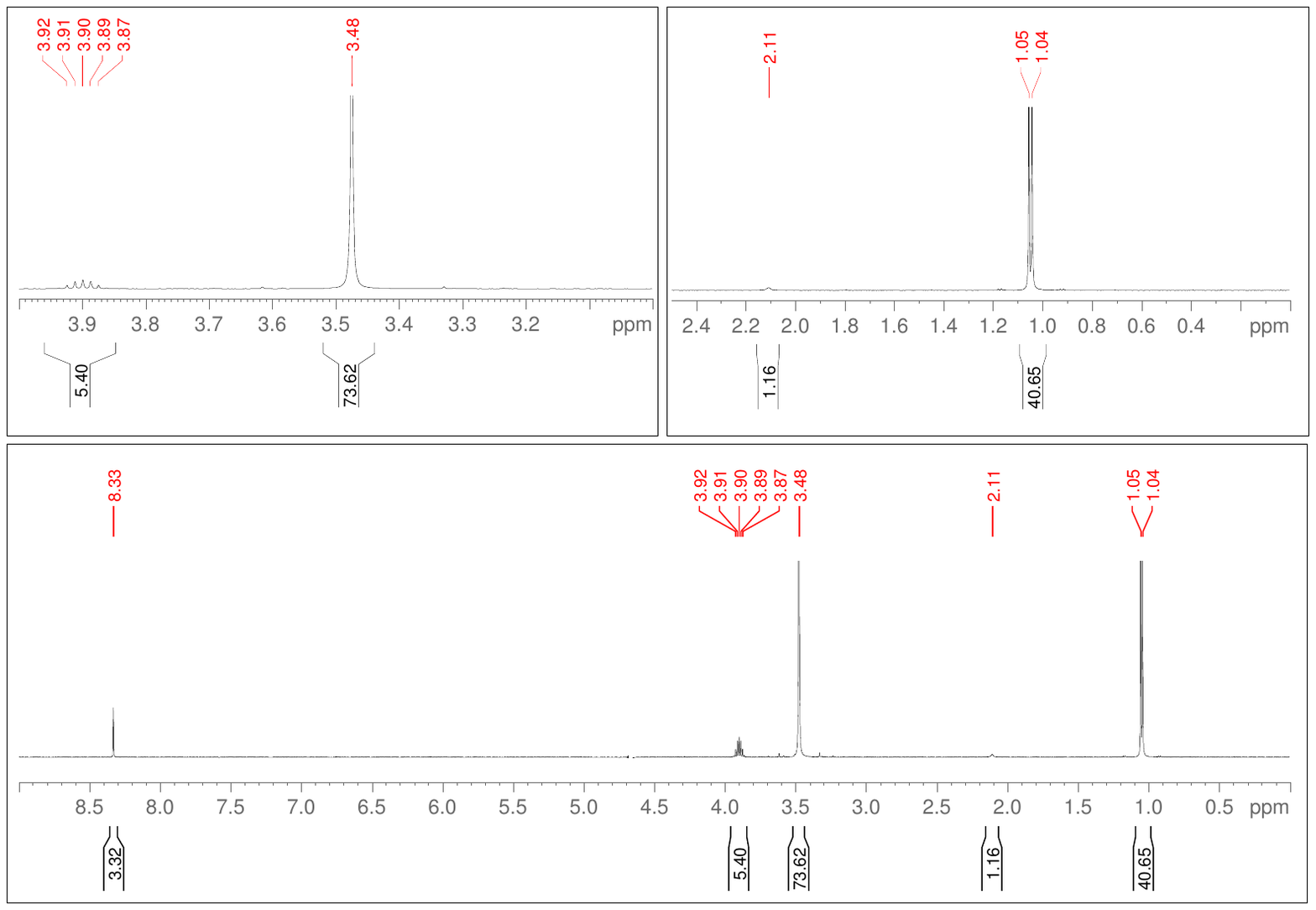}
    \caption{\textbf{Spectra 3.} Determining the Order of Reaction with Respect to Sulfite. The experiment is carried out as in the methods section using 100 mM sodium sulfite and 10 mM sodium bicarbonate. The reaction mixture is irradiated for 6311 seconds. This experiment was repeated twice with three different water suppression techniques used for each run. A single NMR spectrum from the second run is shown here using the noesypr1d water suppression sequence.}
    \label{fig:supp_figure4}
\end{figure}

\begin{figure}[H]
    \centering
    \includegraphics[width=1.0\textwidth]{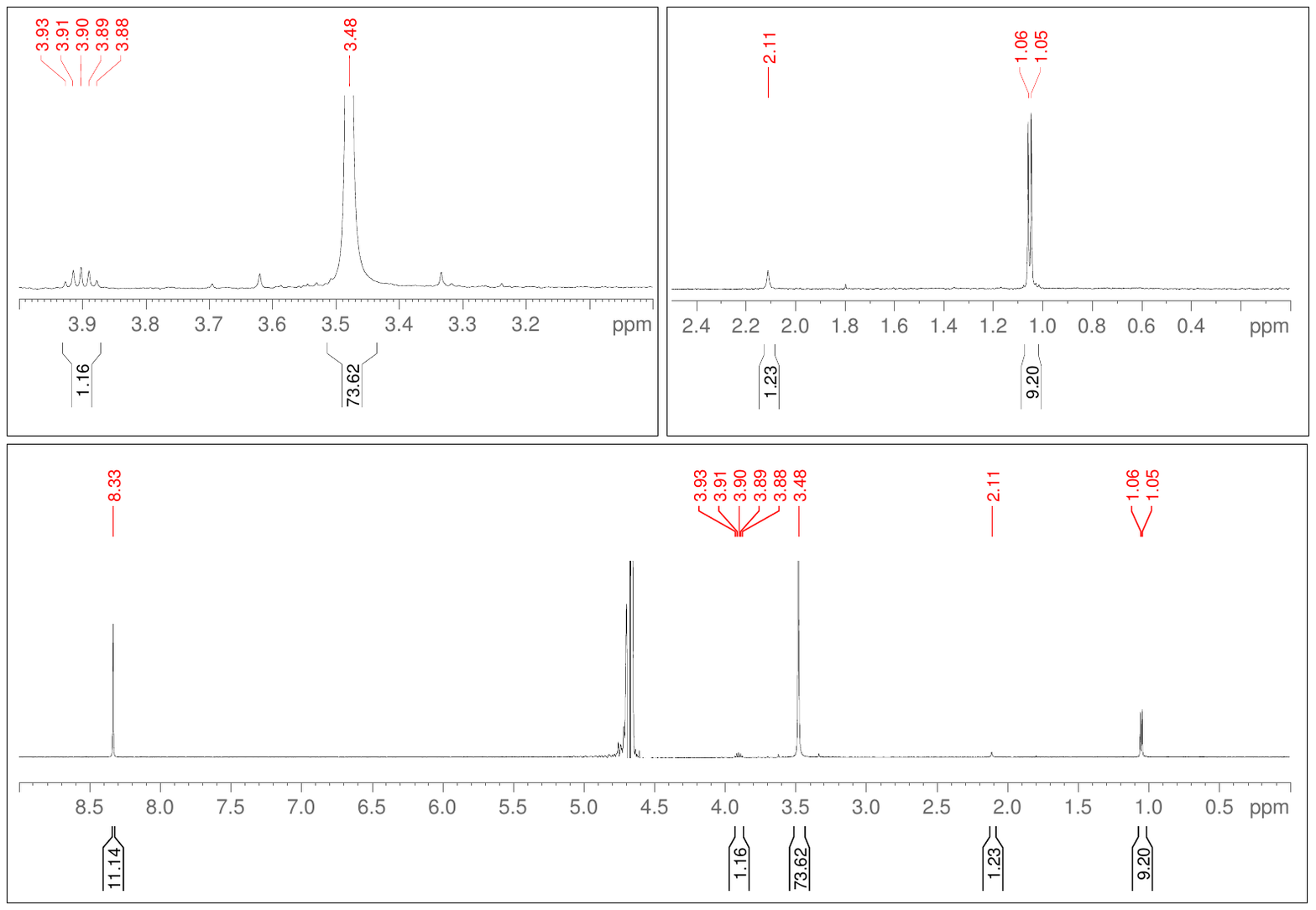}
    \caption{\textbf{Spectra 4.} Determining the Order of Reaction with Respect to Sulfite. The experiment is carried out as in the methods section using 200 mM sodium sulfite and 10 mM sodium bicarbonate. The reaction mixture is irradiated for 90 minutes. This experiment was repeated twice with three different water suppression techniques used for each run. A single NMR spectrum from the second run is shown here using the zgesgppe-cnst12 water suppression sequence.}
    \label{fig:supp_figure5}
\end{figure}

\begin{figure}[H]
    \centering
    \includegraphics[width=1.0\textwidth]{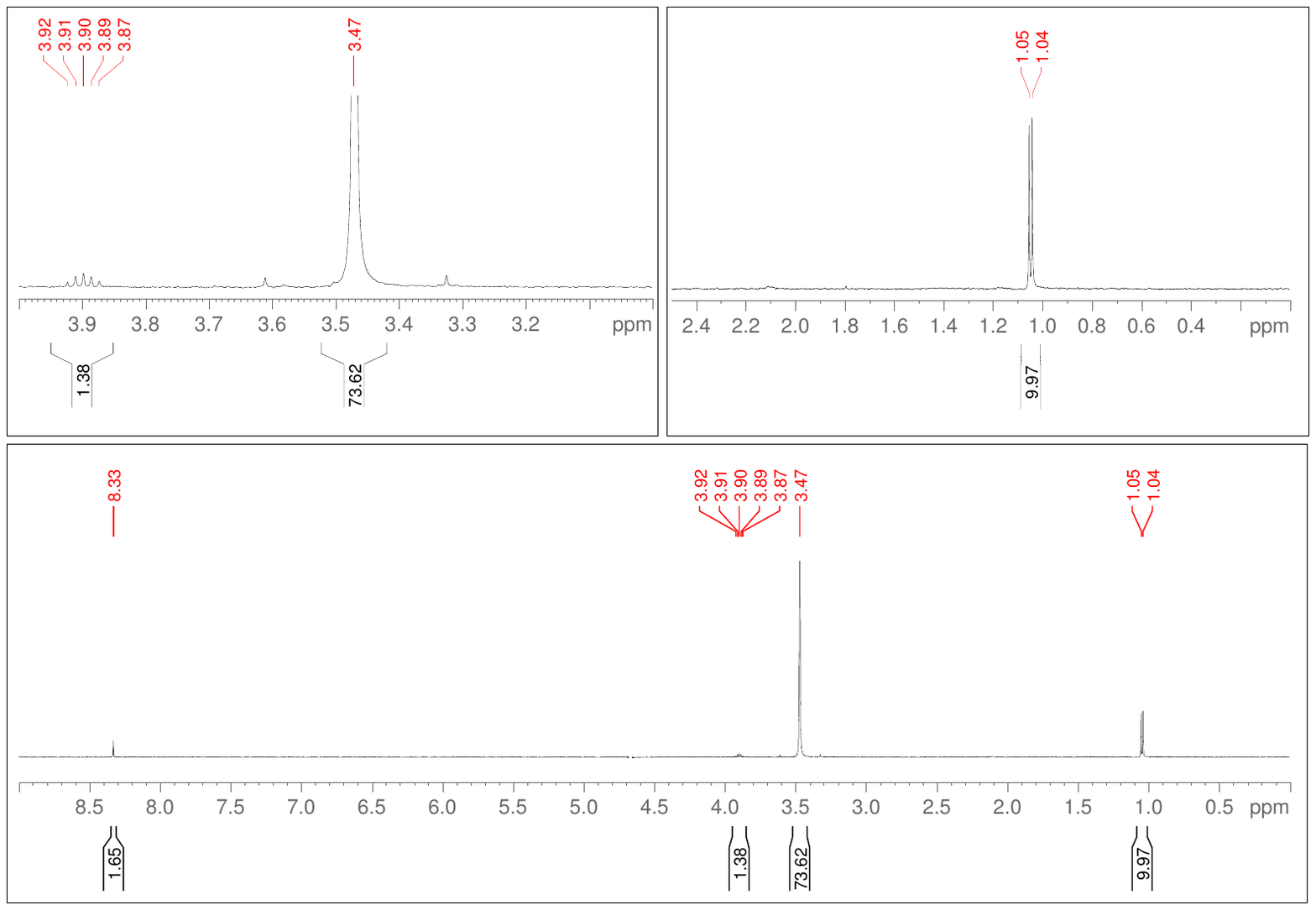}
    \caption{\textbf{Spectra 5.} Determining the Order of Reaction with Respect to Sulfite. The experiment is carried out as in the methods section using 400 mM sodium sulfite and 10 mM sodium bicarbonate. The reaction mixture is irradiated for 120 minutes. This experiment was repeated three times with three different water suppression techniques used for each run. A single NMR spectrum from the second run is shown here using the zgesgppe-cnst12 water suppression sequence.}
    \label{fig:supp_figure6}
\end{figure}

\begin{figure}[H]
    \centering
    \includegraphics[width=1.0\textwidth]{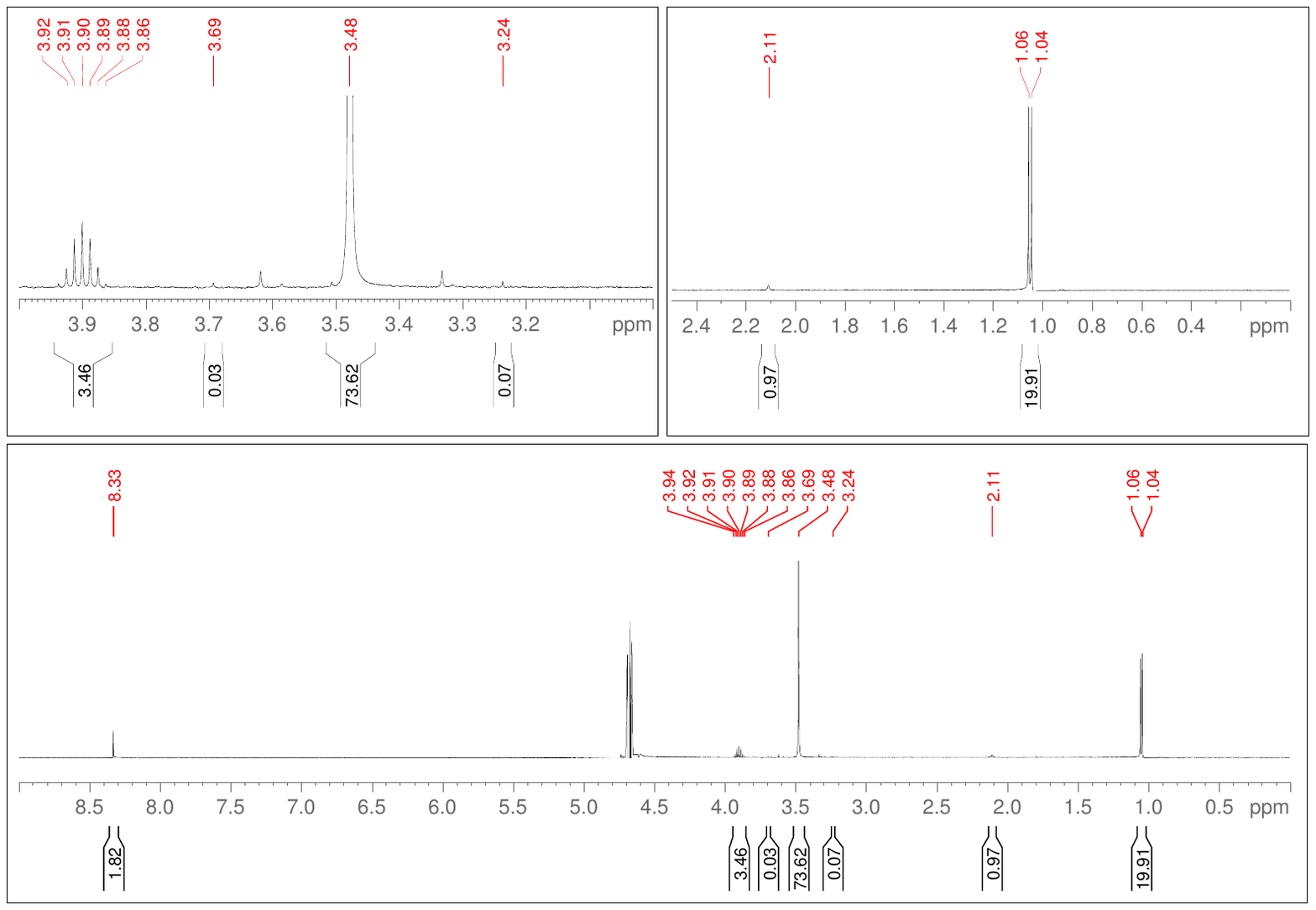}
    \caption{\textbf{Spectra 6.} Determining the Order of Reaction with Respect to Bicarbonate. The experiment is carried out as in the methods section using 100 mM sodium sulfite and 2.5 mM sodium bicarbonate. The reaction mixture is irradiated for 120 minutes. This experiment was repeated twice with three different water suppression techniques used for each run. A single NMR spectrum from the second run is shown here using the noesypr1d water suppression sequence.}
    \label{fig:supp_figure7}
\end{figure}

\begin{figure}[H]
    \centering
    \includegraphics[width=1.0\textwidth]{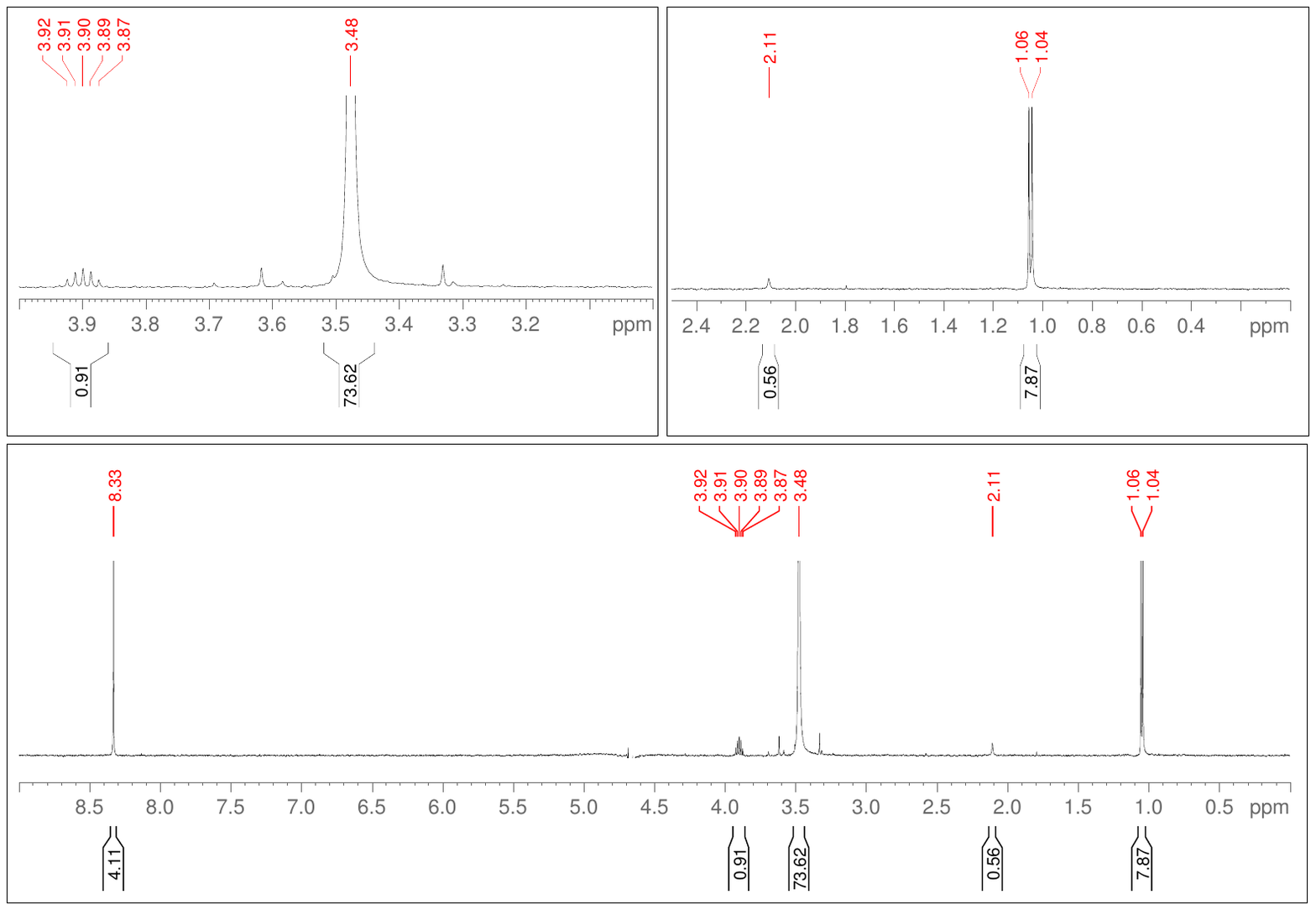}
    \caption{\textbf{Spectra 7.} Determining the Order of Reaction with Respect to Bicarbonate. The experiment is carried out as in the methods section using 100 mM sodium sulfite and 5 mM sodium bicarbonate. The reaction mixture is irradiated for 120 minutes. This experiment was repeated twice with three different water suppression techniques used for each run. A single NMR spectrum from the second run is shown here using the zgesgppe-cnst12 water suppression sequence.}
    \label{fig:supp_figure8}
\end{figure}

\begin{figure}[H]
    \centering
    \includegraphics[width=1.0\textwidth]{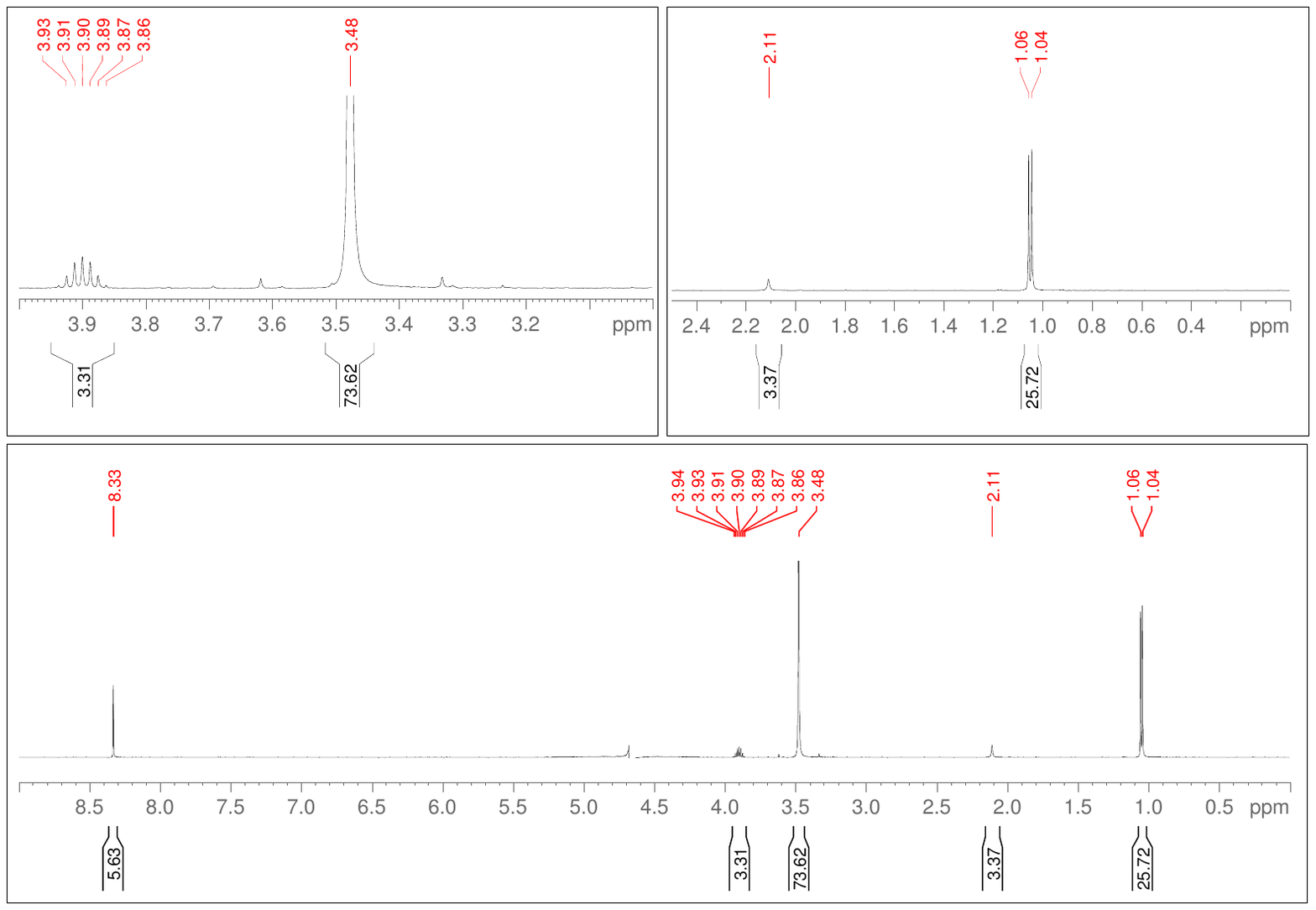}
    \caption{\textbf{Spectra 8.} Determining the Order of Reaction with Respect to Bicarbonate. The experiment is carried out as in the methods section using 100 mM sodium sulfite and 20 mM sodium bicarbonate. The reaction mixture is irradiated for 75 minutes. This experiment was repeated twice with three different water suppression techniques used for each run. A single NMR spectrum from the second run is shown here using the zgesgppe-cnst12 water suppression sequence.}
    \label{fig:supp_figure9}
\end{figure}

\begin{figure}[H]
    \centering
    \includegraphics[width=1.0\textwidth]{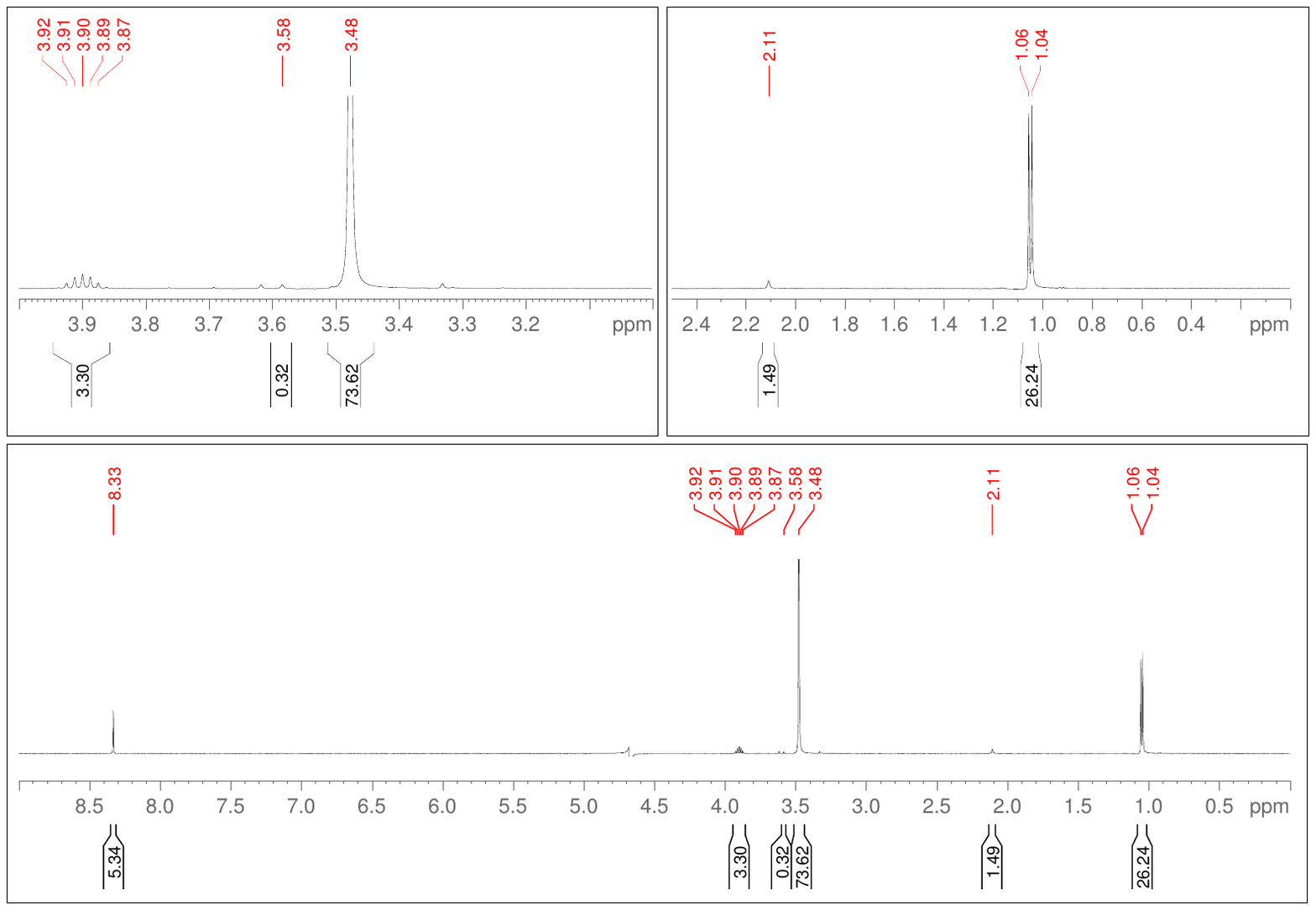}
    \caption{\textbf{Spectra 9.} Determining the Order of Reaction with Respect to Bicarbonate. The experiment is carried out as in the methods section using 100 mM sodium sulfite and 40 mM sodium bicarbonate. The reaction mixture is irradiated for 60 minutes. This experiment was repeated twice with three different water suppression techniques used for each run. A single NMR spectrum from the second run is shown here using the zgesgppe-cnst12 water suppression sequence.}
    \label{fig:supp_figure10}
\end{figure}

\addcontentsline{toc}{subsection}{Determining the Order of Reaction: Supplementary Figures \ref{fig:supp_figure2} - \ref{fig:supp_figure10}}

Here we show NMR spectra for experiments used to determine the order of reaction with respect to both bicarbonate and sulfite.

\subsection{The Rate Constant at pH 6}
\begin{figure}[H]
    \centering
    \includegraphics[width=1.0\textwidth]{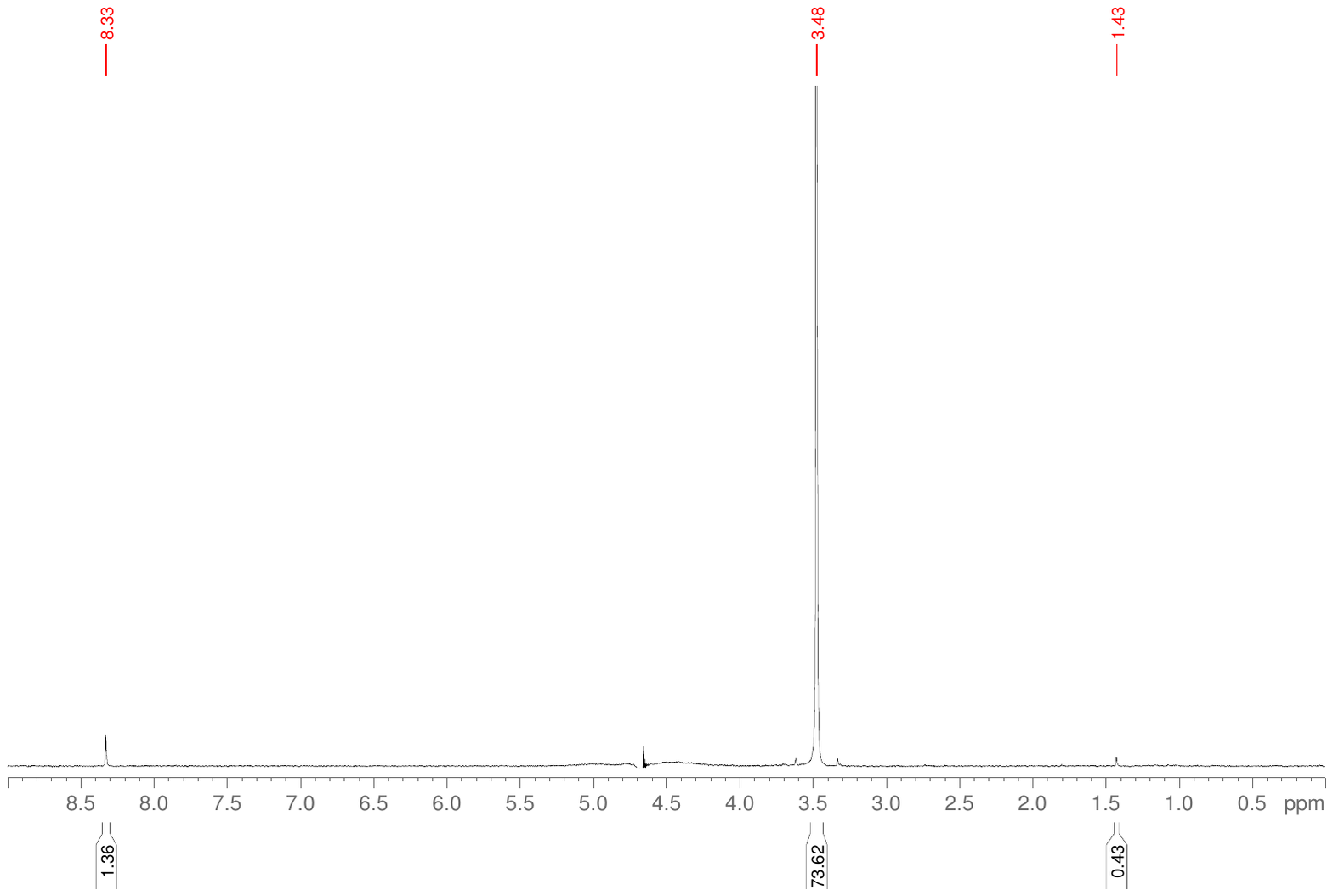}
    \caption{\textbf{Determining the Rate Constant for Formate Production at pH 6, 30 minutes.} The experiment is carried out as in the methods section using 100 mM sodium sulfite and 10 mM sodium bicarbonate. The reaction mixture is irradiated for 30 minutes. This experiment was repeated three times with three different water suppression techniques used for each run. A single NMR spectrum from the first run is shown here using the zgesgppe-cnst12 water suppression sequence.}
    \label{fig:supp_figure11}
\end{figure}

\begin{figure}[H]
    \centering
    \includegraphics[width=1.0\textwidth]{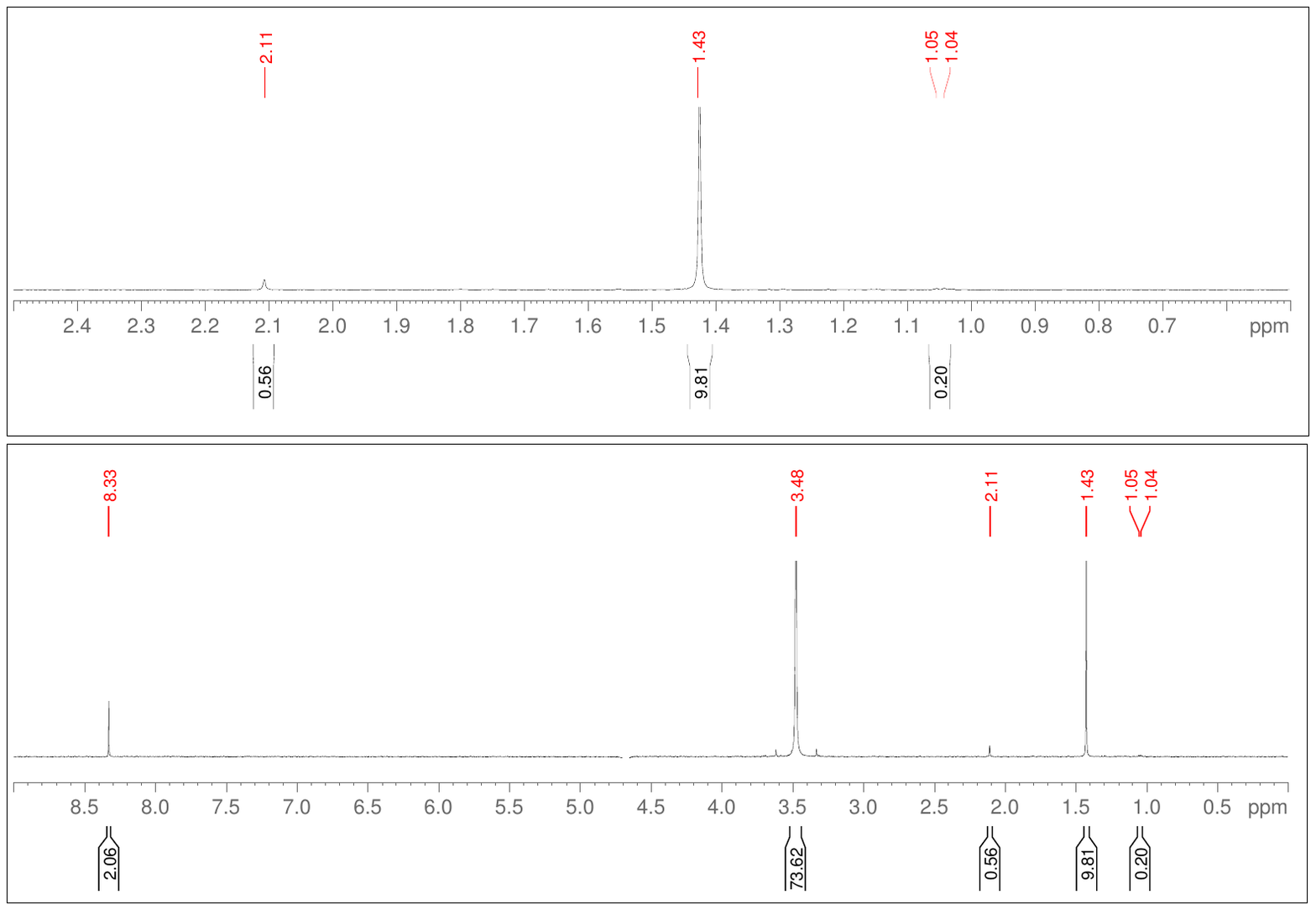}
    \caption{\textbf{Determining the Rate Constant for Formate Production at pH 6, 45 minutes.} The experiment is carried out as in the methods section using 100 mM sodium sulfite and 10 mM sodium bicarbonate. The reaction mixture is irradiated for 45 minutes. This experiment was repeated three times with three different water suppression techniques used for each run. A single NMR spectrum from the second run is shown here using the zgesgppe-cnst12 water suppression sequence.}
    \label{fig:supp_figure12}
\end{figure}

\begin{figure}[H]
    \centering
    \includegraphics[width=1.0\textwidth]{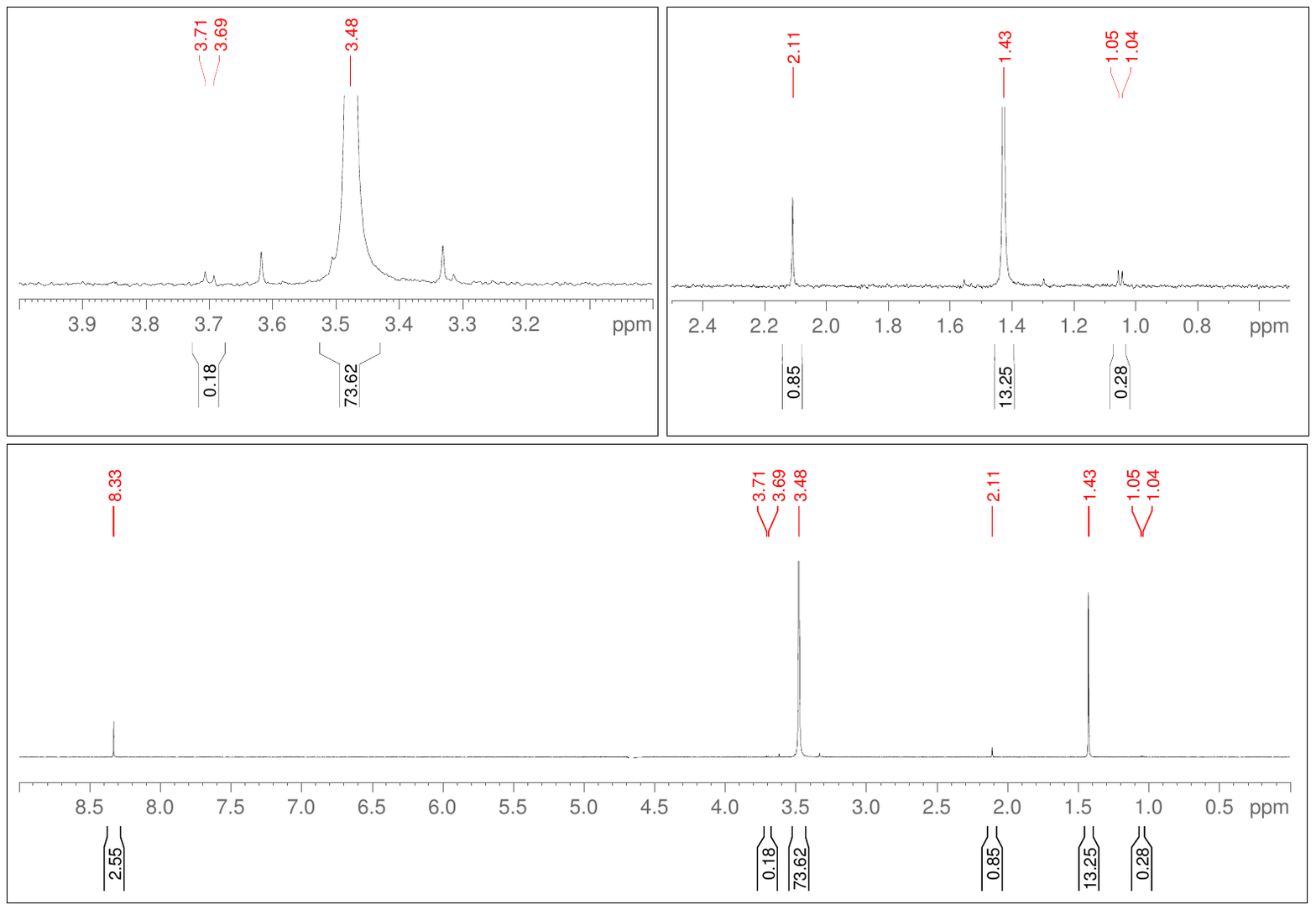}
    \caption{\textbf{Determining the Rate Constant for Formate Production at pH 6, 60 minutes.} The experiment is carried out as in the methods section using 100 mM sodium sulfite and 10 mM sodium bicarbonate. The reaction mixture is irradiated for 60 minutes. This experiment was repeated three times with three different water suppression techniques used for each run. A single NMR spectrum from the third run is shown here using the zgesgppe-cnst12 water suppression sequence.}
    \label{fig:supp_figure13}
\end{figure}

\begin{figure}[H]
    \centering
    \includegraphics[width=1.0\textwidth]{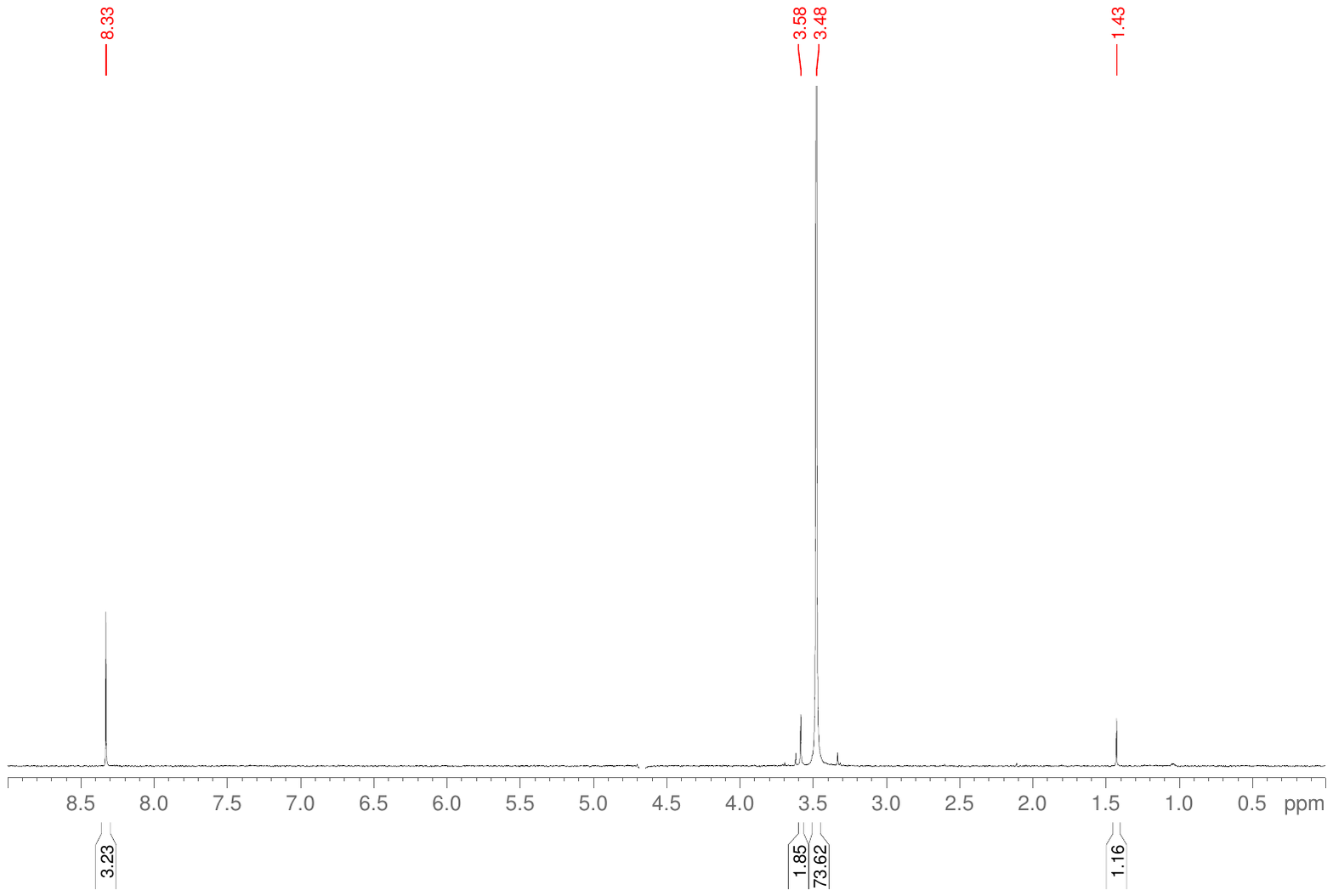}
    \caption{\textbf{Determining the Rate Constant for Formate Production at pH 6, 75 minutes.} The experiment is carried out as in the methods section using 100 mM sodium sulfite and 10 mM sodium bicarbonate. The reaction mixture is irradiated for 75 minutes. This experiment was repeated three times with three different water suppression techniques used for each run. A single NMR spectrum from the third run is shown here using the zgesgppe-cnst12 water suppression sequence.}
    \label{fig:supp_figure14}
\end{figure}

\begin{figure}[H]
    \centering
    \includegraphics[width=1.0\textwidth]{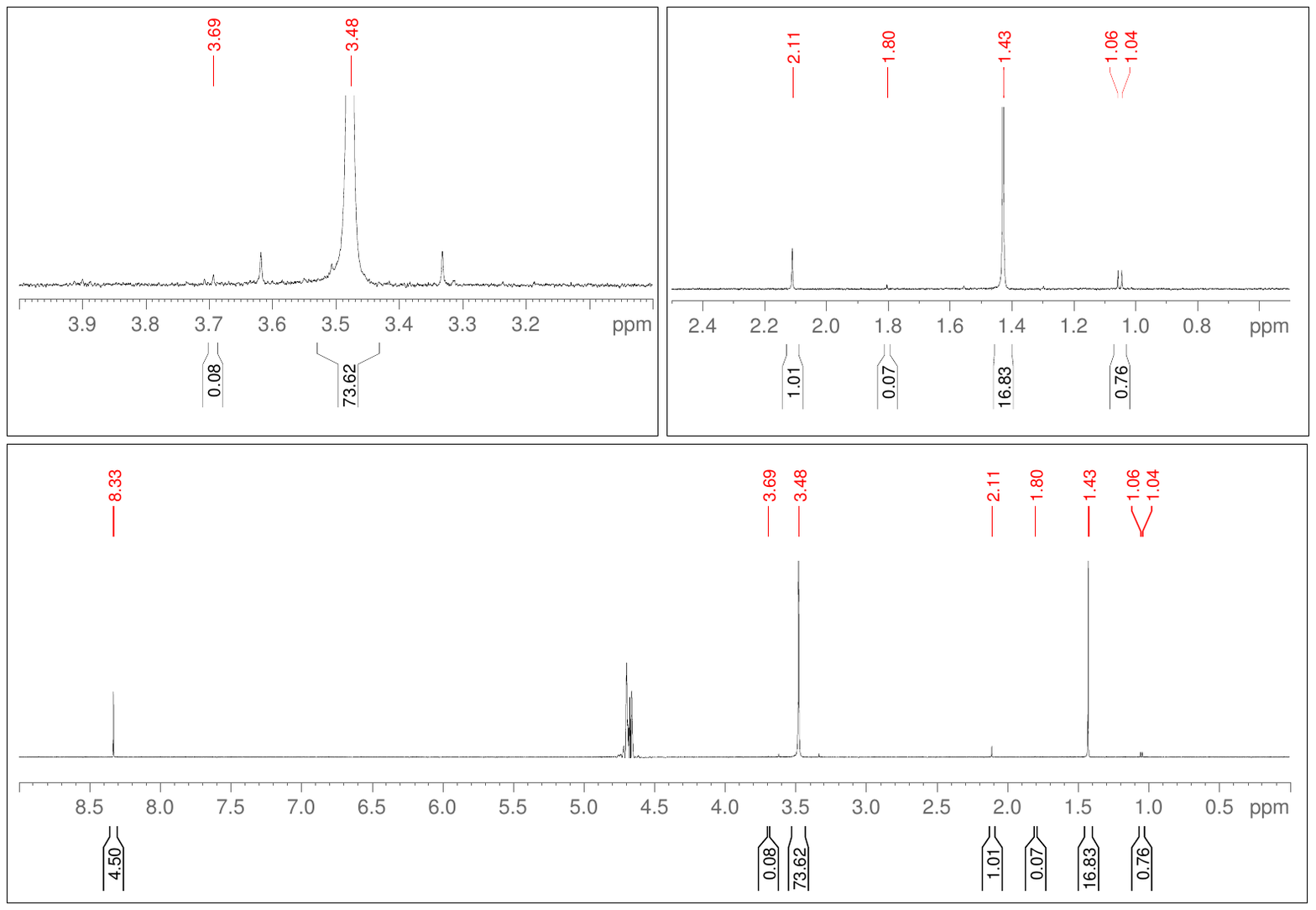}
    \caption{\textbf{Determining the Rate Constant for Formate Production at pH 6, 90 minutes.} The experiment is carried out as in the methods section using 100 mM sodium sulfite and 10 mM sodium bicarbonate. The reaction mixture is irradiated for 90 minutes. This experiment was repeated three times with three different water suppression techniques used for each run. A single NMR spectrum from the first run is shown here using the noesypr1d water suppression sequence.}
    \label{fig:supp_figure15}
\end{figure}

\begin{figure}[H]
    \centering
    \includegraphics[width=1.0\textwidth]{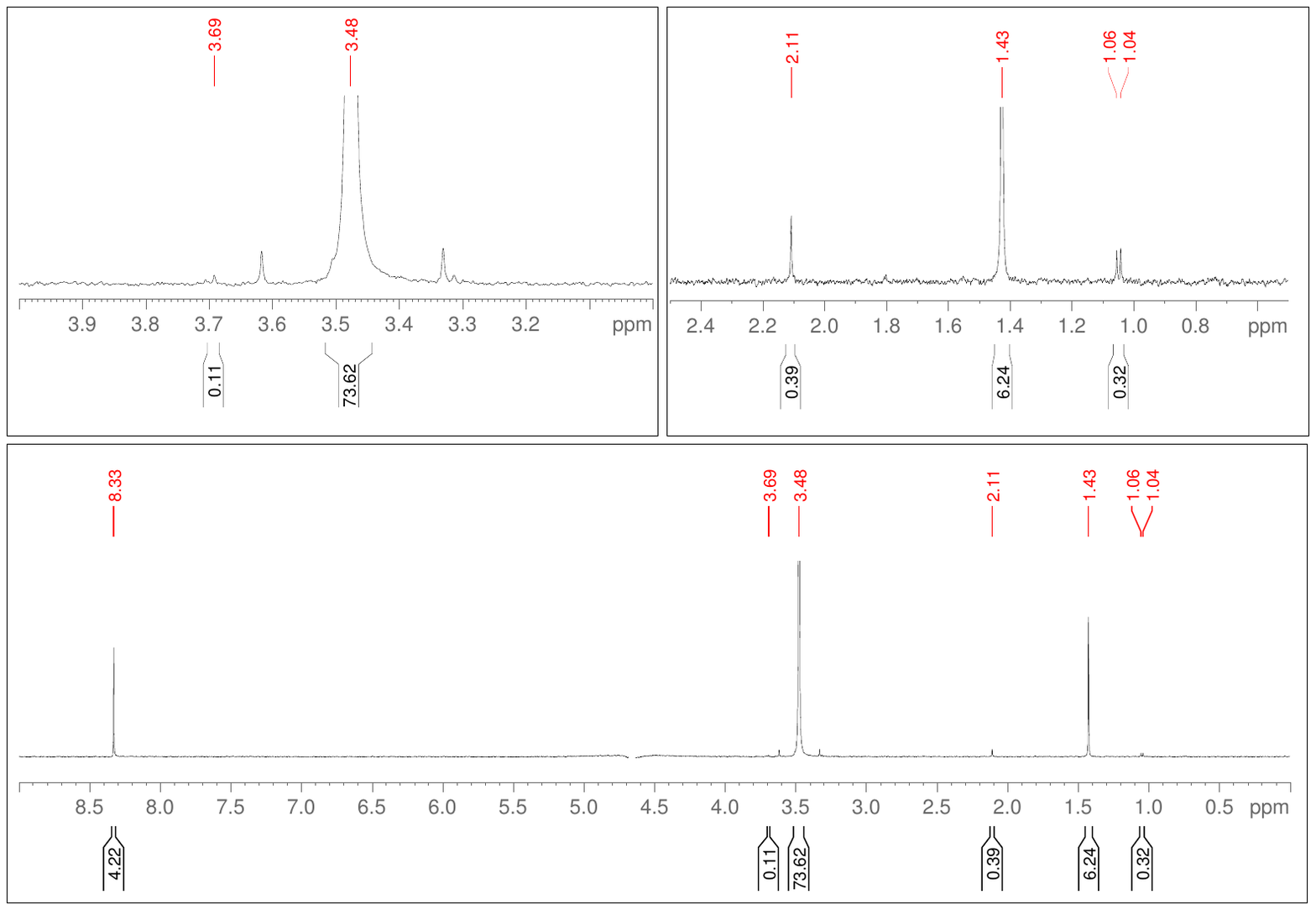}
    \caption{\textbf{Determining the Rate Constant for Formate Production at pH 6, 105 minutes.} The experiment is carried out as in the methods section using 100 mM sodium sulfite and 10 mM sodium bicarbonate. The reaction mixture is irradiated for 105 minutes. This experiment was repeated three times with three different water suppression techniques used for each run. A single NMR spectrum from the third run is shown here using the zgesgppe-cnst12 water suppression sequence.}
    \label{fig:supp_figure16}
\end{figure}

\begin{figure}[H]
    \centering
    \includegraphics[width=1.0\textwidth]{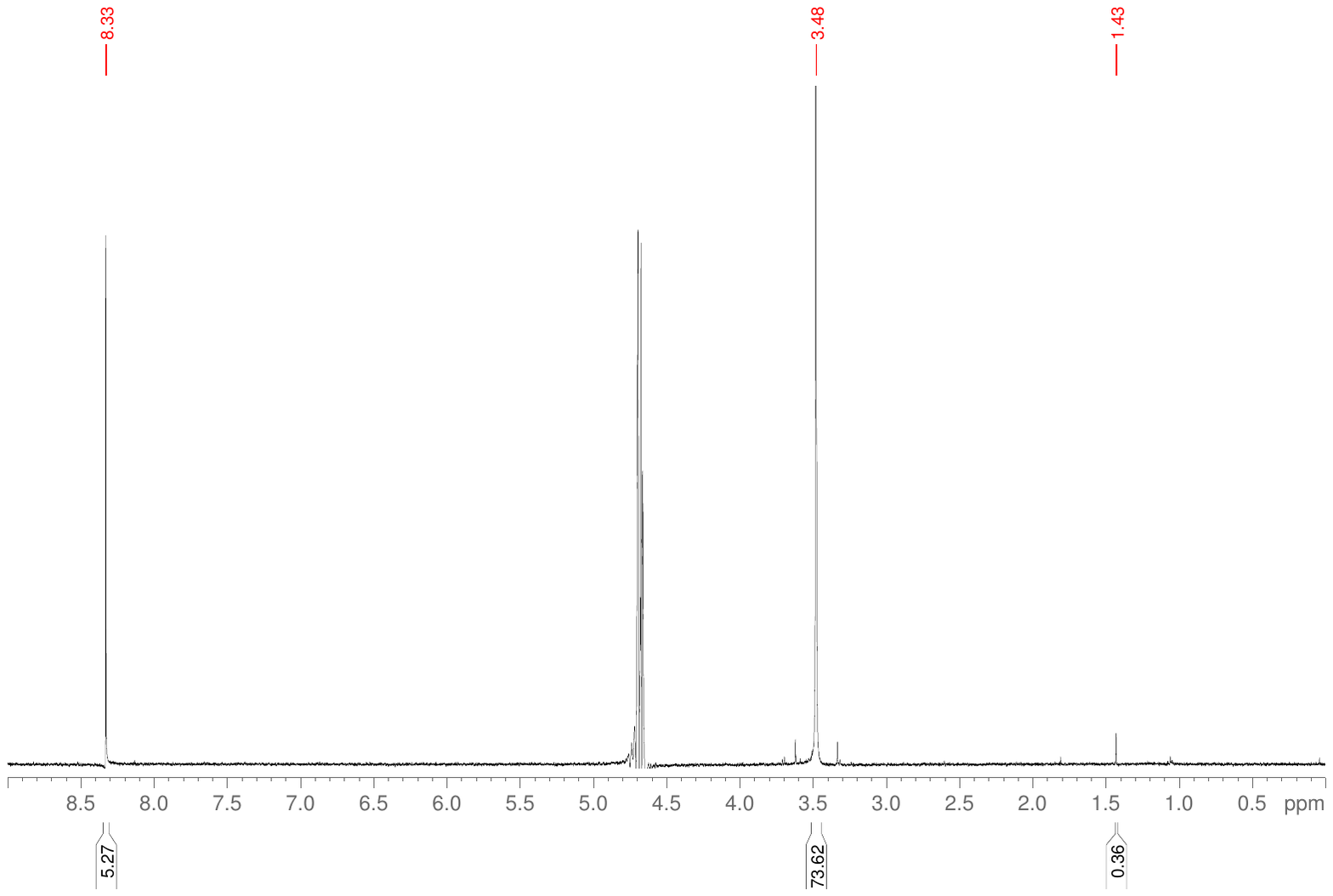}
    \caption{\textbf{Determining the Rate Constant for Formate Production at pH 6, 120 minutes.} The experiment is carried out as in the methods section using 100 mM sodium sulfite and 10 mM sodium bicarbonate. The reaction mixture is irradiated for 120 minutes. This experiment was repeated three times with three different water suppression techniques used for each run. A single NMR spectrum from the third run is shown here using the noesypr1d water suppression sequence.}
    \label{fig:supp_figure17}
\end{figure}

\begin{figure}[H]
    \centering
    \includegraphics[width=1.0\textwidth]{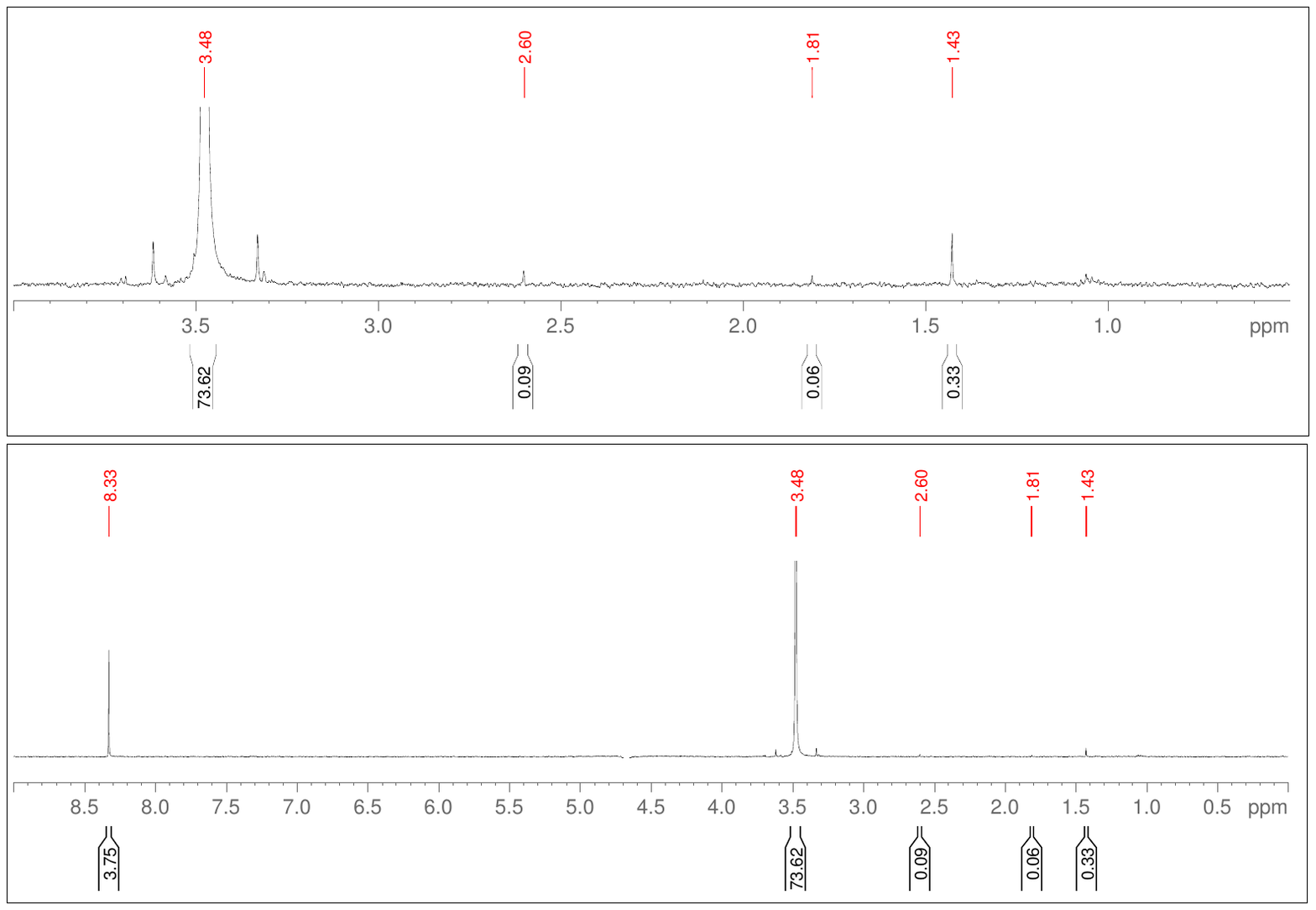}
    \caption{\textbf{Determining the Rate Constant for Formate Production at pH 6, 135 minutes.} The experiment is carried out as in the methods section using 100 mM sodium sulfite and 10 mM sodium bicarbonate. The reaction mixture is irradiated for 135 minutes. This experiment was repeated three times with three different water suppression techniques used for each run. A single NMR spectrum from the third run is shown here using the zgesgppe-cnst12 water suppression sequence.}
    \label{fig:supp_figure18}
\end{figure}

\addcontentsline{toc}{subsection}{The Rate Constant at pH 6: Supplementary Figures \ref{fig:supp_figure11} - \ref{fig:supp_figure18}}

Here we show NMR spectra for experiments used to determine the rate constant for the production of formate at a pH of 6.

\begin{figure}[H]
    \centering
    \includegraphics[width=1.0\textwidth]{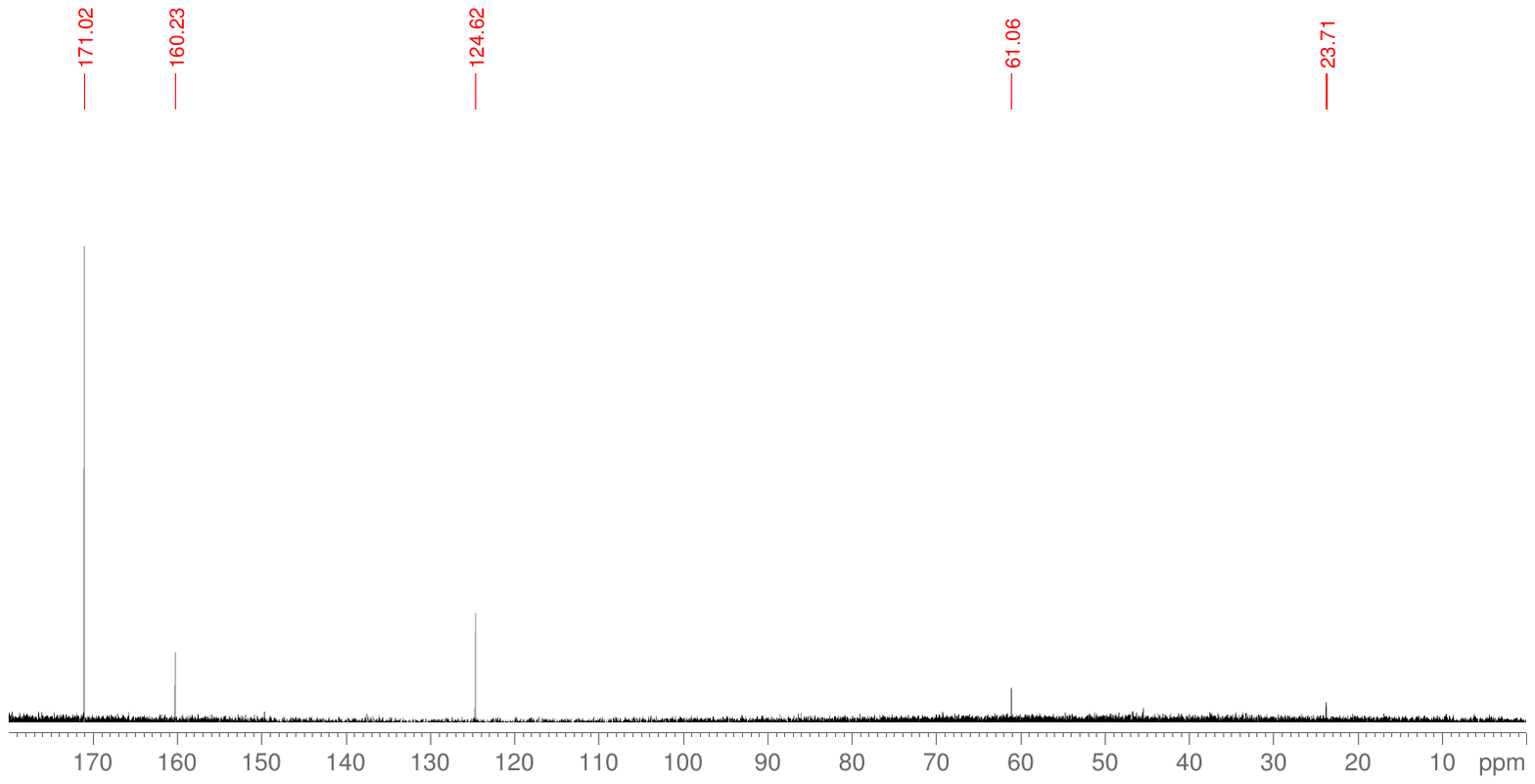}
    \caption{\textbf{Determining the Range of Products Produced by Carboxysulfitic Chemistry at pH 6, $^{13}$C NMR.} The experiment is carried out as in the methods section using 100 mM sodium sulfite and 10 mM $^{13}$C-labelled sodium bicarbonate. The reaction mixture is irradiated for 135 minutes.}
    \label{fig:c13pH6}
\end{figure}

\subsection{The Rate Constant at pH 9}
\begin{figure}[H]
    \centering
    \includegraphics[width=1.0\textwidth]{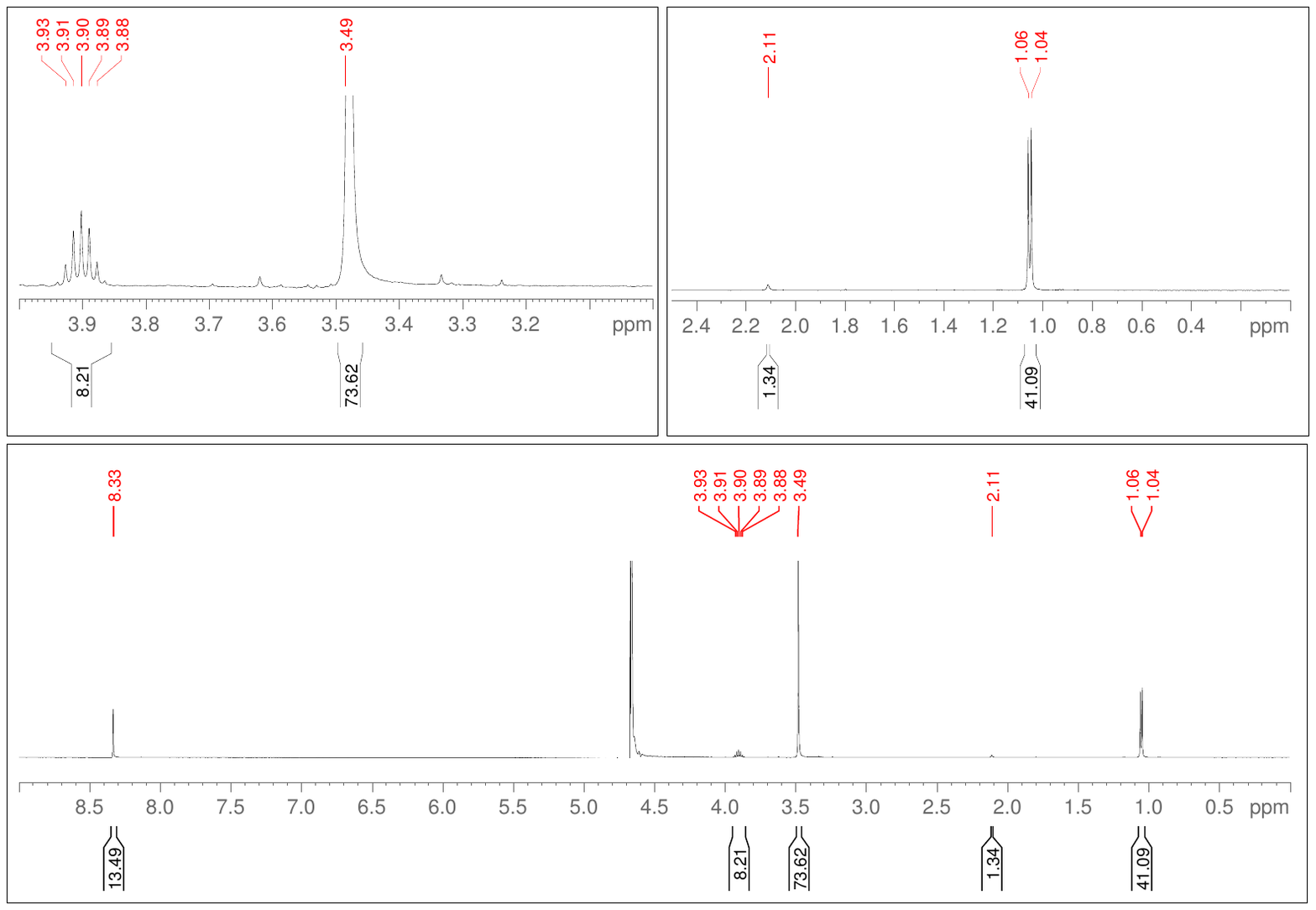}
    \caption{\textbf{Determining the Rate Constant for Formate Production at pH 9, 30 minutes.} The experiment is carried out as in the methods section using 100 mM sodium sulfite and 10 mM sodium bicarbonate. The reaction mixture is irradiated for 30 minutes. This experiment was repeated three times with three different water suppression techniques used for each run. A single NMR spectrum from the first run is shown here using the noesypr1d water suppression sequence.}
    \label{fig:supp_figure19}
\end{figure}

\begin{figure}[H]
    \centering
    \includegraphics[width=1.0\textwidth]{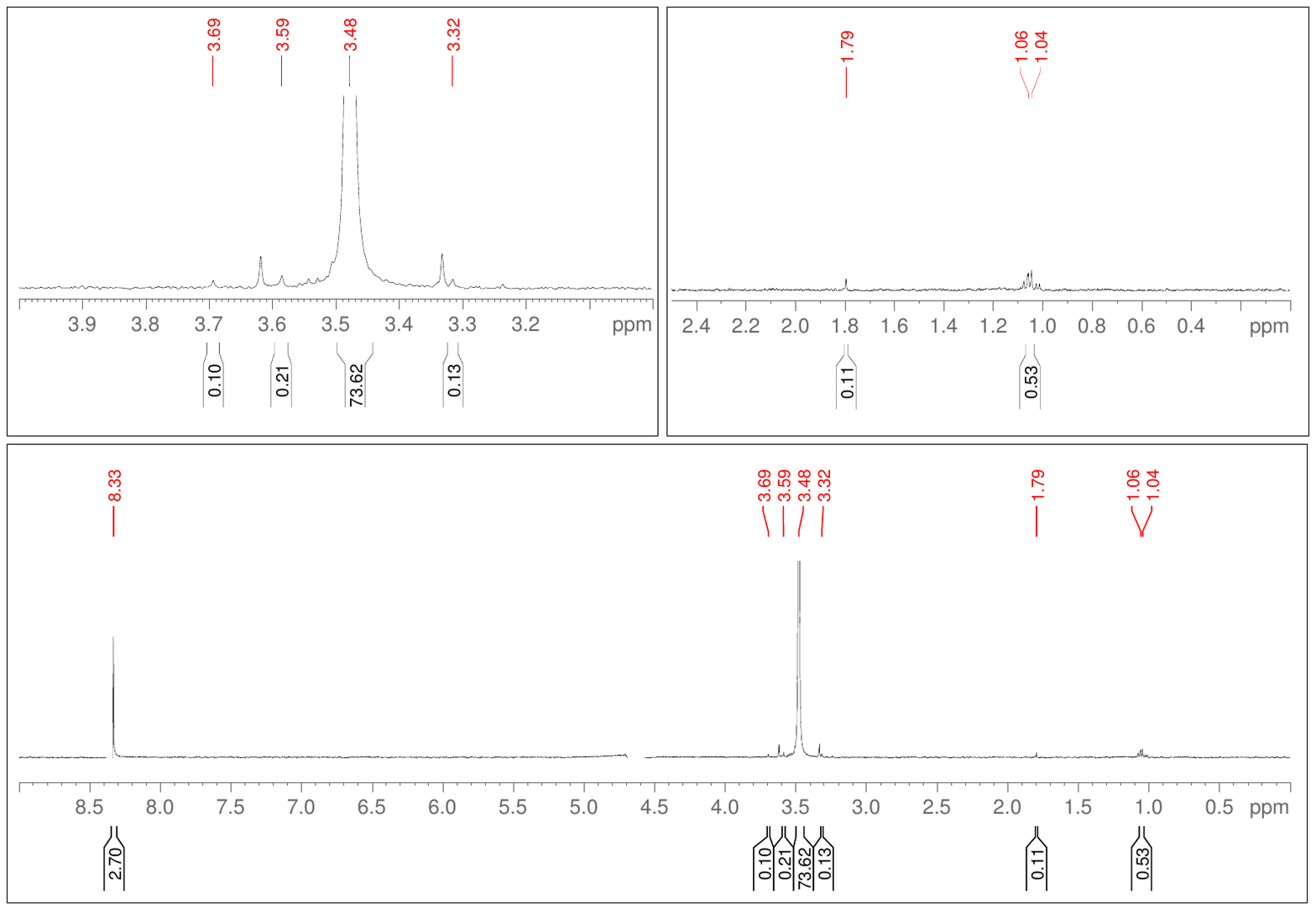}
    \caption{\textbf{Determining the Rate Constant for Formate Production at pH 9, 45 minutes.} The experiment is carried out as in the methods section using 100 mM sodium sulfite and 10 mM sodium bicarbonate. The reaction mixture is irradiated for 45 minutes. This experiment was repeated three times with three different water suppression techniques used for each run. A single NMR spectrum from the first run is shown here using the zgcppr water suppression sequence.}
    \label{fig:supp_figure20}
\end{figure}

\begin{figure}[H]
    \centering
    \includegraphics[width=1.0\textwidth]{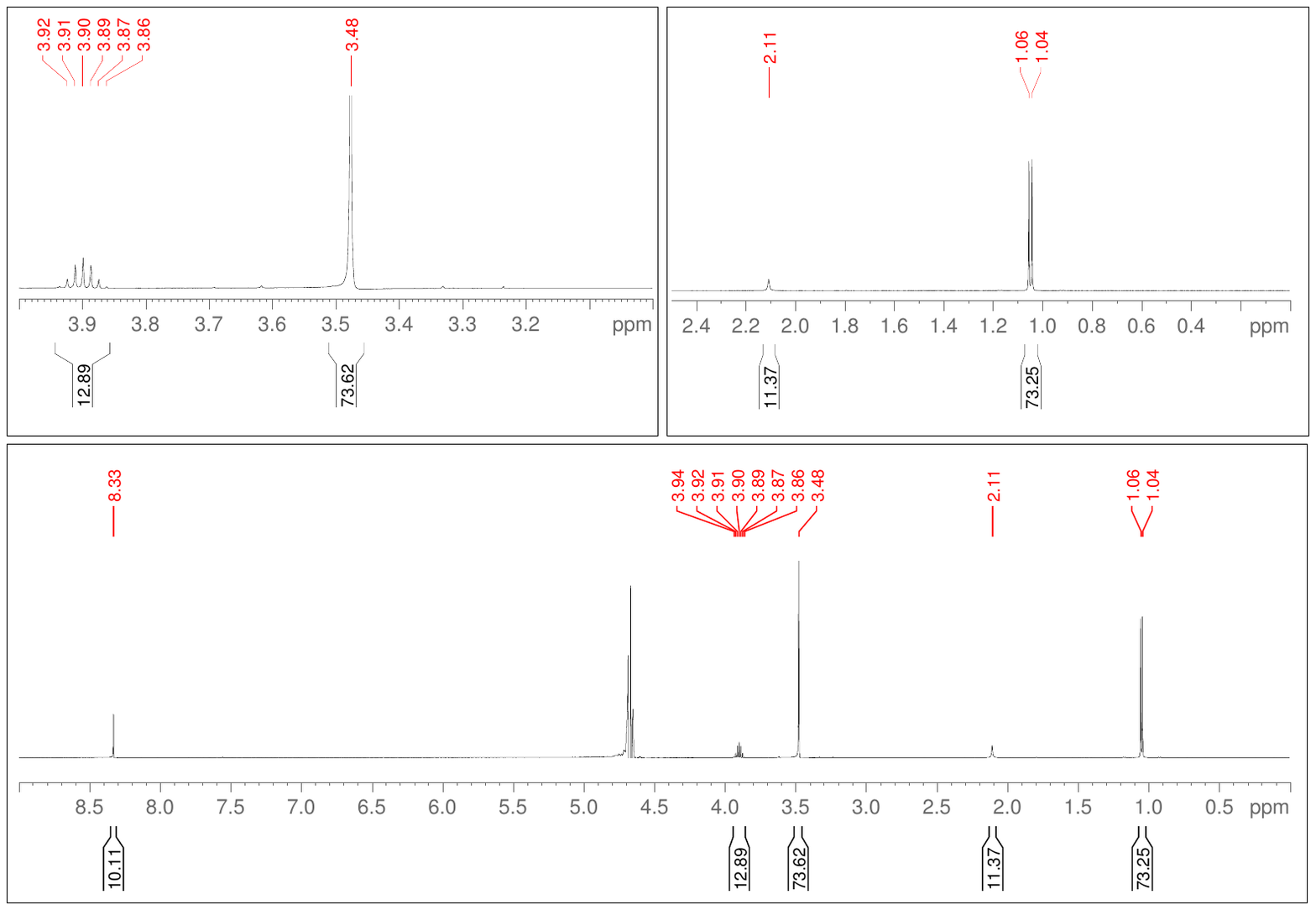}
    \caption{\textbf{Determining the Rate Constant for Formate Production at pH 9, 60 minutes.} The experiment is carried out as in the methods section using 100 mM sodium sulfite and 10 mM sodium bicarbonate. The reaction mixture is irradiated for 60 minutes. This experiment was repeated three times with three different water suppression techniques used for each run. A single NMR spectrum from the second run is shown here using the noesypr1d water suppression sequence.}
    \label{fig:supp_figure21}
\end{figure}

\begin{figure}[H]
    \centering
    \includegraphics[width=1.0\textwidth]{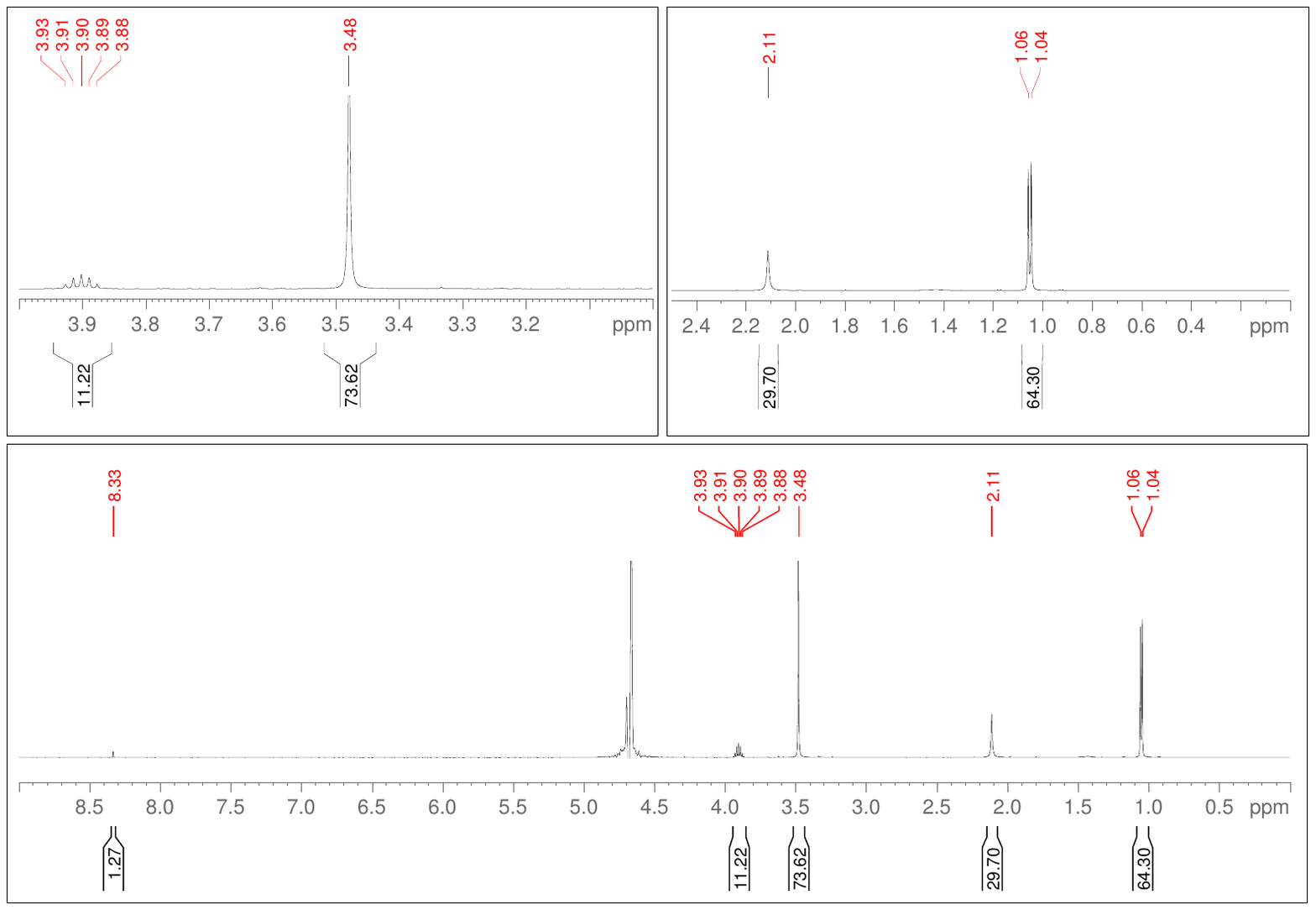}
    \caption{\textbf{Determining the Rate Constant for Formate Production at pH 9, 75 minutes.} The experiment is carried out as in the methods section using 100 mM sodium sulfite and 10 mM sodium bicarbonate. The reaction mixture is irradiated for 75 minutes. This experiment was repeated three times with three different water suppression techniques used for each run. A single NMR spectrum from the first run is shown here using the noesypr1d water suppression sequence.}
    \label{fig:supp_figure22}
\end{figure}

\begin{figure}[H]
    \centering
    \includegraphics[width=1.0\textwidth]{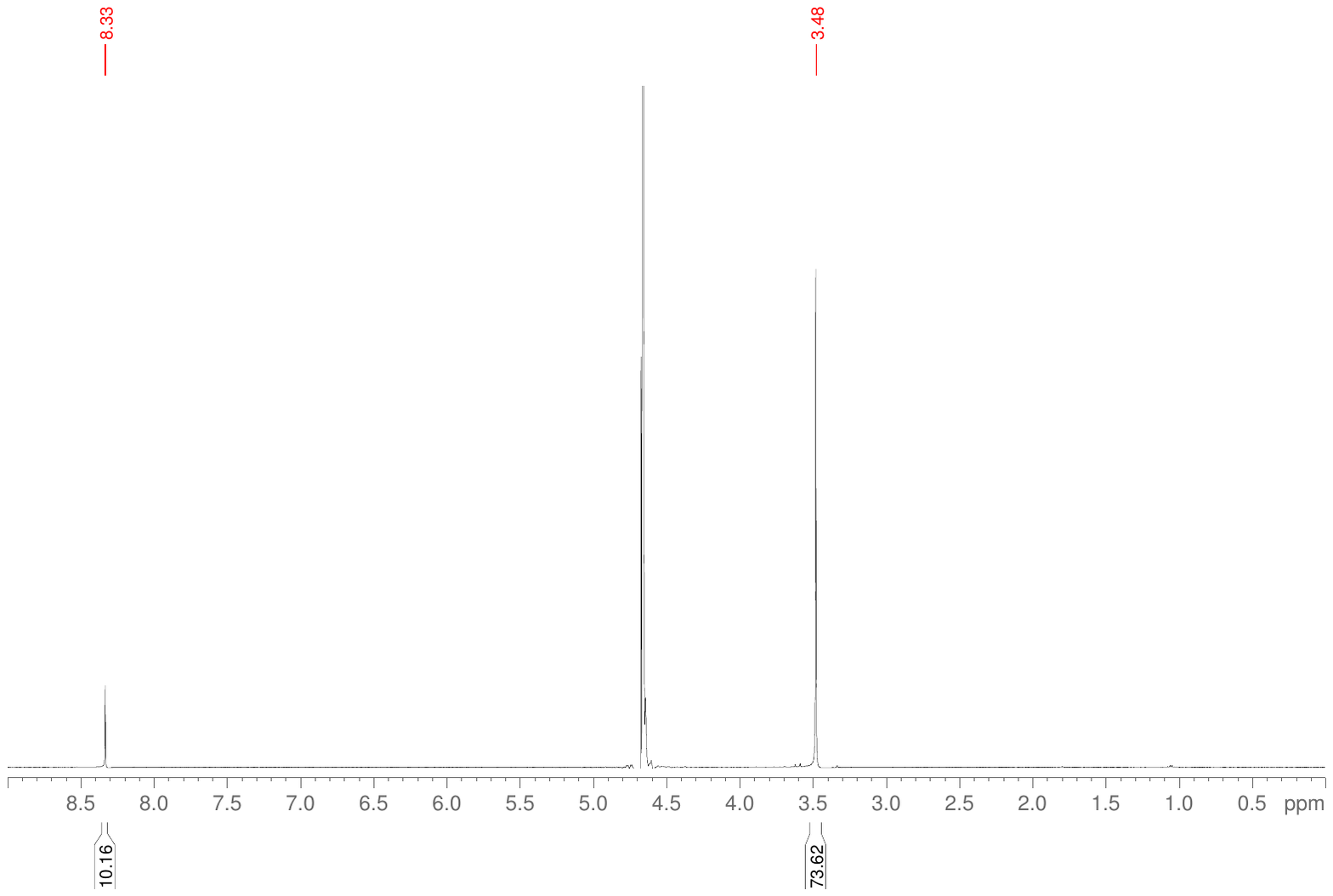}
    \caption{\textbf{Determining the Rate Constant for Formate Production at pH 9, 90 minutes.} The experiment is carried out as in the methods section using 100 mM sodium sulfite and 10 mM sodium bicarbonate. The reaction mixture is irradiated for 90 minutes. This experiment was repeated three times with three different water suppression techniques used for each run. A single NMR spectrum from the first run is shown here using the noesypr1d water suppression sequence.}
    \label{fig:supp_figure23}
\end{figure}

\begin{figure}[H]
    \centering
    \includegraphics[width=1.0\textwidth]{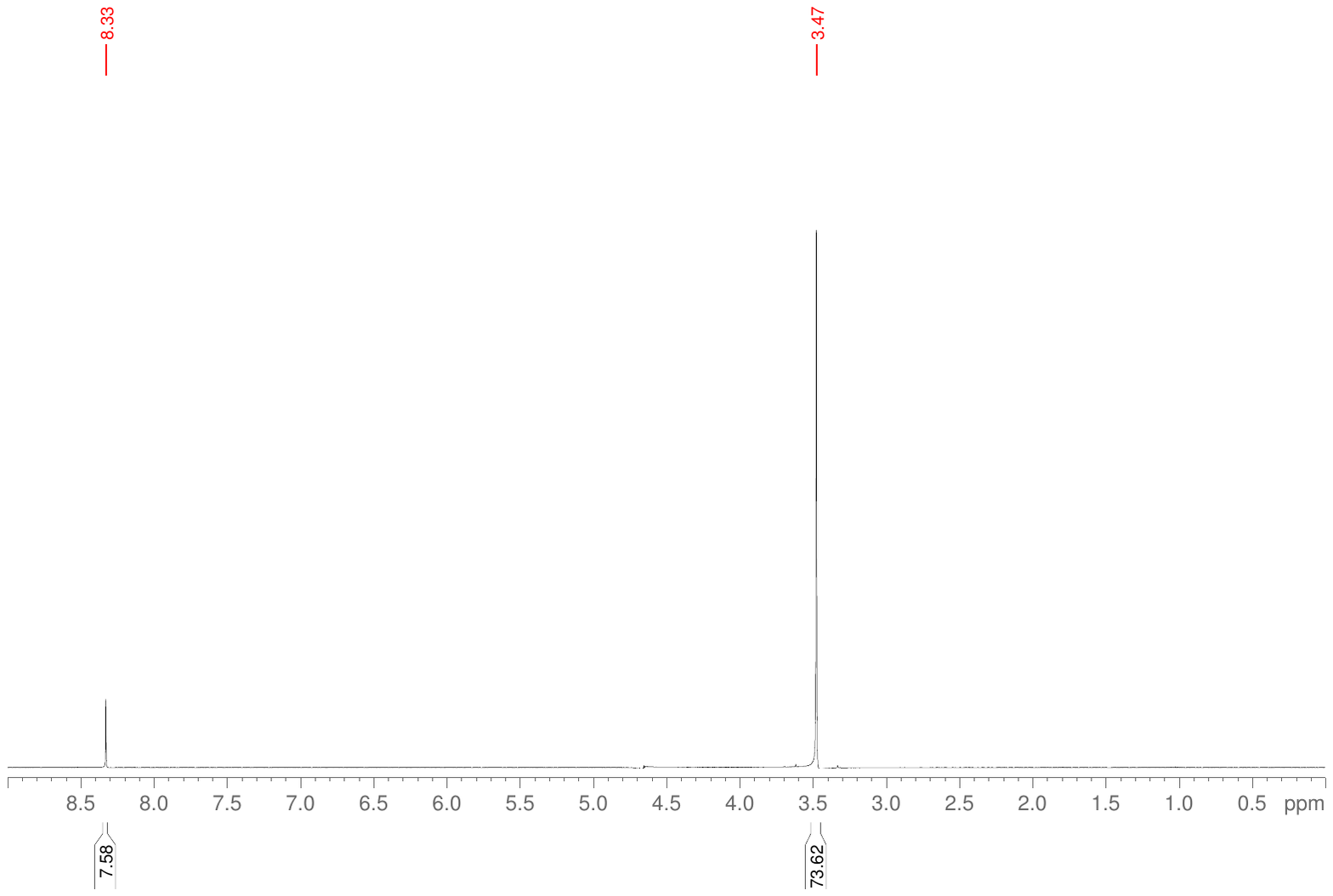}
    \caption{\textbf{Determining the Rate Constant for Formate Production at pH 9, 105 minutes.} The experiment is carried out as in the methods section using 100 mM sodium sulfite and 10 mM sodium bicarbonate. The reaction mixture is irradiated for 105 minutes. This experiment was repeated three times with three different water suppression techniques used for each run. A single NMR spectrum from the third run is shown here using the zgesgppe-cnst12 water suppression sequence.}
    \label{fig:supp_figure24}
\end{figure}

\begin{figure}[H]
    \centering
    \includegraphics[width=1.0\textwidth]{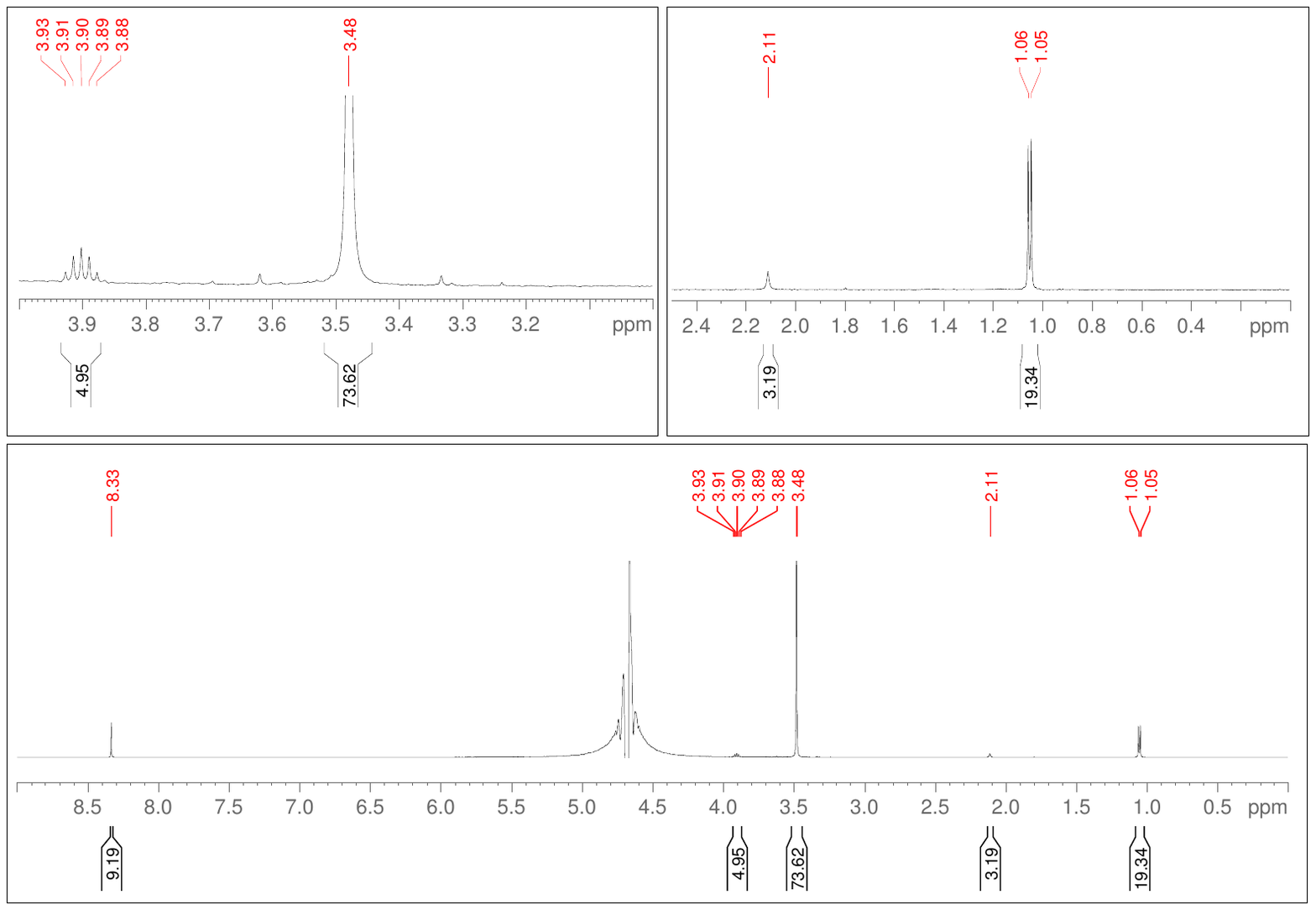}
    \caption{\textbf{Determining the Rate Constant for Formate Production at pH 9, 120 minutes.} The experiment is carried out as in the methods section using 100 mM sodium sulfite and 10 mM sodium bicarbonate. The reaction mixture is irradiated for 120 minutes. This experiment was repeated three times with three different water suppression techniques used for each run. A single NMR spectrum from the first run is shown here using the noesypr1d water suppression sequence.}
    \label{fig:supp_figure25}
\end{figure}

\begin{figure}[H]
    \centering
    \includegraphics[width=1.0\textwidth]{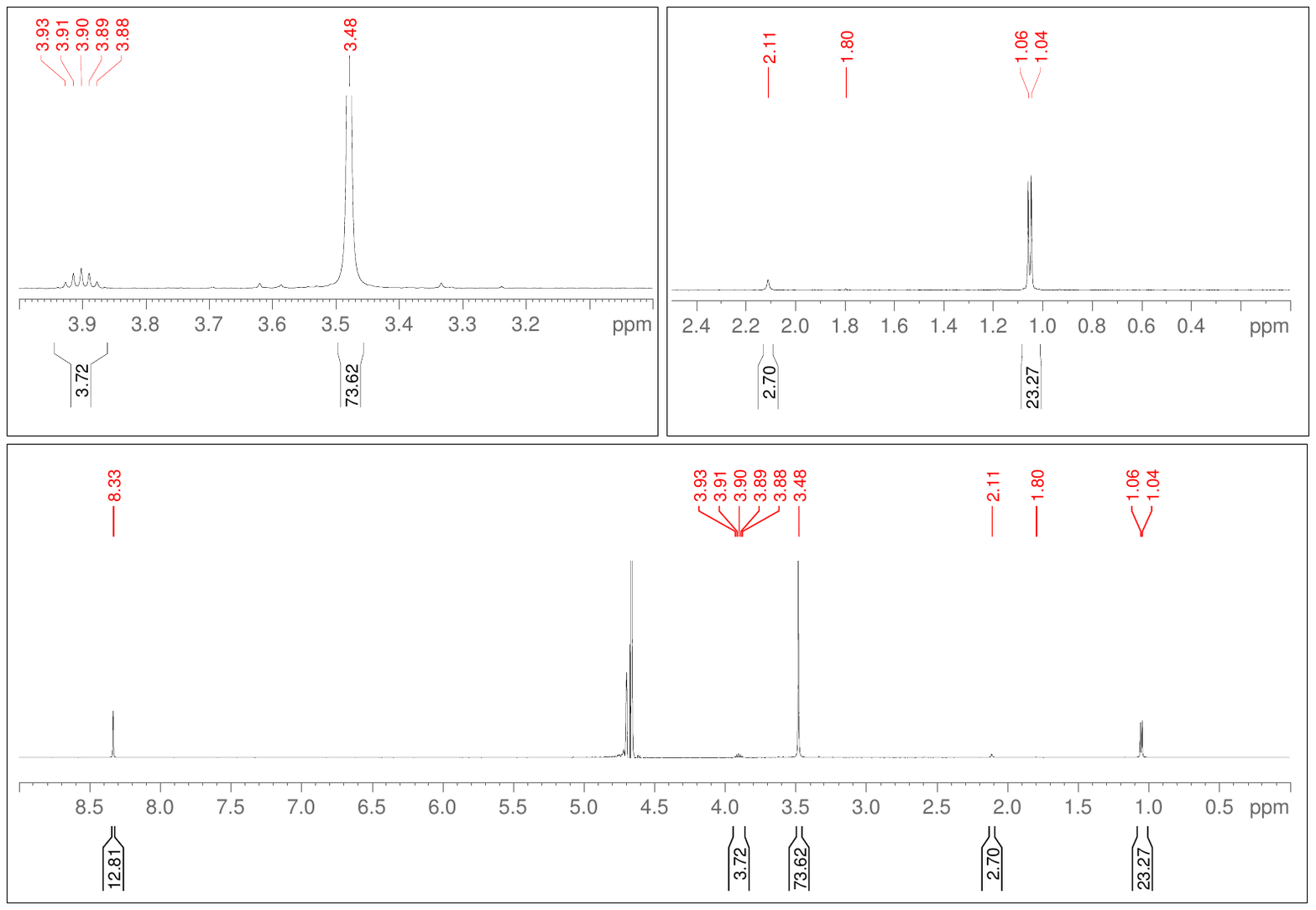}
    \caption{\textbf{Determining the Rate Constant for Formate Production at pH 9, 135 minutes.} The experiment is carried out as in the methods section using 100 mM sodium sulfite and 10 mM sodium bicarbonate. The reaction mixture is irradiated for 135 minutes. This experiment was repeated three times with three different water suppression techniques used for each run. A single NMR spectrum from the first run is shown here using the noesypr1d water suppression sequence.}
    \label{fig:supp_figure26}
\end{figure}

\addcontentsline{toc}{subsection}{The Rate Constant at pH 9: Supplementary Figures \ref{fig:supp_figure19} - \ref{fig:supp_figure26}}

Here we show NMR spectra for experiments used to determine the rate constant for the production of formate at a pH of 9.

\begin{figure}[H]
    \centering
    \includegraphics[width=1.0\textwidth]{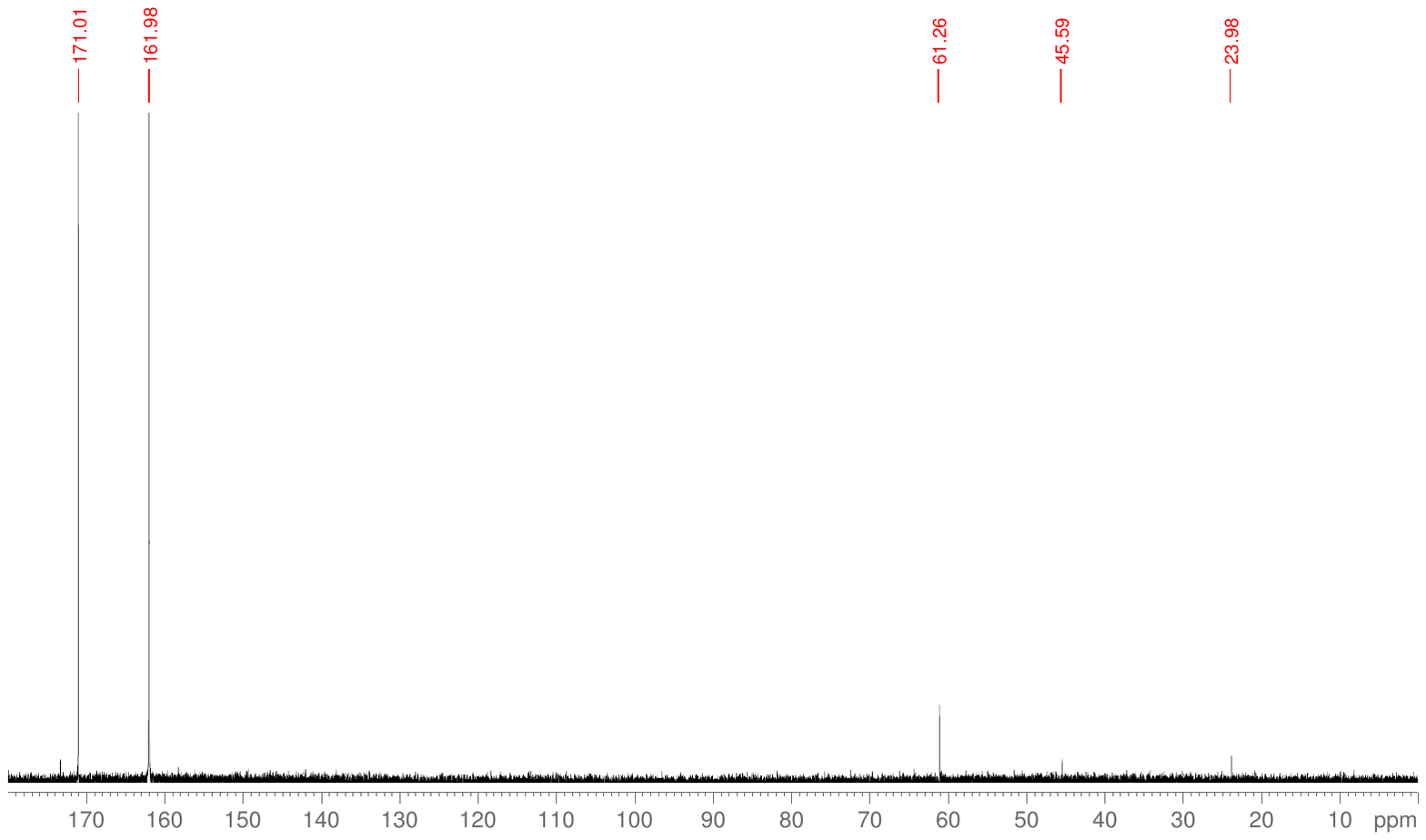}
    \caption{\textbf{Determining the Range of Products Produced by Carboxysulfitic Chemistry at pH 9, $^{13}$C NMR.} The experiment is carried out as in the methods section using 100 mM sodium sulfite and 10 mM $^{13}$C-labelled sodium bicarbonate. The reaction mixture is irradiated for 135 minutes.}
    \label{fig:c13pH9}
\end{figure}

\subsection{The Rate Constant at pH 12}
\begin{figure}[H]
    \centering
    \includegraphics[width=1.0\textwidth]{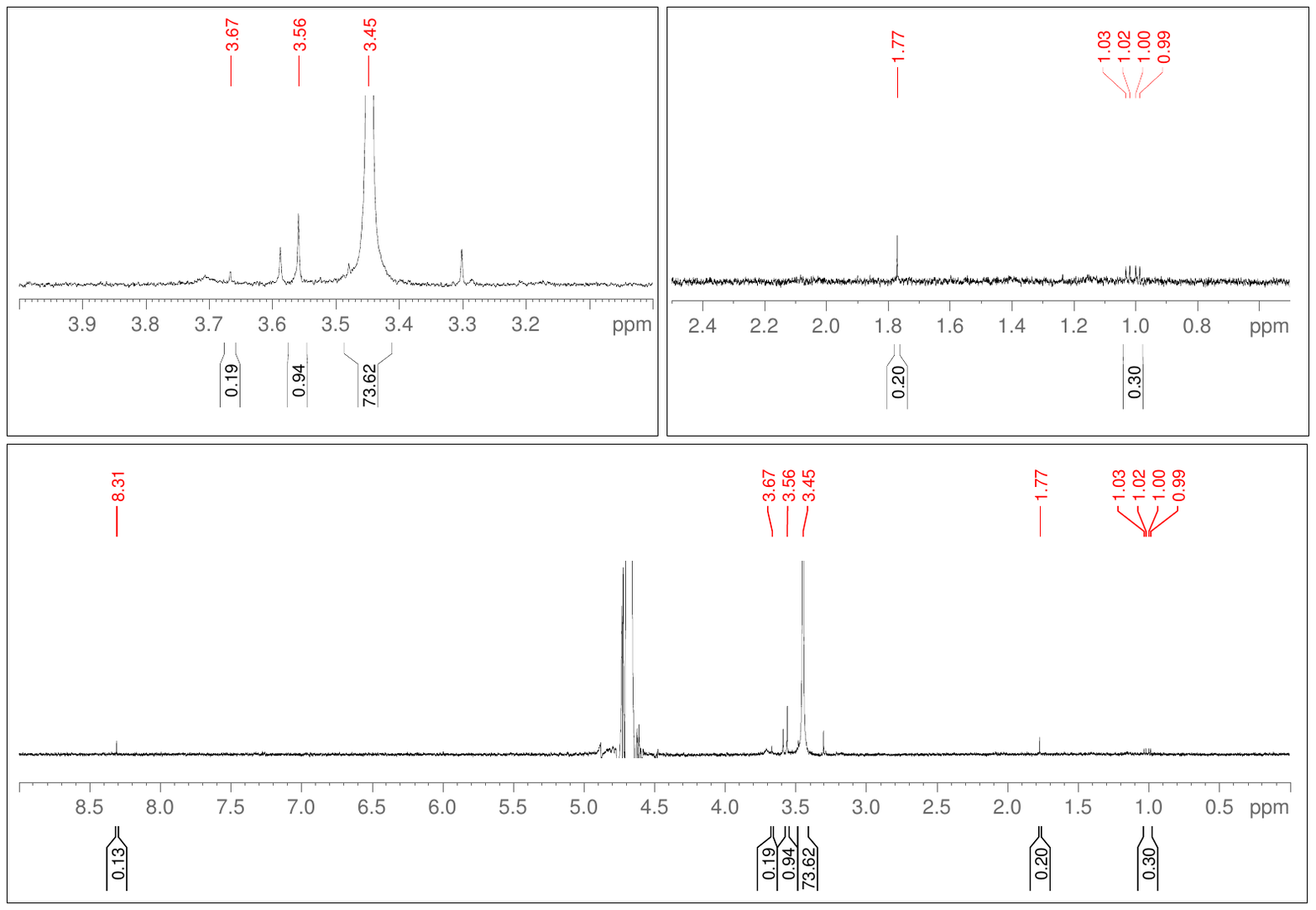}
    \caption{\textbf{Determining the Rate Constant for Formate Production at pH 12, 30 minutes.} The experiment is carried out as in the methods section using 100 mM sodium sulfite and 10 mM sodium bicarbonate. The reaction mixture is irradiated for 30 minutes. This experiment was repeated three times with three different water suppression techniques used for each run. A single NMR spectrum from the third run is shown here using the zgcppr water suppression sequence.}
    \label{fig:supp_figure27}
\end{figure}

\begin{figure}[H]
    \centering
    \includegraphics[width=1.0\textwidth]{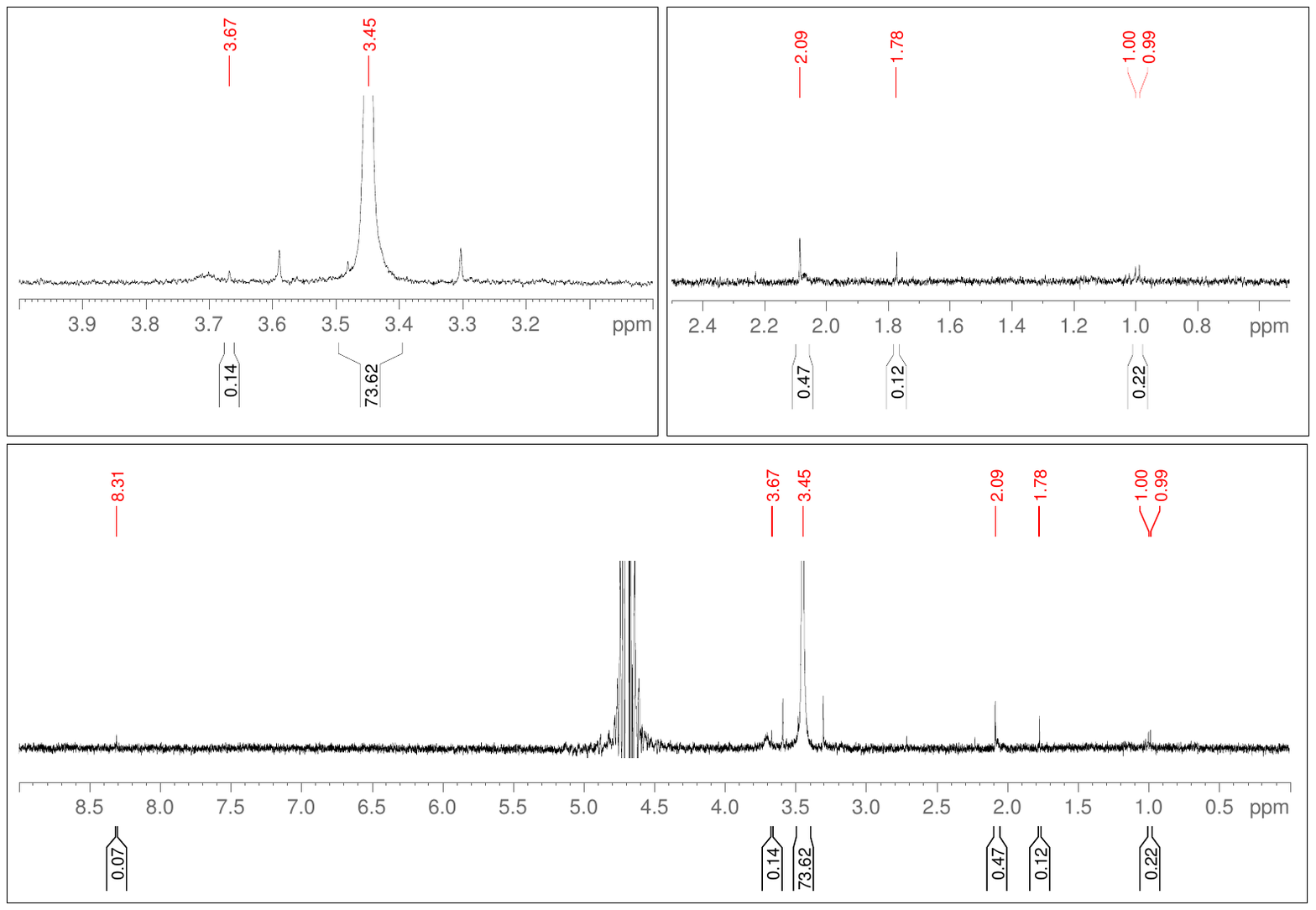}
    \caption{\textbf{Determining the Rate Constant for Formate Production at pH 12, 45 minutes.} The experiment is carried out as in the methods section using 100 mM sodium sulfite and 10 mM sodium bicarbonate. The reaction mixture is irradiated for 45 minutes. This experiment was repeated three times with three different water suppression techniques used for each run. A single NMR spectrum from the first run is shown here using the noesypr1d water suppression sequence.}
    \label{fig:supp_figure28}
\end{figure}

\begin{figure}[H]
    \centering
    \includegraphics[width=1.0\textwidth]{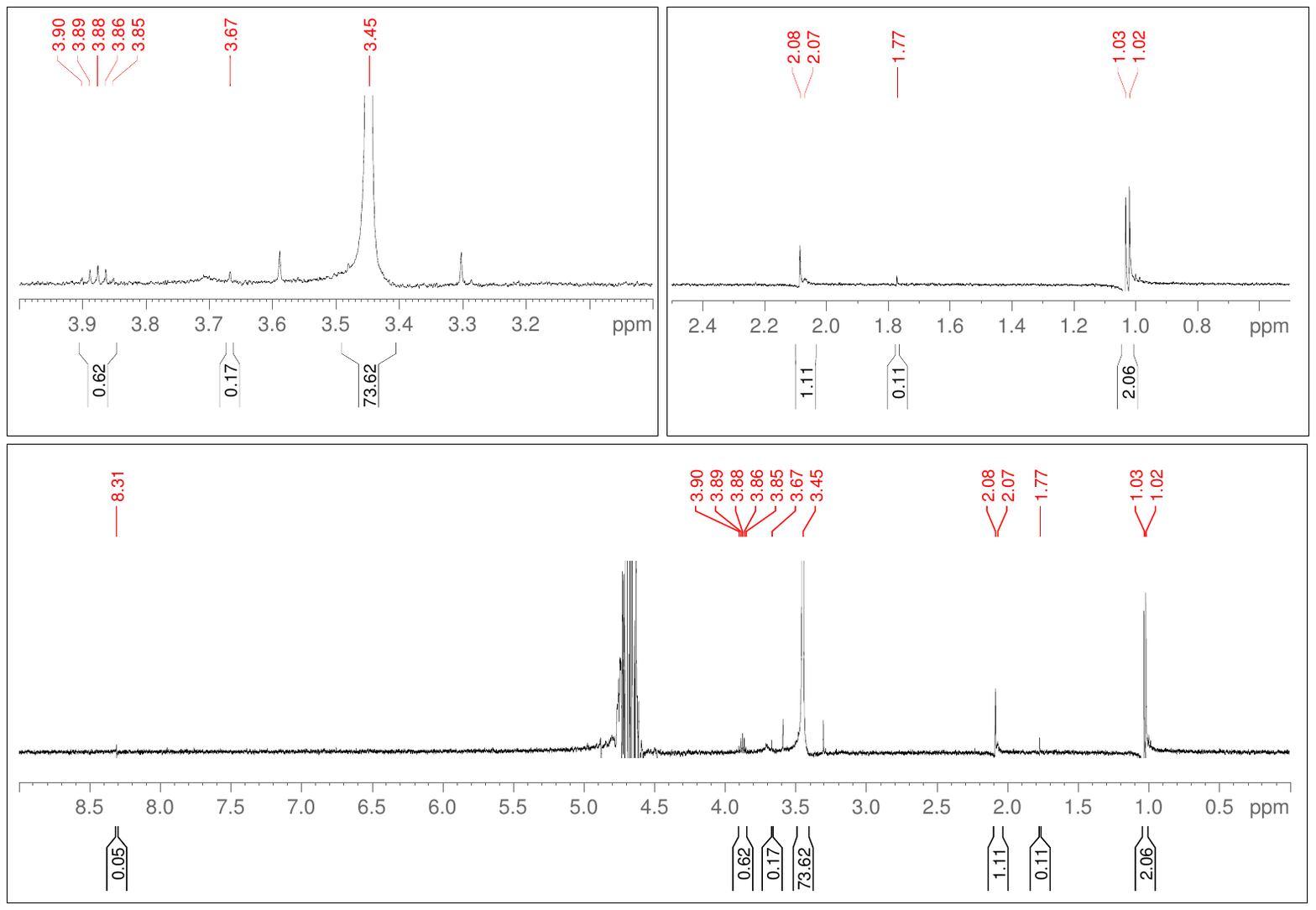}
    \caption{\textbf{Determining the Rate Constant for Formate Production at pH 12, 60 minutes.} The experiment is carried out as in the methods section using 100 mM sodium sulfite and 10 mM sodium bicarbonate. The reaction mixture is irradiated for 60 minutes. This experiment was repeated three times with three different water suppression techniques used for each run. A single NMR spectrum from the second run is shown here using the noesypr1d water suppression sequence.}
    \label{fig:supp_figure29}
\end{figure}

\begin{figure}[H]
    \centering
    \includegraphics[width=1.0\textwidth]{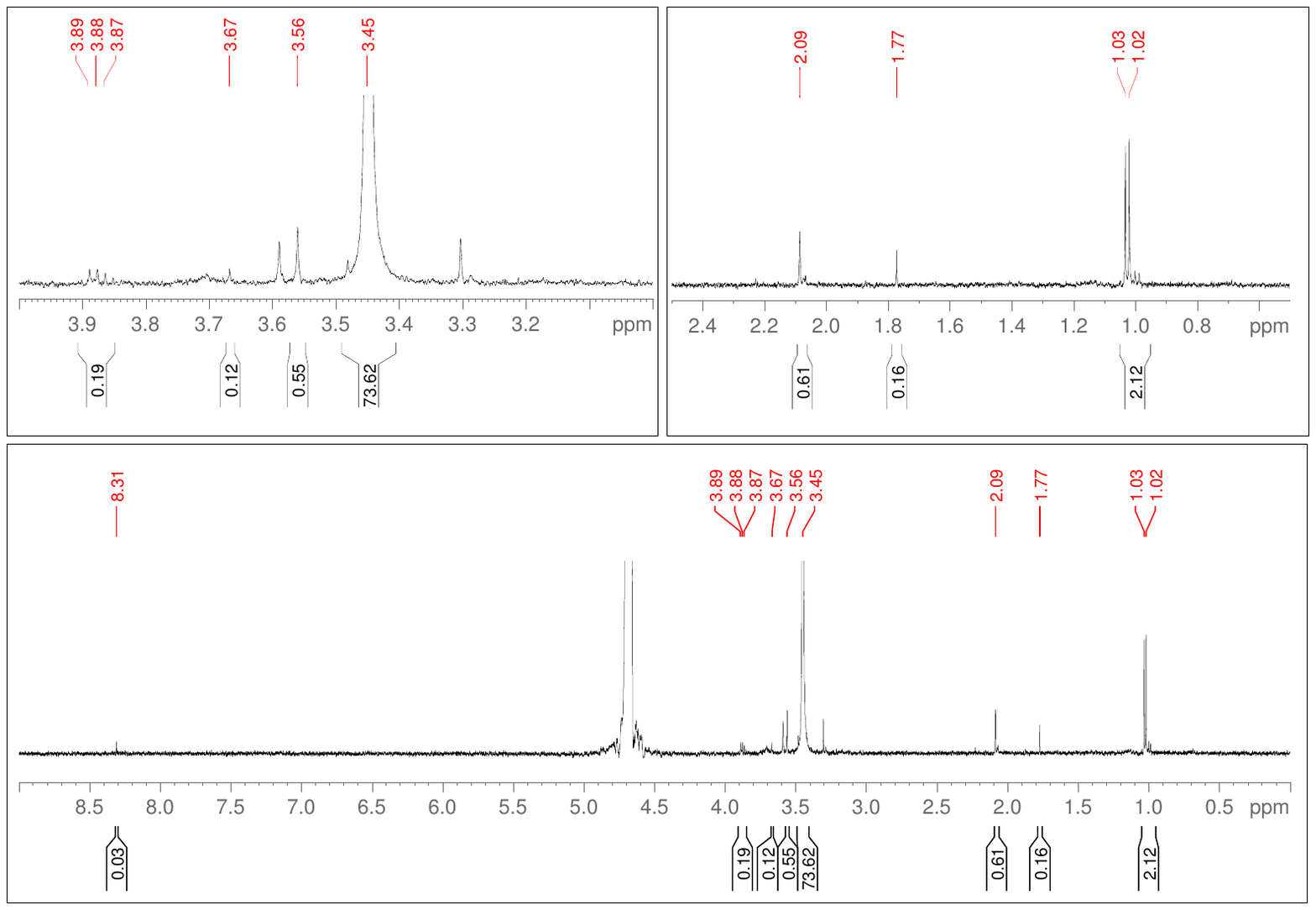}
    \caption{\textbf{Determining the Rate Constant for Formate Production at pH 12, 75 minutes.} The experiment is carried out as in the methods section using 100 mM sodium sulfite and 10 mM sodium bicarbonate. The reaction mixture is irradiated for 75 minutes. This experiment was repeated three times with three different water suppression techniques used for each run. A single NMR spectrum from the first run is shown here using the zgcppr water suppression sequence.}
    \label{fig:supp_figure30}
\end{figure}

\begin{figure}[H]
    \centering
    \includegraphics[width=1.0\textwidth]{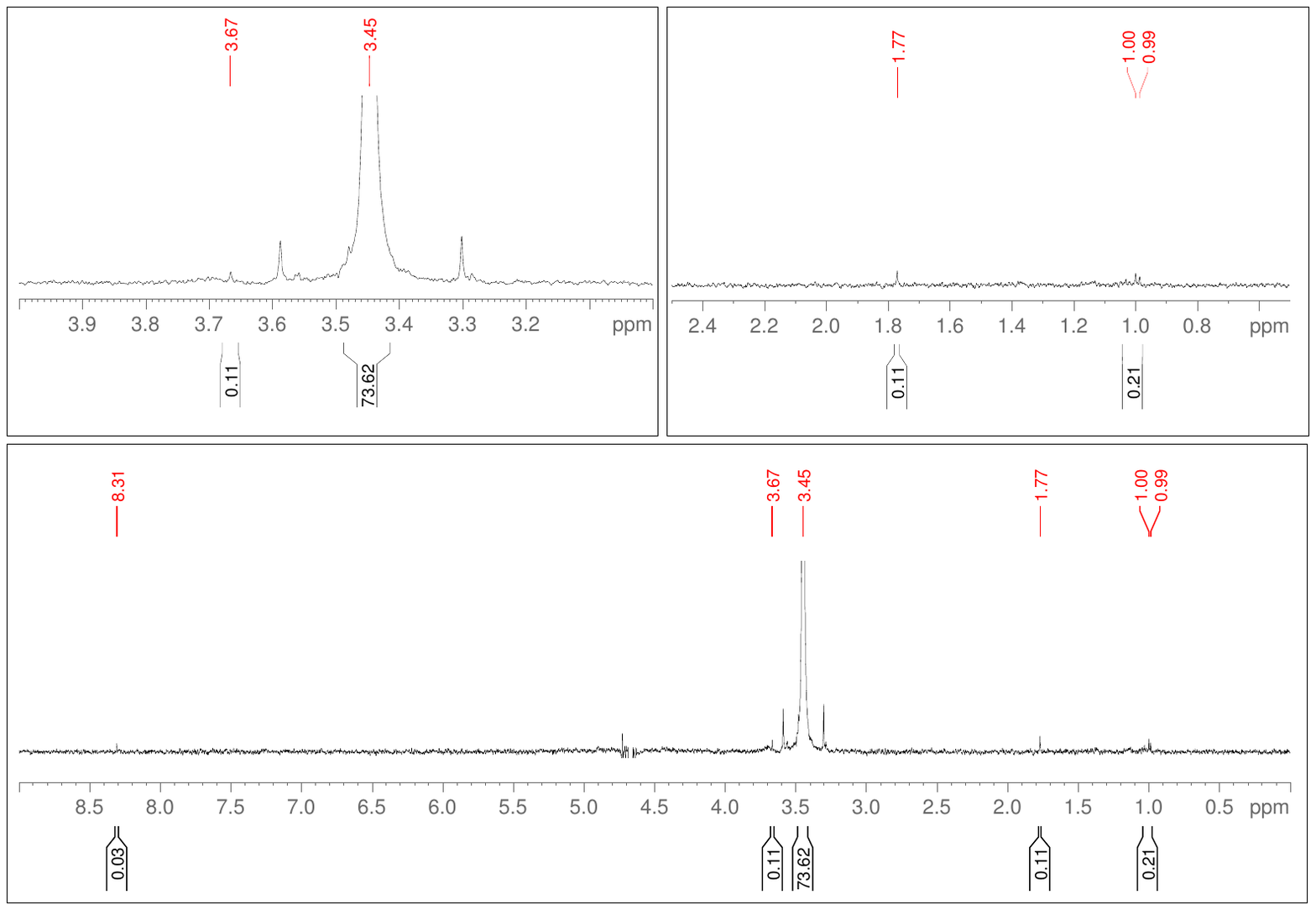}
    \caption{\textbf{Determining the Rate Constant for Formate Production at pH 12, 90 minutes.} The experiment is carried out as in the methods section using 100 mM sodium sulfite and 10 mM sodium bicarbonate. The reaction mixture is irradiated for 90 minutes. This experiment was repeated three times with three different water suppression techniques used for each run. A single NMR spectrum from the second run is shown here using the zgesgppe-cnst12 water suppression sequence.}
    \label{fig:supp_figure31}
\end{figure}

\begin{figure}[H]
    \centering
    \includegraphics[width=1.0\textwidth]{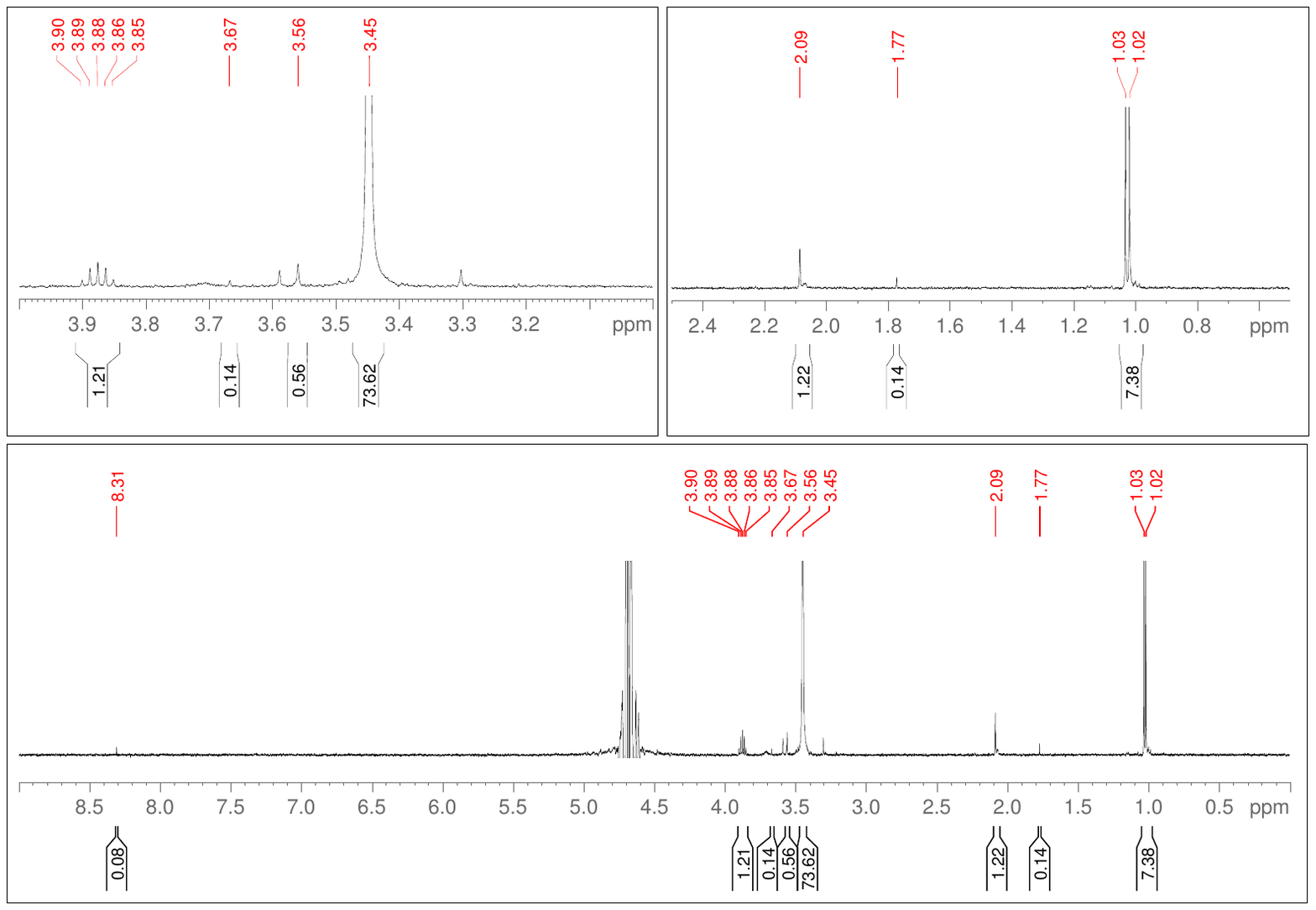}
    \caption{\textbf{Determining the Rate Constant for Formate Production at pH 12, 105 minutes.} The experiment is carried out as in the methods section using 100 mM sodium sulfite and 10 mM sodium bicarbonate. The reaction mixture is irradiated for 105 minutes. This experiment was repeated three times with three different water suppression techniques used for each run. A single NMR spectrum from the first run is shown here using the zgesgppe-cnst12 water suppression sequence.}
    \label{fig:supp_figure32}
\end{figure}

\begin{figure}[H]
    \centering
    \includegraphics[width=1.0\textwidth]{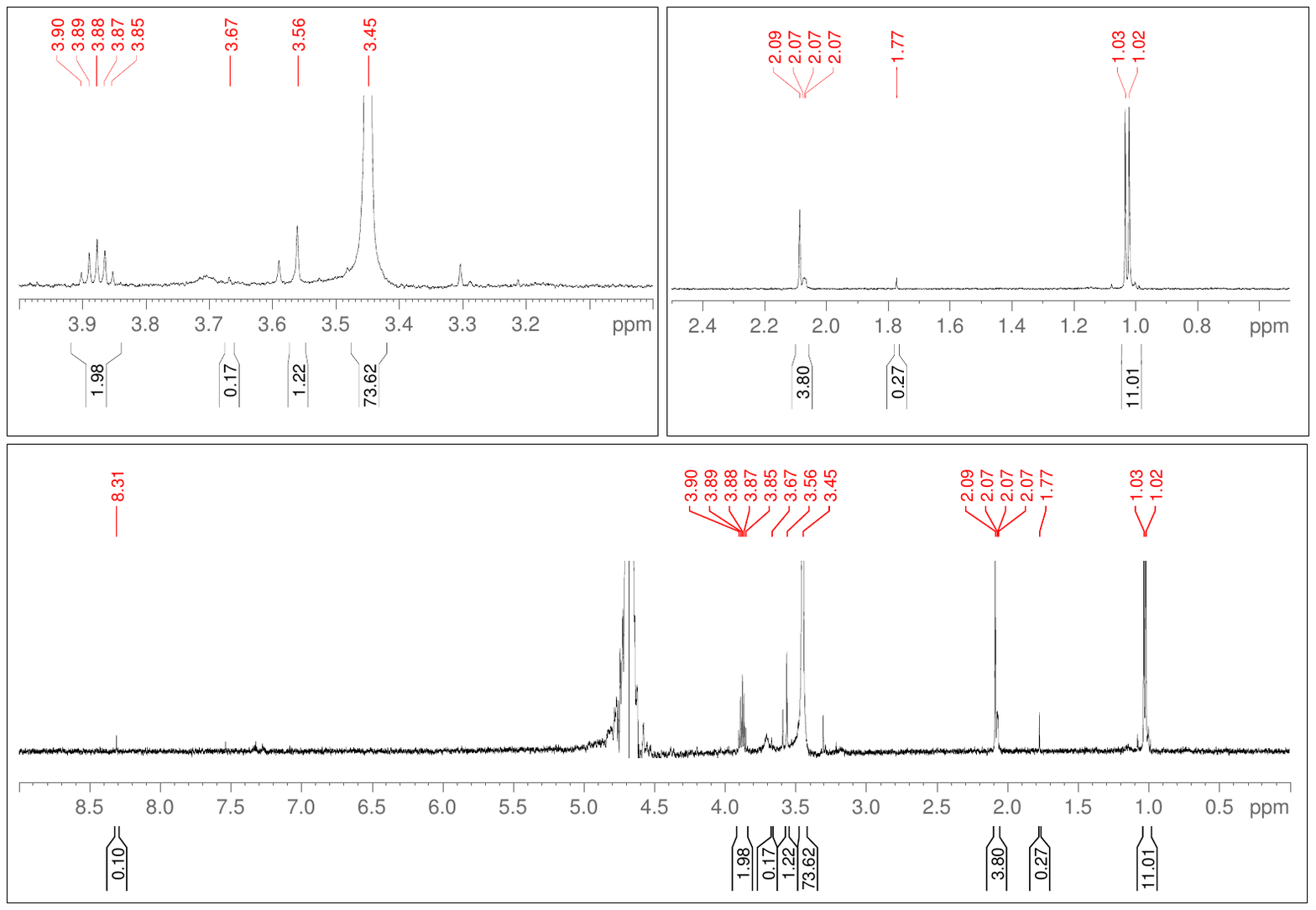}
    \caption{\textbf{Determining the Rate Constant for Formate Production at pH 12, 120 minutes.} The experiment is carried out as in the methods section using 100 mM sodium sulfite and 10 mM sodium bicarbonate. The reaction mixture is irradiated for 120 minutes. This experiment was repeated three times with three different water suppression techniques used for each run. A single NMR spectrum from the first run is shown here using the zgcppr water suppression sequence.}
    \label{fig:supp_figure33}
\end{figure}

\begin{figure}[H]
    \centering
    \includegraphics[width=1.0\textwidth]{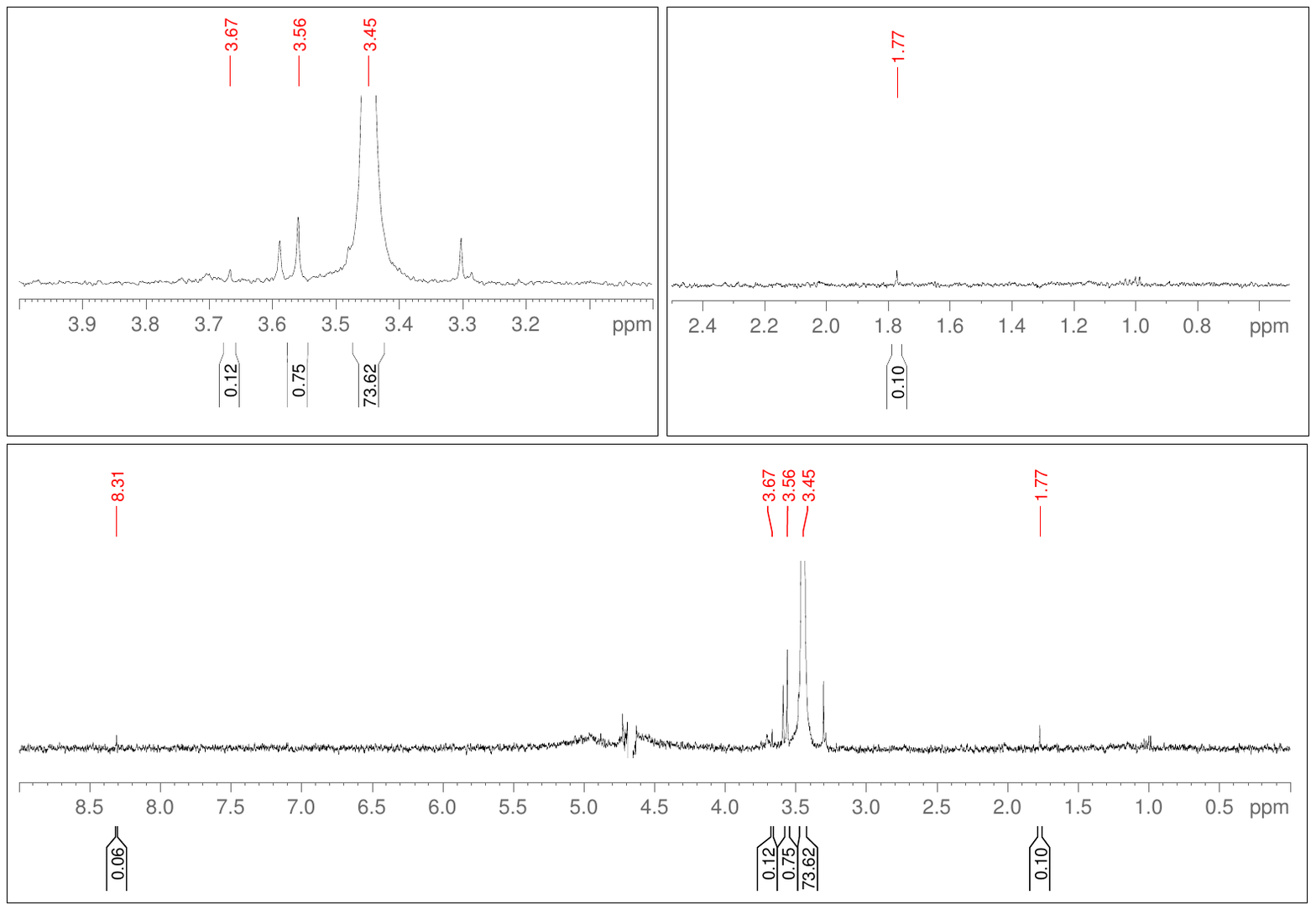}
    \caption{\textbf{Determining the Rate Constant for Formate Production at pH 12, 135 minutes.} The experiment is carried out as in the methods section using 100 mM sodium sulfite and 10 mM sodium bicarbonate. The reaction mixture is irradiated for 135 minutes. This experiment was repeated three times with three different water suppression techniques used for each run. A single NMR spectrum from the second run is shown here using the zgesgppe-cnst12 water suppression sequence.}
    \label{fig:supp_figure34}
\end{figure}

\addcontentsline{toc}{subsection}{The Rate Constant at pH 12: Supplementary Figures \ref{fig:supp_figure27} - \ref{fig:supp_figure34}}

Here we show NMR spectra for experiments used to determine the rate constant for the production of formate at a pH of 12.

\begin{figure}[H]
    \centering
    \includegraphics[width=1.0\textwidth]{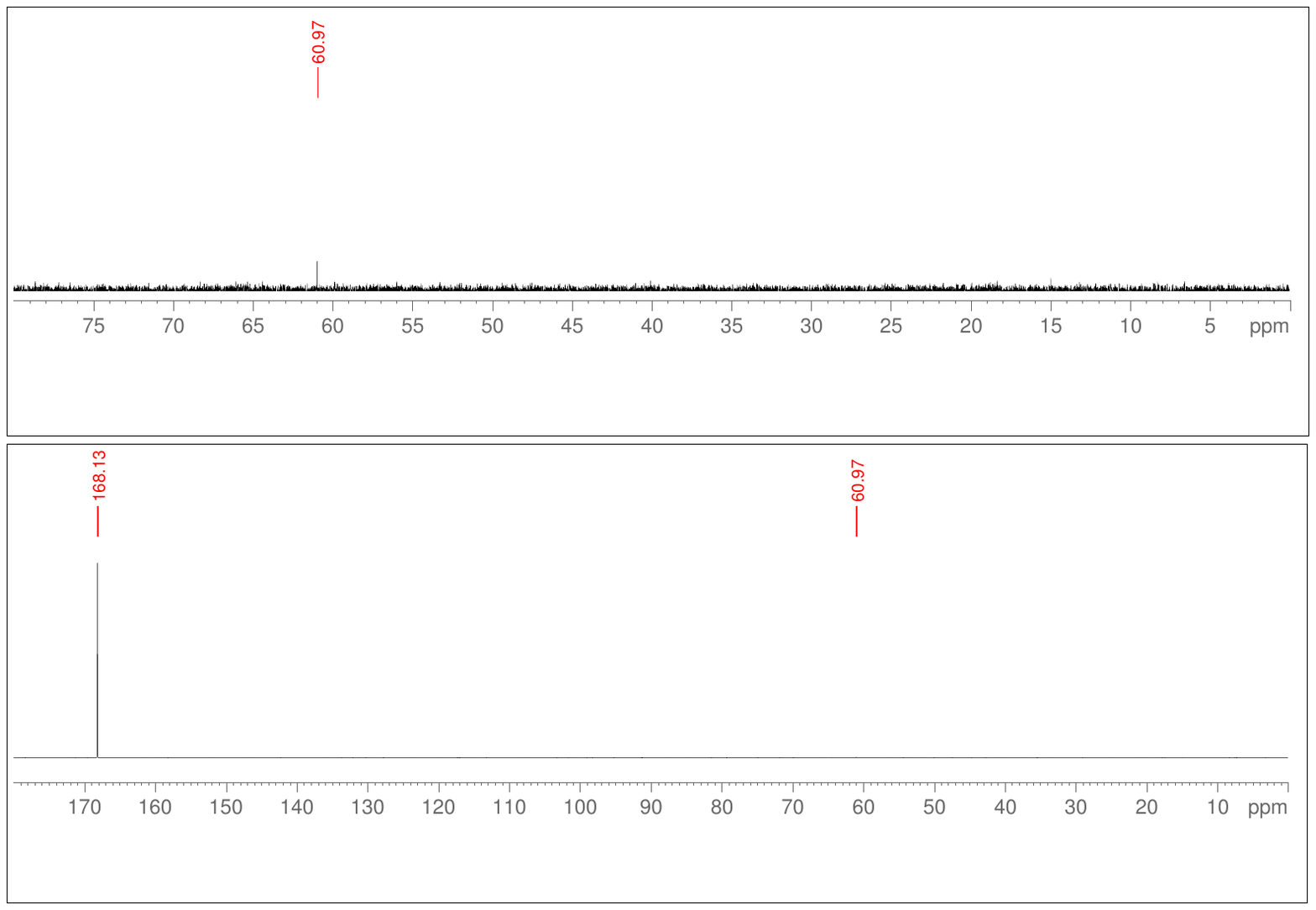}
    \caption{\textbf{Determining the Range of Products Produced by Carboxysulfitic Chemistry at 12, $^{13}$C NMR.} The experiment is carried out as in the methods section using 100 mM sodium sulfite and 10 mM $^{13}$C-labeled sodium bicarbonate. The reaction mixture is irradiated for 135 minutes.}
    \label{fig:c13pH12}
\end{figure}

\section{The Predicted Order of Reaction with Respect to Sulfite in the Production of Formate due to Self-Shielding}

\addcontentsline{toc}{section}{The Predicted Order of Reaction with Respect to Sulfite in the Production of Formate due to Self-Shielding}

\setcounter{equation}{0}

In this section, we develop an analytical model for the electron photodetachment and reduction of carbon species. We cast this in the most general form we can, to incorporate both of our proposed mechanisms for carbon reduction, and the possible role of ion-clustering in affecting the order of the reaction.\\

We start with a minimal network that encapsulates both of our proposed mechanisms for the production of formate. This network includes the chemical species sulfite (\ce{SO3^{2-}}), the radical (\cdot\ce{SO3-}), solvated electrons ($e^-$), carbon dioxide (\ce{CO2}), dithionate (\ce{S2O6^{2-}}) and a combination of reduced carbon and sulfur, represented by $\ce{CO2^-} \cdot j \, \ce{SO3^{2-}}$. The reactions are:

\begin{align}
\ch{SO3^{2-}} + h\nu &\ch{-> \cdot SO3^-} + e^-, \label{eqn:detach} \tag{R1} \\
\cdot\ch{SO3^-} + e^- + \ell \, \ch{SO3^{2-}} &\ch{->} \big(\ell + 1\big) \, \ch{SO3^{2-}}, \label{eqn:reform} \tag{R2} \\
\ch{CO2} + e^- + j \, \ch{SO3^{2-}} &\ch{-> CO2^-} \cdot j \, \ch{SO3^{2-}}, \label{eqn:formate} \tag{R3} \\
\ch{2 \cdot SO3^- &-> S2O6^{2-}}; \label{eqn:adduct} \tag{R4}
\end{align}
with rate constants $k_1$ to $k_4$, respectively. The concentrations are denoted by $n_s \, {\rm [M]}$ for \ce{SO3^{2-}}, $n_{s'} \, {\rm [M]}$ for $\cdot\ce{SO3-}$, $n_e \, {\rm [M]}$ for $e^-$, $n_c \, {\rm [M]}$ for \ce{CO2}, $n_{s2} \, {\rm [M]}$ for \ce{S2O6^{2-}}, and $n_f \, {\rm [M]}$ for $\ce{CO2^-} \cdot j \, \ce{SO3^{2-}}$. The rate constants have units of:
\begin{align}
k_1 & \; \; : \; \; {\rm s^{-1}}\notag\\
k_2 & \; \; : \; \; {\rm M^{-\ell-1}\,s^{-1}}\notag\\
k_3 & \; \; : \; \; {\rm M^{-j-1}\,s^{-1}}\notag\\
k_4 & \; \; : \; \; {\rm M^{-1}\,s^{-1}}\notag
\end{align}
As long as species are in equilibrium with their dissociation products and the activity of a species is approximately equal to its concentration, the following calculations themselves will not be affected if \ce{SO3^{2-}} is replaced by \ce{HSO3^-}, or if \ce{CO2} is replaced by \ce{HCO3-}. To see this, consider that:
\begin{align}
K_{a,{\rm CO2}} &= \dfrac{[\ce{HCO3-}][\ce{H+}]}{[\ce{CO2}][\ce{H2O}]},\notag\\
K_{a,{\rm SO3}} &= \dfrac{[\ce{SO3^{2-}}][\ce{H+}]}{[\ce{HSO3-}]};\notag
\end{align}
where $[\ce{X}]$ represents the activity of the given species \ce{X}. Since $[\ce{H+}] = 10^{\rm-pH}$, in any rate equation where the concentration of \ce{HCO3^-} or \ce{HSO3-} appears, it could be replaced by $10^{\rm-pH} \, n_c / K_{a,{\rm CO2}}$ or $10^{\rm-pH} \, n_s / K_{a,{\rm SO3}}$, respectively. For both mechanisms we consider, Reactions (\ref{eqn:detach}), (\ref{eqn:reform}) and (\ref{eqn:adduct}) are as written. For Reaction (\ref{eqn:formate}) given Mechanism 1, $j=1$, $\ce{SO3^{2-}} \rightarrow \ce{HSO3-}$, and $\ce{CO2^-} \cdot j \, \ce{SO3^{2-}} \rightarrow \ce{HCO2-} + \ce{SO3-}$. For Reaction (\ref{eqn:formate}) given Mechanism 2, $j=1$, $\ce{CO2} \rightarrow \ce{HCO3-}$, $\ce{SO3^{2-}} \rightarrow \ce{HSO3-}$, and $\ce{CO2^-} \cdot j \, \ce{SO3^{2-}} \rightarrow \ce{HCO2-} + \ce{OH-} + \ce{SO3-}$. For our two mechanisms, $j$ is always equal to unity, but there may well be other mechanisms not considered in this paper, for which $j \ne 1$.\\

Note the inclusion of $\ell \, \ce{SO3^{2-}}$ on both the right-hand and left-hand sides of Reaction (\ref{eqn:reform}). This is to represent the possible role of ion clustering on the recombination rate, and therefore the hypothetical dependence of $\ce{SO3^{2-}} + e^-$ recombination on the concentration of \ce{SO3^{2-}}. The order of that dependence is, for now, treated as a free parameter, $\ell \ge 0$.\\

We can now write out the rate equations for our reactions. We consider the rate equation for the concentration of solvated electrons:
\begin{align}
\dfrac{dn_e}{dt} &= k_1 \, n_s - k_2 n_s^{\ell} n_{s'} n_e - k_3 n_s^j n_c n_e, \label{eqn:electron-rate} \tag{E1}\\
\dfrac{dn_f}{dt} &= k_3 n_s^j n_c n_e. \label{eqn:formate-rate} \tag{E2}
\end{align}
Here we assume that we are at an early stage of the reaction, and so can treat $n_s$ and $n_c$ as constant. We also assume that the $\cdot \ce{SO3-}$ radical concentration is approximately equal to the concentration of solvated electrons, and so $n_{s'} = n_e$. When the rate of recombination with $e^-$ and $\cdot \ce{SO3-}$ dominates the destruction of $e^-$, $n_{s'} \approx n_e$, because the dominant source and sink of electrons is a source and sink of equal amounts of $\cdot \ce{SO3-}$. When the rate of recombination with $e^-$ and $\cdot \ce{SO3-}$ does not dominate, we will see that the calculation becomes independent of $n_{s'}$. Finally, we assume that the concentration of solvated electrons is in steady state, this is reasonable given the expected short lifetime of solvated electrons.\\

Given these assumptions, we can now set Equation (\ref{eqn:electron-rate}) equal to zero, giving:
\begin{equation}
k_2n_s^{\ell}n_e^2 + k_3 n_s^j n_c n_e - k_1 n_s = 0,
\label{eqn:E1-rewritten} \tag{E3}
\end{equation}
with a solution of the form:
\begin{equation}
n_e = \sqrt{A^2 + B^2} - A, \label{eqn:quadratic-e} \tag{E4}
\end{equation}
where:
\begin{align}
A &= \dfrac{k_3 n_c}{2k_2} \, n_s^{j - \ell},\notag\\[12pt]
B &= \sqrt{\dfrac{k_1}{k_2}}n_s^{(1-\ell)/2}.\notag
\end{align}
The natural limits for Equation (\ref{eqn:quadratic-e}) are $A \ll B$ and $A \gg B$. Both $A$ and $B$ have units of M. It is instructive to look first at a comparison of $A n_s$ and $B^2$, given that $A \ll B$ implies $A n_s \ll B^2$. We see that $A n_s \ll B^2$ when
\begin{equation}
k_3 n_c n_s \ll k_1 n_s^{1-j} n_w.\notag
\end{equation}
Both of the mechanisms we consider have $j = 1$, so in that case:
\begin{equation}
k_3 n_c n_s \ll k_1.\notag
\end{equation}
This is a relation between timescales. The timescale $\tau_f \equiv 1/(k_3 n_c n_s)$ is the average time it takes for a solvated electron to react with dissolved carbon dioxide/bicarbonate and bisulfite to make formate. The timescale $\tau_s \equiv 1/k_1$ is the average time it takes for an electron to photodetach from a given sulfite anion. $A \ll B$ is therefore consistent with $\tau_f \gg \tau_s$, the timescale of formate production from a given solvated electron being much longer than the timescale of sulfite photodetachment. $A \gg B$, on the other hand, is consistent with $\tau_f \ll \tau_s$, and here the timescale of sulfite photodetachment is much longer than the timescale of formate production from a given solvated electron.\\

When $A \ll B$, Equation (\ref{eqn:quadratic-e}) can be represented by the Taylor series:
\begin{equation}
n_e = B - A + \mathcal{O}\Bigg(\dfrac{A^2}{B}\Bigg) = \sqrt{\dfrac{k_1}{k_2}} \, n_s^{(1-\ell)/2} + \dotsm
\label{eqn:taylor-series} \tag{E5}
\end{equation}
And the production rate of formate is:
\begin{equation}
\dfrac{dn_f}{dt} = k_3 n_s^j n_c n_e = k_3 \sqrt{\dfrac{k_1}{k_2}} \, n_c n_s^{-1/2-\ell/2 + j} + \dotsm
\label{eqn:rate-formate-production-taylor} \tag{E6}
\end{equation}
When $A \gg B$, Equation (\ref{eqn:quadratic-e}) can be represented by the Laurent series:
\begin{equation}
n_e = \dfrac{B^2}{2A} + \mathcal{O}\Bigg(\dfrac{B^4}{A^3}\Bigg) = \dfrac{k_1}{k_3n_c}n_s^{1-j} + \dotsm
\label{eqn:laurent-series} \tag{E7}
\end{equation}
And the production rate of formate is:
\begin{equation}
\dfrac{dn_f}{dt} = k_3 n_s^j n_c n_e = k_3 n_s^j n_c \dfrac{k_1}{k_3 n_c}n_s^{1-j} + \dotsm = k_1 n_s + \dotsm
\label{eqn:rate-formate-production-laurent} \tag{E8}
\end{equation}
This will provide us with the expected functional dependence of the rate on sulfite concentration once we determine the functional dependence of $k_1$ on $n_s$.\\

\begin{figure}[H]
\centering
\includegraphics[width=0.7\textwidth,trim={2cm 4.5cm 2cm 4cm},clip]{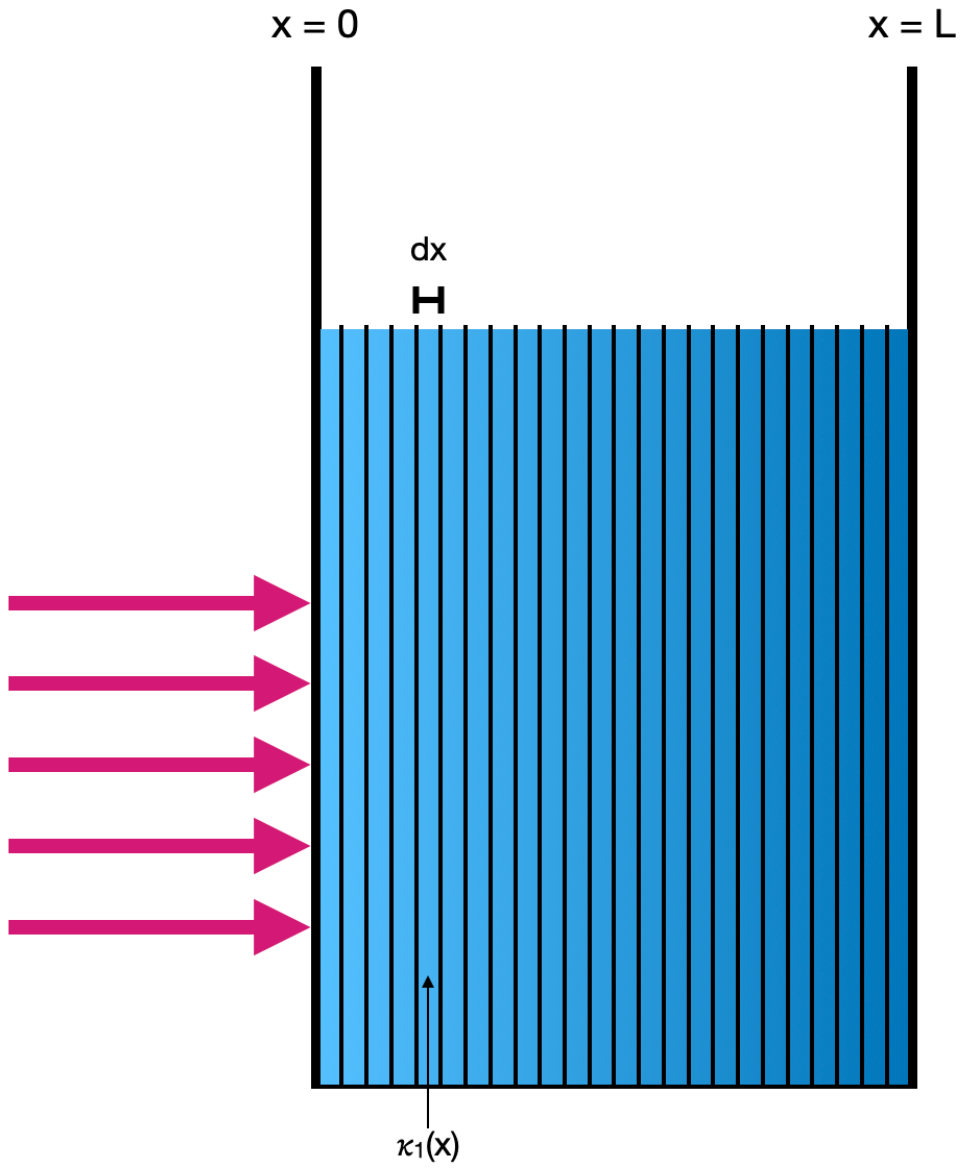}
\caption{``Infinite slab'' representation of UV light (pink arrows) impinging on a cuvette of width $L$, with the rate constant, $\kappa(x)$, for a photochemical reaction at a segment of infinitesimal width $dx$ at $0 \leq x leq L$.}\label{fig:radtrans}
\end{figure}

The rate constant $k_1$ is an average of the rate constants for slices of water through the cuvette. See Fig. \ref{fig:radtrans}. At each depth, $x$ (cm), we have a rate constant $\kappa(x)$ (s$^{-1}$) equal to:
\begin{equation}
\kappa(x) = A_G \int_{0}^{\infty} \phi(\lambda) \, \sigma(\lambda) \, F(\lambda) \, e^{-N_a n_s \sigma(\lambda) x} \, d\lambda.
\label{eqn:rate-constant} \tag{E9}
\end{equation}
where $\lambda$ (nm) is wavelength, $A_G$ is a unitless geometric factor, $\phi(\lambda)$ is the quantum yield of Reaction (\ref{eqn:detach}), $\sigma(\lambda)$ (cm$^2$) is the absorption cross-section, $F(\lambda)$ (cm$^{-2}$ s$^{-1}$ nm$^{-1}$) is the actinic flux, $N_a = 6.022 \times 10^{23}$ is Avagardo's number and $n_s$ (M) is the concentration of sulfite.

We will define:
\begin{equation}
P \equiv A_G \int_{0}^{\infty} \phi(\lambda) \, F(\lambda) \, d\lambda,
\label{eqn:P} \tag{E10}
\end{equation}
and note that:
\begin{equation}
\kappa(0) = A_G \int_{0}^{\infty} \phi(\lambda) \, \sigma(\lambda) \, F(\lambda) \, d\lambda.
\label{eqn:rate-constant0} \tag{E11}
\end{equation}

\begin{figure}[H]
\centering
\includegraphics[width=0.9\textwidth]{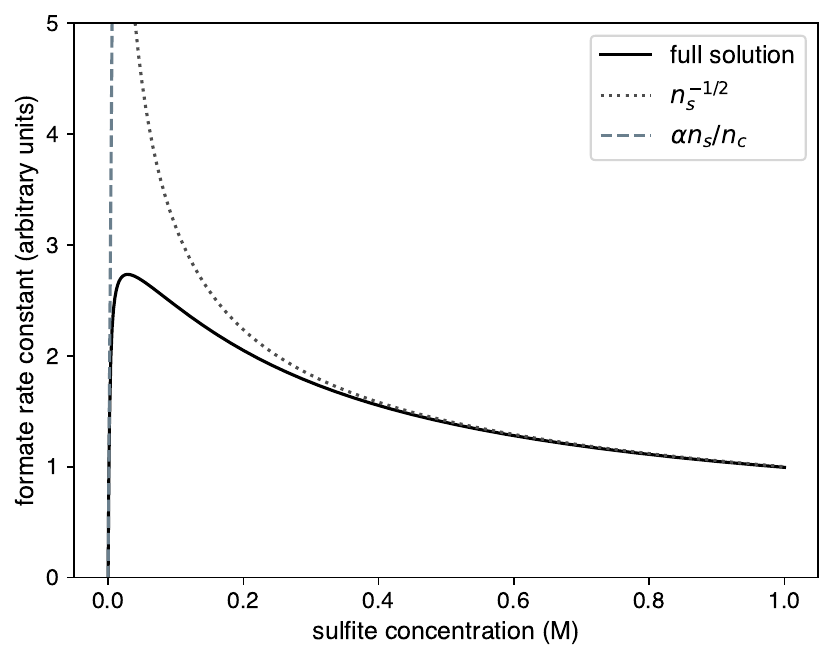}
\caption{The rate constant for the production of formate from sulfite and carbon species (in arbitrary units) as a function of sulfite concentration (M) is represented by a solid black line. The linear approximation at low sulfite concentrations is shown as a dashed slate line and the $1/n_s^{1/2}$ approximation at high concentrations is dipicted as a dotted dark grey line. 
\label{fig:sulfite-order}}
\end{figure}

We now approximate $k_1$ as the average of $\kappa(x)$ over $x$ from $0$ to $L$. This means: 
\begin{equation}
k_1 \approx \dfrac{\int_{0}^{L}\kappa(x)\, dx}{\int_{0}^{L} dx} = \dfrac{A_G}{N_a n_s L}\Bigg(\int_0^{\infty} \phi(\lambda) \, F(\lambda) \, d\lambda - \int_0^{\infty} \phi(\lambda) \, F(\lambda) \, e^{-N_a n_s \sigma(\lambda) L} \, d\lambda \Bigg). \label{eqn:photo-constant} \tag{E12}
\end{equation}
In the limit where the optical depth of the sample is small, $n_s \ll 1/\big(N_a \sigma(\lambda) L\big)$, Equation (\ref{eqn:photo-constant}) can be Taylor expanded in $n_s$ to:
\begin{align}
 k_1 &= \dfrac{A_G}{N_a n_s L}\Bigg(\int_0^{\infty} \phi(\lambda) \, F(\lambda) \, d\lambda - \int_0^{\infty} \phi(\lambda) \, F(\lambda) \, d\lambda + N_a n_s L \int_0^{\infty} \phi(\lambda) \, \sigma(\lambda) \, F(\lambda) \, d\lambda \Bigg), \notag\\
 &= A_G \int_0^{\infty} \phi(\lambda) \, \sigma(\lambda) \, F(\lambda) \, d\lambda, \notag \\
 & = \kappa(0).\notag
 \label{eqn:E12-taylor expansion} \tag{E13}
\end{align}
In the limit where the optical depth is large, $n_s \gg 1/\big(N_a \sigma(\lambda) L\big)$, and the last term on the R.H.S. of Equation (\ref{eqn:photo-constant}) can be set to zero,
\begin{align}
 k_1 &= \dfrac{A_G}{N_a n_s L}\int_0^{\infty} \phi(\lambda) \, F(\lambda) \, d\lambda \notag \\
 & = \dfrac{P}{N_a n_s L}.\notag
\end{align}
We have found that, when $n_s \gg 1/\big(N_a \sigma(\lambda) L\big)$, $k_1 \propto 1/n_s$ and therefore $\tau_s \propto n_s$. We also know that $\tau_f \propto 1/n_s$, and therefore when the solution has enough $n_s$ to be optically thick, $A \gg B$, and when $n_s$ is small enough that the solution is optically thin, $A \ll B$. We can now combine the two limits we have found. In the optically thin case:
\begin{equation}
\dfrac{dn_f}{dt} \approx \kappa(0) \, n_s,
\label{eqn:formate-opticallythin} \tag{E14}
\end{equation}
and in the optically thick case:
\begin{equation}
\dfrac{dn_f}{dt} \approx k_3 \sqrt{\dfrac{P}{k_2 N_a L}}n_c \, n_s^{j - \ell/2 - 1}.
\label{eqn:formate-opticallythick} \tag{E15}
\end{equation}
We find that, if the recombination rate, for Reaction (\ref{eqn:reform}), depends linearly on the sulfite concentration, then $j = \ell = 1$ and:
\begin{equation}
\dfrac{dn_f}{dt} \approx k_3 \sqrt{\dfrac{P}{k_2 N_a L}}n_c \, n_s^{-1/2}.
\label{eqn:formate-general} \tag{E15}
\end{equation}
It is important to note that we do not know (\textit{a priori}) what value to set for $\ell$. This value needs to be informed by either experiments or quantum chemical models. In our case, the measurements are consistent, within experimental error, with $n_s^{-1/2}$, and so the value of $\ell = 1$ is supported by our rate law measurements. This functional dependence, within these two regimes, is shown in Figure \ref{fig:sulfite-order}.

\section{Defining the Rate Equations for the Production of Formate}
\addcontentsline{toc}{section}{Defining the Rate Equations for the Production of Formate}

In this section we outline two potential mechanisms, Mechanism 1 and Mechanism 2, that we use to explain the reduction of carbon species to formate. We derive rate laws based on each of these mechanisms. We begin by outlining the reactions relevant to our system:

\begin{align}
\ch{CO2 + H2O &<=> HCO3- + H+ <=> 2 H+ + CO3^{2-}} \label{eqn:equHCO3-} \tag{R1} \\
\ch{HSO3- &<=> H+ + SO3^{2-}} \label{eqn:equHSO3-} \tag{R2} \\
\ch{SO3^{2-} + h$\nu$ &-> .SO3- + e-} \label{eqn2:detach} \tag{R3} \\
\ch{.SO3- + .SO3- &-> S2O6^{2-}} \label{eqn2:adduct} \tag{R4} \\
\ch{CO2 + e- &-> .CO2-} \label{eqn:crad} \tag{R5} \\
\ch{.CO2- + HSO3- &-> HCO2- + .SO3^{-}} \label{eqn:formatemech1} \tag{R6} \\
\ch{HCO3- + e- &-> HCO3^{2-}} \label{eqn:red1} \tag{R7} \\
\ch{HCO3^{2-} + HSO3- &-> HCO2- + OH- + .SO3-} \label{eqn:red2} \tag{R8}
\end{align}

Reactions (\ref{eqn2:detach}) - (\ref{eqn:red2}) have rate constants $k_3$ to $k_8$. We do not assign rate constants to the equilibrium speciation reactions denoted by (\ref{eqn:equHCO3-}) and (\ref{eqn:equHSO3-}). We treat these reactions as being sufficiently fast to maintain equilibrium concentrations of \ch{CO2}, \ch{HCO3-}, \ch{CO3^{2-}}, \ch{HSO3-} and \ch{SO3^{2-}}.

\subsection{Mechanism 1}

Mechanism 1 can be described by Reactions (\ref{eqn:crad}) and (\ref{eqn:formatemech1}) with the electrons being provided by (\ref{eqn2:detach}).

\subsubsection{Rate Law 1}

\paragraph{Step 1:}

\begin{align}
    \ch{CO2 + e- &-> .CO2-} \tag{R5}
\end{align}

\begin{align}
    \frac{d[\cdot\text{CO}_2^-]}{dt} = k_5 [\text{CO}_2] [e^-] \label{eqn:E1} \tag{E1}
\end{align}

\paragraph{Step 2:}

\begin{align}
    \ch{.CO2- + HSO3- &-> HCO2- + .SO3^{-}} \tag{R6} 
\end{align}

\begin{align}
    \frac{d[\text{HCO}_2^-]}{dt} = k_6 [\cdot\text{CO}_2^-] [\text{HSO}_3^-] \label{eqn:2} \tag{E2}
\end{align}

\paragraph{Step 3:}

\begin{align}
    \ch{SO3^{2-} + h$\nu$ &-> .SO3- + e-} \tag{R3}
\end{align}

\begin{align}
    \frac{d[\cdot\text{SO}_3^-]}{dt} = k_3 [\text{SO}_3^{2-}] \label{eqn:3} \tag{E3}
\end{align}

\paragraph{Step 4:}

\begin{align}
    \ch{.SO3- + .SO3- &-> S2O6^{2-}} \tag{R4}
\end{align}

\begin{align}
    \frac{d[\text{S}_2\text{O}_6^{2-}]}{dt} = k_4 [\cdot\text{SO}_3^-]^2 \label{eqn:4} \tag{E4}
\end{align}

We are interested in the rate of change of \(\text{HCO}_2^-\) concentration, \(\frac{d[\text{HCO}_2^-]}{dt}\). From the reaction mechanism we see that \(\text{HCO}_2^-\) is produced in Reaction (\ref{eqn:formatemech1}). Therefore, the rate of production of \(\text{HCO}_2^-\) is given by the rate of Reaction (\ref{eqn:formatemech1}) as shown in the Rate Equation \ref{eqn:2}.\\

If we assume that the intermediate \(\cdot\text{CO}_2^-\) reaches a steady state:

\begin{align}
    \frac{d[\cdot\text{CO}_2^-]}{dt} \approx 0 \label{eqn:5} \tag{E5}
\end{align}

From step 1 and step 2, we have:

\begin{align}
    k_5 [\text{CO}_2] [e^-] = k_6 [\cdot\text{CO}_2^-] [\text{HSO}_3^-] \label{eqn:6} \tag{E6}
\end{align}

Solving for \([\cdot\text{CO}_2^-]\):

\begin{align}
    [\cdot\text{CO}_2^-] = \frac{k_5 [\text{CO}_2] [e^-]}{k_6 [\text{HSO}_3^-]} \label{eqn:7} \tag{E7}
\end{align}

Substituting the expression for \([\cdot\text{CO}_2^-]\) into the rate law for \(\frac{d[\text{HCO}_2^-]}{dt}\):

\begin{align}
    \frac{d[\text{HCO}_2^-]}{dt} = k_6 \left( \frac{k_5 [\text{CO}_2] [e^-]}{k_6 [\text{HSO}_3^-]} \right) [\text{HSO}_3^-] \label{eqn:8} \tag{E8}
\end{align}

Therefore, the rate of change of the concentration of \(\text{HCO}_2^-\) is:

\begin{align}
    \frac{d[\text{HCO}_2^-]}{dt} = k_5 [\text{CO}_2] [e^-] \label{eqn:9} \tag{E9}
\end{align}

We can replace [\ch{e-}] with [\ch{SO3^{2-}}] as  [\ch{e-}] depends on [\ch{SO3^{2-}}] with the expected relation such that: 

\begin{align}
    \frac{d[\ch{HCO2^-}]}{dt} \sim [\ch{SO3^{2-}}]^{-0.5} \label{eqn:10} \tag{E10}
\end{align}

 This expected relation comes from our analytical model for electron photodetachment, as described in further detail from page $S44$ to page $S52$ of the Supporting Information.\\

Therefore, the rate of change of the concentration of \(\text{HCO}_2^-\) is:

\begin{align}
    \frac{d[\text{HCO}_2^-]}{dt} = k_{eff} [\text{CO}_2]^{0.71 \pm 0.12} [\ch{SO3^{2-}}]^{-0.60 \pm 0.10} \label{eqn:11} \tag{E11}
\end{align}

where the order of reaction with respect to each reactant was determined experimentally.

\subsection{Mechanism 2}

Mechanism 2 can be described by Reactions (\ref{eqn:red1}) and (\ref{eqn:red2}) with the electrons being provided by (\ref{eqn2:detach}).

\subsubsection{Rate Law 2}

\paragraph{Step 1:}

\begin{align}
    \ch{HCO3- + e- &-> HCO3^{2-}} \tag{R7}
\end{align}

\begin{align}
\frac{d[\text{HCO}_3^{2-}]}{dt} = k_7 [\text{HCO}_3^-][e^-] \label{eqn:12} \tag{E12}
\end{align}

\paragraph{Step 2:}

\begin{align}
    \ch{HCO3^{2-} + HSO3- &-> HCO2- + OH- + .SO3-} \tag{R8}
\end{align}

\begin{align}
    \frac{d[\text{HCO}_2^{-}]}{dt} = k_8 [\text{HCO}_3^{2-}] [\text{HSO}_3^-] \label{eqn:13} \tag{E13}
\end{align}

In this mechanism, steps 3 and 4 are identical to those in Mechanism 1 (refer to Mechanism 1 for a detailed description of these steps).\\

Once again, we are interested in the rate of change of \(\text{HCO}_2^-\) concentration, \(\frac{d[\text{HCO}_2^-]}{dt}\). From the reaction mechanism, we see that \(\text{HCO}_2^-\) is produced in Reaction (\ref{eqn:red2}). Therefore, the rate of production of \(\text{HCO}_2^-\) is given by the Rate Equation \ref{eqn:13}.\\

To derive the overall rate law, we use the steady-state approximation for the intermediate \(\text{HCO}_3^{2-}\) where the rate of formation of \(\text{HCO}_3^{2-}\) is equal to the rate of consumption of \(\text{HCO}_3^{2-}\).\\

At steady state:
\begin{align}
    k_7 [\text{HCO}_3^-] [e^-] = k_8 [\text{HCO}_3^{2-}] [\text{HSO}_3^-] \label{eqn:14} \tag{E14}
\end{align}

Solving for \([ \text{HCO}_3^{2-} ]\):
\begin{align}
[ \text{HCO}_3^{2-} ] = \frac{k_7 [ \text{HCO}_3^- ] [ e^- ]}{k_8 [ \text{HSO}_3^- ]} \label{eqn:15} \tag{E15}
\end{align}

Thus, the overall rate of production of \(\text{HCO}_2^-\) is given by:
\begin{align}
\frac{d[\text{HCO}_2^{-}]}{dt} = k_8 [ \text{HCO}_3^{2-} ] [ \text{HSO}_3^- ] \label{eqn:16} \tag{E16}
\end{align}

Substituting \([ \text{HCO}_3^{2-} ]\) into the expression for $\frac{d[\text{HCO}_2^{-}]}{dt}$:

\begin{align}
\frac{d[\text{HCO}_2^{-}]}{dt} = k_8 \left( \frac{k_7 [ \text{HCO}_3^- ] [ e^- ]}{k_8 [ \text{HSO}_3^- ]} \right) [ \text{HSO}_3^- ] \label{eqn:17} \tag{E17}
\end{align}

Thus, the overall rate law for the production of \(\text{HCO}_2^-\) is:
\begin{align}
\frac{d[\text{HCO}_2^{-}]}{dt} = k_7 [ \text{HCO}_3^- ] [ e^- ] \label{eqn:18} \tag{E18}
\end{align}

Once again, we can replace [\ch{e-}] with [\ch{SO3^{2-}}] as  [\ch{e-}] depends on [\ch{SO3^{2-}}] with the expected relation such that: 

\begin{align}
\frac{d[\ch{HCO2^-}]}{dt} \sim [\ch{SO3^{2-}}]^{-0.5} \label{eqn:19} \tag{E19}
\end{align}

This expected relation comes from our analytical model for electron photodetachment (see pages $S44$ to $S52$ of the Supporting Information for further details).\\

Therefore, the rate of change of the concentration of \(\text{HCO}_2^-\) is:
\begin{align}
\frac{d[\text{HCO}_2^-]}{dt} = k_{eff} [\text{HCO}_3^-]^{0.71 \pm 0.12} [\ch{SO3^{2-}}]^{-0.60 \pm 0.10} \label{eqn:20} \tag{E20}
\end{align}

where the order of reaction with respect to each reactant was determined experimentally.

\section{References}

(1) Adams, R. W.; Holroyd, C. M.; Aguilar, J. A.; Nilsson, M.; Morris, G. A. “Perfecting”
WATERGATE: clean proton NMR spectra from aqueous solution. \emph{Chemical communi-
cations} \textbf{2013}, \emph{49}, 358–360.